\documentclass[lettersize,journal]{IEEEtran}
\usepackage{amsmath,amsfonts}
\usepackage{algorithmic}
\usepackage{algorithm}
\usepackage{array}
\usepackage[caption=false,font=normalsize,labelfont=sf,textfont=sf]{subfig}
\usepackage{textcomp}
\usepackage{stfloats}
\usepackage{placeins}
\usepackage{makecell}
\usepackage{xurl}
\usepackage{longtable}
\usepackage{ragged2e}
\usepackage{verbatim}
\usepackage{graphicx}
\usepackage{cite}
\usepackage{multirow}
\usepackage{adjustbox}
\usepackage{tikz}
\usepackage{forest}

\usepackage{xspace}
\usepackage{tcolorbox}
\useforestlibrary{edges}
\usetikzlibrary{arrows.meta, positioning, shapes, fit, calc}
\usepackage{booktabs}

\usepackage[table]{xcolor}
\definecolor{linkblue}{HTML}{003366}

\usepackage[
    colorlinks=true,
    linkcolor=linkblue,
    citecolor=linkblue,
    urlcolor=linkblue
]{hyperref}
\usepackage{cleveref}

\crefname{figure}{fig.}{figs.}
\crefname{table}{tab.}{tabs.}
\crefname{section}{sec.}{secs.}
\crefname{subsection}{sec.}{secs.}
\crefname{subsubsection}{sec.}{secs.}
\crefname{equation}{eq.}{eqs.}

\Crefname{figure}{Figure}{Figures}
\Crefname{table}{Table}{Tables}
\Crefname{section}{Section}{Sections}
\Crefname{subsection}{Section}{Sections}
\Crefname{subsubsection}{Section}{Sections}
\Crefname{equation}{Equation}{Equations}

\newcolumntype{L}[1]{>{\raggedright\arraybackslash}p{#1}}

\newcolumntype{L}[1]{>{\RaggedRight\arraybackslash}p{#1}}

\definecolor{hidden-draw}{RGB}{0,0,0}
\definecolor{hidden-blue}{RGB}{194,232,247}
\definecolor{hidden-orange}{RGB}{243,202,120}
\definecolor{hidden-yellow}{RGB}{242,244,193}
\definecolor{hidden-red}{RGB}{205, 44, 36}
\definecolor{hidden-blue}{RGB}{194,232,247}
\definecolor{hidden-orange}{RGB}{243,202,120}
\definecolor{hidden-green}{RGB}{34,139,34}
\definecolor{hidden-pink}{RGB}{255,245,247}
\definecolor{hidden-black}{RGB}{20,68,106}
\definecolor{purple}{RGB}{144,153,196}
\definecolor{yellow}{RGB}{255,228,123}
\definecolor{hidden-yellow}{RGB}{255,248,203}
\definecolor{tkcolor}{RGB}{224,223,255}
\definecolor{darkblue}{rgb}{0, 0.40, 0.75}
\definecolor{tree-level-1}{RGB}{245,20,85}
\definecolor{tree-level-2}{RGB}{246,86,118}
\definecolor{tree-level-3}{RGB}{248,177,193}
\definecolor{tree-leaf}{RGB}{176,230,198}

\tcbset{
  takeawaysbox/.style={
    colback=lightblue!80,
    colframe=black,
    fonttitle=\bfseries\small,
    coltitle=white,
    colbacktitle=black,
    enhanced,
    attach boxed title to top left={xshift=2.5mm,yshift=-2.5mm},
    boxed title style={rounded corners, size=small, colframe=black, colback=black},
    width=\linewidth,
    arc=3.5mm
  }
}

\begin{document}

\title{Toward Multimodal Emotion Recognition and Generation with \\[-2pt]Large Language Models}

\author{Daiqing Wu, Hongrui Zhang, Yuman Wang, Siyuan Liu, Hangshek Lau, Shuo Lu, Guiguang Ding, \IEEEmembership{Senior Member, IEEE}, Björn Schuller, \IEEEmembership{Fellow, IEEE}, Sicheng Zhao, \IEEEmembership{Senior Member, IEEE}

\IEEEcompsocitemizethanks{\IEEEcompsocthanksitem D. Wu, S. Liu, H. Lau, S. Lu and S. Zhao are with the Department of Psychological and Cognitive Sciences, Tsinghua University, Beijing 100084, China. (e-mail: wudaiqing@mail.tsinghua.edu.cn, liusiyuan26@mails.tsinghua.edu.cn, liuhs26@mails.tsinghua.edu.cn, shuolucs@gmail.com, schzhao@tsinghua.edu.cn).\protect
\IEEEcompsocthanksitem H. Zhang is with the School of Data Science, The Chinese University of Hong Kong, Shenzhen, Shenzhen 518172, China (e-mail: smilingweeping@gmail.com).\protect
\IEEEcompsocthanksitem Y. Wang is with the School of Computer Science and Engineering, University of Electronic Science and Technology of China, Chengdu 611731, China (e-mail: yuman.mail16@gmail.com).\protect
\IEEEcompsocthanksitem G. Ding is with the School of Software, Tsinghua University, Beijing 100084, China (e-mail: dinggg@tsinghua.edu.cn).\protect
\IEEEcompsocthanksitem B. Schuller is with the Chair of Health Informatics, Technical University of Munich, Munich 81675, Germany, and also with the Department of Computing, Imperial College London, London SW7 2AZ, U.K. (e-mail: schuller@ieee.org).
}}

% \markboth{Toward Multimodal Emotion Recognition and Generation with Large Language Models}%
% {Anonymous Authors}

% ldq\IEEEpubid{0000-0000/00\$00.00~\copyright~2021 IEEE}

\maketitle

\begin{abstract}
Multimodal affective computing aims to understand and generate human affect through language, voice, facial expressions, and other behavioral signals. This long-standing research field is undergoing a profound paradigm shift as Large Language Models (LLMs) bring extensive world knowledge, flexible instruction following, and powerful reasoning capabilities to affective modeling, fueling rapidly growing research on recognition and generation. This survey organizes these developments into two overarching paradigms by consolidating related tasks around affective understanding and expression. We adopt a challenge-oriented perspective centered on \textit{Subjective and Contextual Variability}, the \textit{Multimodal Affective Gap}, and \textit{Affective Process Opacity}. Within this framework, methods are grouped by the challenges they address and the mechanisms they employ, revealing methodological connections across task boundaries. The review examines shared design principles and methodological differences within and across categories, alongside the data resources supporting their development. A field-wide analysis places these approaches in a broader context, examining temporal trends, task coverage, and the relationship between recognition and generation. Through this synthesis, we aim to clarify an increasingly diverse research landscape and help researchers navigate existing approaches, assess their connections, and identify opportunities for further investigation.
\end{abstract}


\section{Introduction}
\label{sec:1}

\IEEEPARstart{A}{s} Marvin Minsky observed, ``\textit{The question is not whether intelligent machines can have any emotions, but whether machines can be intelligent without any emotions}''~\cite{minsky86mind}. This perspective underscores the central role of emotion in human intelligence, shaping how we interpret events, make decisions, and interact with others~\cite{lerner15emo}. These capabilities unfold in a multimodal world, where emotional perception, understanding, and expression draw on the interplay of language, voice, facial expressions, bodily behavior, and physiological signals~\cite{poria17review}. Multimodal affective computing seeks to bring such capabilities to machines, enabling them to infer emotions from heterogeneous signals and express affect through multiple modalities~\cite{picard97affect}. Situated at the intersection of human cognition and computational intelligence, this field combines fundamental questions about emotion with opportunities for more responsive and human-centered technologies. Its scientific breadth, technical depth, and enduring challenges have sustained research across successive generations.

\begin{figure*}[t]
    \centering
    %\vskip -0.04in
    \includegraphics[width=1\linewidth]{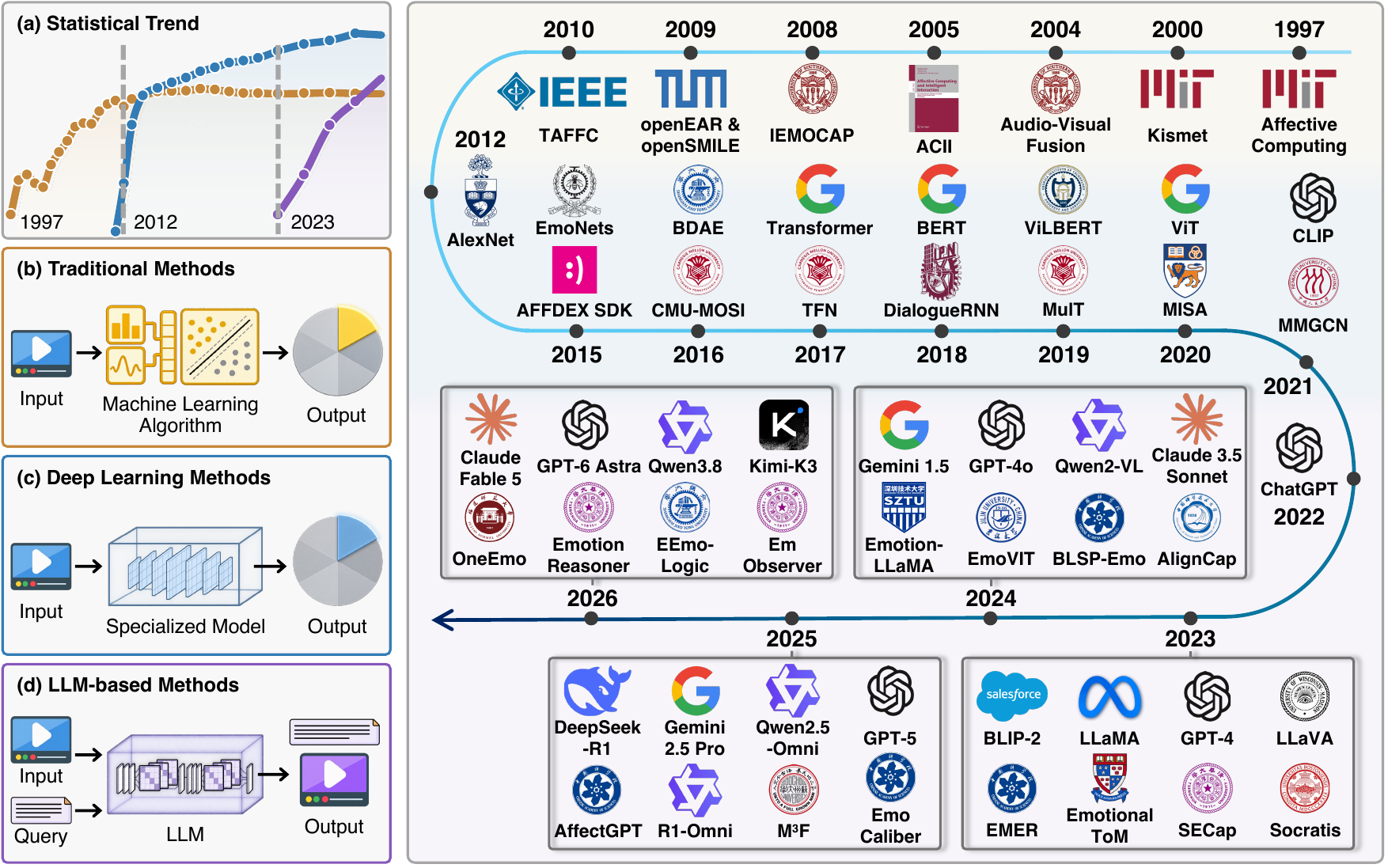}
    \vskip -0.1in
    \caption{Development of multimodal affective computing over the past two decades. On the left, panel (a) illustrates publication trends for traditional (yellow, b), deep learning (blue, c), and LLM-based methods (purple, d). Each marker indicates an approximate annual publication count informed by keyword-based searches in Scopus\protect\footnotemark, plotted on a logarithmic vertical scale with a nonuniform time axis. The values for 2026 are projected from preceding trends. The timeline on the right highlights selected milestones and representative studies, accompanied by logos of their principal contributing institutions or organizations.}
    \label{fig:1-1}
    \vskip -0.02in
\end{figure*}

The evolution of multimodal affective computing mirrors broader advances in Artificial Intelligence (AI), progressing from traditional machine learning to deep learning and, more recently, approaches based on Large Language Models (LLMs), as reflected in \cref{fig:1-1} (a). At the outset of this progression, traditional methods extracted handcrafted features from individual modalities and combined them through feature- or decision-level fusion to predict emotion categories (\cref{fig:1-1} (b))~\cite{busso04fusion,calvo10affect}. Although domain-informed feature design supported relatively compact and interpretable models, their capacity to represent subtle affective variations and complex cross-modal relationships remained limited. As deep learning matured and gained widespread adoption, the field increasingly shifted toward data-driven representation learning. These methods used neural encoders to transform modality-specific inputs into latent representations, which were integrated through trainable fusion layers and mapped to predictions under task-specific supervision (\cref{fig:1-1} (c))~\cite{zadeh17tfn}. Despite gains in representational capacity and predictive performance, many shared a limitation of traditional approaches: their supported inputs and prediction formats remained closely tied to their training configurations~\cite{zhang26merllm}. Consequently, stronger benchmark performance did not necessarily translate into flexible, context-aware, and broadly generalizable affective intelligence.

The emergence of LLMs has opened a path beyond these task-specific constraints. Generative pre-training on web-scale data equips these models with extensive knowledge, while subsequent instruction tuning and alignment strengthen their ability to follow instructions and perform diverse tasks autoregressively using shared parameters~\cite{brown20gpt3,ouyang22instru}. Initially developed for text understanding and generation, LLMs have rapidly evolved into Multimodal Large Language Models (MLLMs) through the integration of modality-specific encoders~\cite{liu23llava}, with additional generation components enabling Unified Multimodal Models (UMMs) that support both multimodal understanding and generation~\cite{wu24nextgpt}. With appropriate prompting or adaptation, LLM-based methods have achieved performance competitive with specialized affective models while supporting a broader range of tasks and output formats~\cite{xie24emovit,wang24blspemo}. These capabilities bring affective computing closer to real-world needs by extending its scope to nuanced emotional interpretation~\cite{lian23emer}, contextual reasoning~\cite{zhao25r1omni}, and affective content generation~\cite{wang24blspemo}. Such opportunities have fueled a surge in LLM-based research on multimodal affective computing (\cref{fig:1-1} (a)). To harness this growing momentum, this paper systematically reviews recent developments, organizes the field's technical approaches, and examines emerging trends, synthesizing current advances into a coherent perspective to inform and inspire future research toward more capable and broadly applicable multimodal affective intelligence.

\footnotetext{Scopus searches conducted on September 27, 2026 used task-specific terms such as ``facial expression recognition'', ``emotional speech synthesis'', etc., in titles, abstracts, and keywords. Deduplicated records were classified by methodological cues, with targeted manual checks.}

Specifically, we distinguish two complementary branches according to their technical objectives: Multimodal Emotion Recognition (MER), which infers affective states and meanings from multimodal evidence~\cite{busso04fusion}, and Multimodal Emotion Generation (MEG), which produces content that expresses or evokes intended emotions across modalities~\cite{wang20mead}. Following the MER-with-LLMs terminology established in prior work~\cite{zhang26merllm}, we refer to these branches as the MER-with-LLMs and MEG-with-LLMs paradigms, respectively. For the relatively few methods that integrate both, we examine their recognition mechanisms under the former paradigm and their coordination with generation under the latter. 

Within each paradigm, methods may share design principles despite differences in input and output modalities, task objectives, and evaluation protocols. To look beyond these task-specific differences and uncover deeper methodological insights, we adopt a challenge-oriented perspective, first identifying the field's core technical challenges and then examining how each paradigm addresses them. The remainder of this survey follows this perspective. \Cref{sec:2} establishes the conceptual background, including emotion models and task formulations. \Cref{sec:3} presents the core challenges and explains how they shape methodological directions. \Cref{sec:4} introduces the data resources underpinning these developments, while \cref{sec:5} and \cref{sec:6} review MER-with-LLMs and MEG-with-LLMs, respectively. \Cref{sec:7} synthesizes current trends and future research directions, and \cref{sec:8} concludes the survey with an overarching discussion.

\definecolor{surveyGroupGray}{RGB}{235,235,235}
\definecolor{surveyOursPurple}{RGB}{230,230,255}

\begin{table*}[t]
\centering
\caption{Overview of representative surveys on affective computing. I, A, V, P, and T denote image, audio, video, physiological signals, and text, respectively. U and M denote unimodal and multimodal content.}
\label{tab:0}
\vskip -0.1in
\begin{adjustbox}{max width=\textwidth}
\scriptsize
\renewcommand{\arraystretch}{1.15}
\setlength{\tabcolsep}{3pt}
\begin{tabular}{@{}llc*{5}{c}*{6}{>{\centering\arraybackslash}p{0.41in}}cc@{}}
\toprule
\multirow{2}{*}{\textbf{Survey}} & \multirow{2}{*}{\textbf{Venue}} & \multirow{2}{*}{\textbf{Year}} & \multicolumn{5}{c}{\textbf{Modality}} & \multicolumn{2}{c}{\textbf{Recognition}} & \multicolumn{4}{c}{\textbf{Generation}} & \multirow{2}{*}{\textbf{Datasets}} & \multirow{2}{*}{\textbf{Methods}} \\
\cmidrule(lr){4-8}\cmidrule(lr){9-10}\cmidrule(lr){11-14}
& & & \textbf{I} & \textbf{A} & \textbf{V} & \textbf{P} & \textbf{T} & \textbf{U$\rightarrow$Label} & \textbf{M$\rightarrow$Label} & \textbf{U$\rightarrow$U} & \textbf{U$\rightarrow$M} & \textbf{M$\rightarrow$U} & \textbf{M$\rightarrow$M} & & \\
\midrule
\rowcolor{surveyGroupGray}\multicolumn{16}{c}{\textbf{Traditional Machine Learning--Centered Surveys}} \\
Pang and Lee~\cite{pang08opinion} & Found. Trends Inf. Retr. & 2008 & -- & -- & -- & -- & $\checkmark$ & $\checkmark$ & -- & -- & -- & -- & -- & $\checkmark$ & $\checkmark$ \\
Zeng et al.~\cite{zeng09affect} & IEEE TPAMI & 2009 & $\checkmark$ & $\checkmark$ & $\checkmark$ & -- & $\checkmark$ & $\checkmark$ & $\checkmark$ & -- & -- & -- & -- & $\checkmark$ & $\checkmark$ \\
Calvo and D'Mello~\cite{calvo10affect} & IEEE TAFFC & 2010 & $\checkmark$ & $\checkmark$ & $\checkmark$ & $\checkmark$ & $\checkmark$ & $\checkmark$ & $\checkmark$ & -- & -- & -- & -- & -- & $\checkmark$ \\
D'Mello and Kory~\cite{dmello15multimodalreview} & ACM CSUR & 2015 & -- & $\checkmark$ & $\checkmark$ & $\checkmark$ & $\checkmark$ & -- & $\checkmark$ & -- & -- & -- & -- & -- & $\checkmark$ \\
Poria et al.~\cite{poria17review} & Information Fusion & 2017 & $\checkmark$ & $\checkmark$ & $\checkmark$ & $\checkmark$ & $\checkmark$ & $\checkmark$ & $\checkmark$ & -- & -- & -- & -- & $\checkmark$ & $\checkmark$ \\
\addlinespace[2pt]
\rowcolor{surveyGroupGray}\multicolumn{16}{c}{\textbf{Deep Learning--Centered Surveys}} \\
Zhao et al.~\cite{zhao_acmmm_tomm_2019} & ACM TOMM & 2019 & $\checkmark$ & $\checkmark$ & $\checkmark$ & $\checkmark$ & $\checkmark$ & $\checkmark$ & $\checkmark$ & -- & -- & -- & -- & $\checkmark$ & $\checkmark$ \\
Zhao et al.~\cite{zhao22aica} & IEEE TPAMI & 2022 & $\checkmark$ & -- & -- & -- & -- & $\checkmark$ & -- & -- & -- & -- & -- & $\checkmark$ & $\checkmark$ \\
Li and Deng~\cite{li22deepfer} & IEEE TAFFC & 2022 & $\checkmark$ & -- & $\checkmark$ & -- & -- & $\checkmark$ & -- & -- & -- & -- & -- & $\checkmark$ & $\checkmark$ \\
Zhao et al.~\cite{zhao_pieee_2023} & Proc. IEEE & 2023 & $\checkmark$ & $\checkmark$ & $\checkmark$ & $\checkmark$ & $\checkmark$ & $\checkmark$ & $\checkmark$ & -- & -- & -- & -- & -- & $\checkmark$ \\
Wang et al.~\cite{wang_pieee_2023} & Proc. IEEE & 2023 & $\checkmark$ & $\checkmark$ & $\checkmark$ & -- & $\checkmark$ & $\checkmark$ & $\checkmark$ & -- & -- & -- & -- & $\checkmark$ & $\checkmark$ \\
Gandhi et al.~\cite{gandhi2023inffus} & Information Fusion & 2023 & $\checkmark$ & $\checkmark$ & $\checkmark$ & -- & $\checkmark$ & -- & $\checkmark$ & -- & -- & -- & -- & $\checkmark$ & $\checkmark$ \\
Triantafyllopoulos et al.~\cite{triantafyllopoulos23affectivespeech} & Proc. IEEE & 2023 & -- & $\checkmark$ & -- & -- & $\checkmark$ & -- & -- & $\checkmark$ & -- & $\checkmark$ & -- & $\checkmark$ & $\checkmark$ \\
Raamkumar and Yang~\cite{raamkumar23empathetic} & IEEE TAFFC & 2023 & -- & -- & -- & -- & $\checkmark$ & -- & -- & $\checkmark$ & -- & -- & -- & $\checkmark$ & $\checkmark$ \\
Wu et al.~\cite{wu2025emnlp} & Findings of EMNLP & 2025 & -- & $\checkmark$ & $\checkmark$ & -- & $\checkmark$ & -- & $\checkmark$ & -- & -- & -- & -- & $\checkmark$ & $\checkmark$ \\
Ma et al.~\cite{Ma2025TAFFC} & IEEE TAFFC & 2025 & $\checkmark$ & $\checkmark$ & $\checkmark$ & -- & $\checkmark$ & -- & -- & $\checkmark$ & -- & $\checkmark$ & -- & $\checkmark$ & $\checkmark$ \\
Lescano et al.~\cite{Lescano2026TAFFC} & IEEE TAFFC & 2026 & -- & -- & -- & -- & $\checkmark$ & -- & -- & $\checkmark$ & -- & -- & -- & $\checkmark$ & $\checkmark$ \\
Mobbs et al.~\cite{mobbs26unifiedaffective} & Applied Sciences & 2026 & $\checkmark$ & $\checkmark$ & $\checkmark$ & -- & $\checkmark$ & $\checkmark$ & $\checkmark$ & $\checkmark$ & $\checkmark$ & $\checkmark$ & $\checkmark$ & $\checkmark$ & $\checkmark$ \\
\addlinespace[2pt]
\rowcolor{surveyGroupGray}\multicolumn{16}{c}{\textbf{Large Language Model--Centered Surveys}} \\
Yang et al.~\cite{yang25textmsa} & Sci. China Inf. Sci. & 2025 & $\checkmark$ & $\checkmark$ & $\checkmark$ & -- & $\checkmark$ & -- & $\checkmark$ & -- & -- & -- & -- & $\checkmark$ & $\checkmark$ \\
Kim and Bartolini~\cite{kim26reframing} & Artif. Intell. Rev. & 2026 & $\checkmark$ & $\checkmark$ & $\checkmark$ & -- & $\checkmark$ & $\checkmark$ & $\checkmark$ & -- & -- & -- & -- & $\checkmark$ & $\checkmark$ \\
Zhang et al.~\cite{zhang26merllm} & IJCAI & 2026 & $\checkmark$ & $\checkmark$ & $\checkmark$ & $\checkmark$ & $\checkmark$ & $\checkmark$ & $\checkmark$ & -- & -- & -- & -- & $\checkmark$ & $\checkmark$ \\
Zhang et al.~\cite{zhang26affectivecomputing} & Knowledge-Based Syst. & 2026 & $\checkmark$ & $\checkmark$ & $\checkmark$ & -- & $\checkmark$ & $\checkmark$ & $\checkmark$ & $\checkmark$ & -- & -- & -- & $\checkmark$ & $\checkmark$ \\
\rowcolor{surveyOursPurple}\multicolumn{1}{@{}>{\columncolor{surveyOursPurple}[0pt][\tabcolsep]}l}{\textbf{Ours}} & -- & \textbf{2026} & $\checkmark$ & $\checkmark$ & $\checkmark$ & $\checkmark$ & $\checkmark$ & $\checkmark$ & $\checkmark$ & $\checkmark$ & $\checkmark$ & $\checkmark$ & $\checkmark$ & $\checkmark$ & \multicolumn{1}{>{\columncolor{surveyOursPurple}[\tabcolsep][0pt]}c@{}}{$\checkmark$} \\
\bottomrule
\end{tabular}
\end{adjustbox}
\vskip -0.1in
\end{table*}

As a longstanding research area, multimodal affective computing has seen numerous surveys over the past two decades, with representative examples summarized in \cref{tab:0}. Deep learning-centered approaches have received extensive coverage, while comparatively few surveys address the rapidly expanding LLM-based paradigm. Multimodal affective generation remains particularly underrepresented, with most reviews concentrating on recognition or text-based generation. Two closely related surveys help situate the present review. Our earlier survey of MER-with-LLMs~\cite{zhang26merllm} introduced a challenge-oriented framework for organizing advances in affective recognition. The present work builds on that foundation, substantially expanding the recognition analysis and extending its scope to affective generation. Zhang et al.~\cite{zhang26affectivecomputing} review both affective understanding and generation from an NLP perspective, organizing the literature around adaptation techniques such as instruction tuning, prompt engineering, and reinforcement learning. Our review places greater emphasis on the multimodal dimensions of affective intelligence, encompassing facial, acoustic, visual, physiological, and linguistic evidence, alongside generation in textual, acoustic, and visual forms. Beyond this broader coverage, we organize MER-with-LLMs and MEG-with-LLMs within a unified, challenge-oriented taxonomy, connecting methods across tasks through the problems they address. Together, these choices distinguish our review in both scope and perspective, revealing shared methodological principles and complementary advances across recognition and generation.

\subsection{Survey Scope and Methodology}

\textbf{Literature Search.}
To retrieve relevant papers, we searched Google Scholar, arXiv, IEEE Xplore, and the ACM Digital Library, supplemented by conference proceedings, official project pages, and technical reports. Our queries combined model-related terms, such as ``LLM'' and ``MLLM'', with affect-related terms, including ``emotion'', ``affect'', and ``empathy''. The retrieved studies then served as seeds for expanding the candidate pool through their references, experimental baselines, and subsequent citations. We screened titles, abstracts, and full texts for substantive LLM involvement in MER or MEG and recorded associated data resources. Deduplication was conducted on a case-by-case basis. The inclusion cutoff was October 2026.

\textbf{Research Categorization.}
We organized the final corpus into datasets, MER-with-LLMs methods, and MEG-with-LLMs methods, then further categorized them within the corresponding taxonomies. Since a paper may contribute both data and methods, and a method may support multiple tasks, we allowed overlapping assignments to capture distinct contributions separately. Given the field's rapid pace of development, we used the earliest public release of each work to date its contribution in subsequent analyses.

\section{Background}
\label{sec:2}
Before reviewing specific studies, we introduce the preliminaries on emotion models and the various subtasks covered by MER-with-LLMs and MEG-with-LLMs.

\begin{figure*}[t]
    \centering
    %\vskip -0.04in
    \includegraphics[width=1\linewidth]{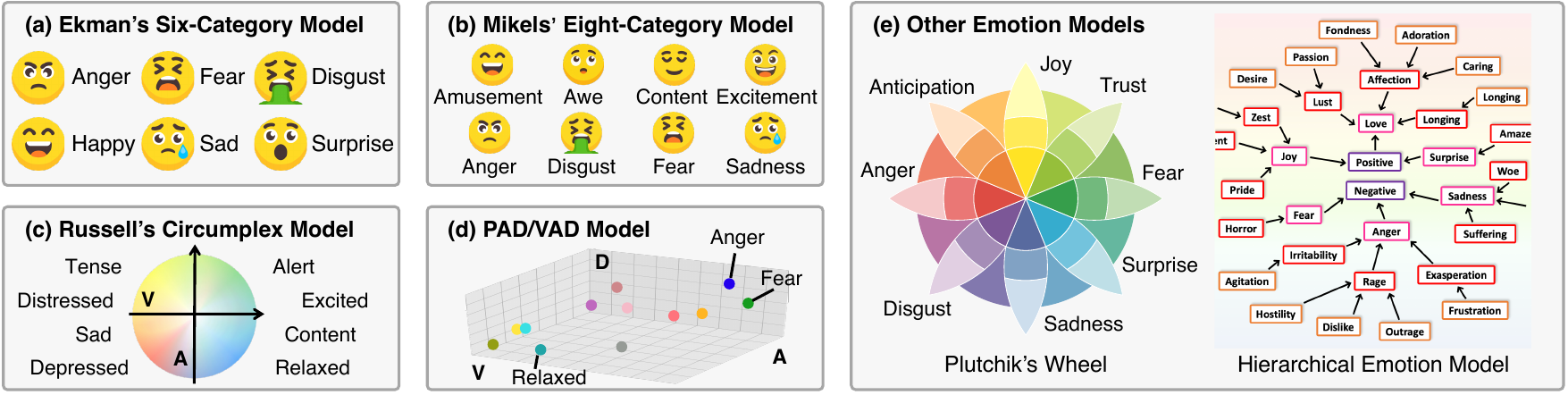}
    \vskip -0.1in
    \caption{Illustration of well-established emotion models. PAD and VAD denote Pleasure-Arousal-Dominance and Valence-Arousal-Dominance, respectively.}
    \label{fig:2-1}
    \vskip -0.1in
\end{figure*}

\subsection{Emotion Model}
Over the past decades, psychological research has developed various emotion models. These models organize abstract emotional concepts into structured representational spaces, providing a conceptual foundation for multimodal affective systems, with representative ones illustrated in \cref{fig:2-1}.

\textbf{Ekman's Six-Category Model.}
Ekman's model~\cite{ekman1992basic} describes families of emotions with characteristic expressions and other distinguishing features of emotional responding. Its commonly adopted six-category formulation comprises anger, disgust, fear, happiness, sadness, and surprise, which serve as discrete prediction targets for diverse annotation units.

\textbf{Mikels' Eight-Category Model.}
Mikels' model~\cite{mikels2005emotional} provides a greater partition of positive affect, distinguishing amusement, awe, contentment, and excitement alongside anger, disgust, fear, and sadness. Unlike Ekman's model, this taxonomy was originally developed from observers' responses to images~\cite{lang99iaps} and serves a consistent role in later research.

\textbf{Russell's Circumplex Model.}
Whereas categorical accounts distinguish named emotions, Russell's circumplex model~\cite{russell1980circumplex} describes their relationships within a continuous Valence-Arousal (VA) space 
based on Wilhelm Wundt's foundations. 
Valence ranges from unpleasant to pleasant, and arousal from deactivated to activated. Excitement and relaxation, for example, share positive valence but differ in arousal. The circular organization places similar affective states nearby and contrasting states in opposite regions, expressing relationships previously left implicit.

\textbf{PAD/VAD Model.}
Pleasure/Valence-Arousal-Dominance (PAD/VAD) model~\cite{mehrabian1974environment,warriner13norms} further incorporates dominance to describe perceived control versus submission. This additional dimension differentiates states with similar pleasantness and activation but different senses of control. The resulting three-dimensional representation supports continuous affect annotation and prediction in subsequent datasets.

\textbf{Other Emotion Models.}
Beyond categorical and dimensional representations, Plutchik's wheel~\cite{plutchik01nature} organizes eight primary emotions by similarity, opposition, intensity, and combination, while Shaver's hierarchical taxonomy~\cite{shaver87emotion} differentiates broad families into finer concepts. Other theories explain emotion elicitation: the Ortony–Clore–Collins (OCC) model~\cite{ortony88cognitive} links emotions to appraisals of events, actions, and objects, whereas the Component Process Model (CPM)~\cite{scherer05emotions} emphasizes coordinated changes in appraisal, physiology, action tendencies, expression, and feeling.

\subsection{Task Formulation}
Building on these emotion models, researchers proposed and developed various tasks to pursue affective intelligence.

\begin{figure*}[t]
    \centering
    %\vskip -0.04in
    \includegraphics[width=1\linewidth]{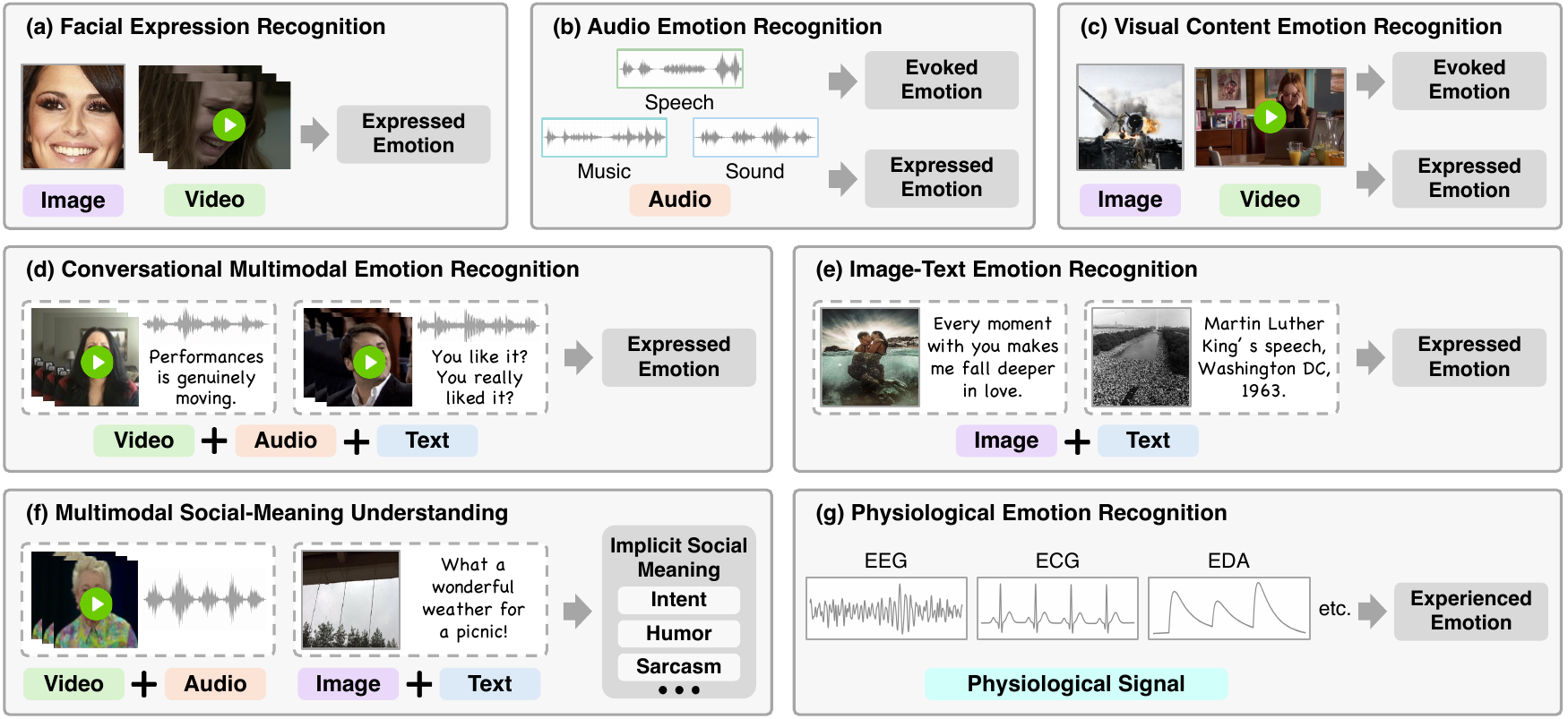}
    \vskip -0.1in
    \caption{Illustration of the seven MER-with-LLMs subtasks, with representative examples drawn from the corresponding datasets~\cite{mollah19affect,wang22ferv39k,lee19caer,zadeh16mosi,yang21tumemo,hasan19urfunny,cai19hfm} or synthesized.}
    \label{fig:2-2}
    \vskip -0.1in
\end{figure*}

\textbf{MER-with-LLMs}.
MER identifies human emotions by integrating affective cues from multiple modalities~\cite{zeng09affect}, including images ($\mathrm{I}$), video ($\mathrm{V}$), audio ($\mathrm{A}$), physiological signals ($\mathrm{P}$), and text ($\mathrm{T}$). Given the text-centric nature of LLMs, the MER-with-LLMs paradigm inherently incorporates the text modality. Therefore, we extend the scope of this survey to certain traditionally unimodal tasks that take non-text input, such as facial expression recognition.

Let $\mathrm{X}$ denote a sample, comprising at least one of $\mathrm{I}$, $\mathrm{V}$, $\mathrm{A}$, or $\mathrm{P}$, optionally accompanied by $\mathrm{T}$. Given a textual query $\mathrm{Q}$ specifying the task, the MLLM $\pi_{\theta}$ typically embeds $\{\mathrm{I}, \mathrm{V}, \mathrm{A}, \mathrm{P}\}$ through modality-specific encoders and adapters, and $\{\mathrm{T}, \mathrm{Q}\}$ through tokenizers. We distinguish $\mathrm{Q}$ from $\mathrm{T}$ according to whether the text specifies what the model should do or provides affective evidence about the sample. Conditioned on these inputs, the model generates a textual response $\mathrm{R}=[r_1,r_2,\cdots,r_n]$ autoregressively, where the $i$-th token is computed as:
\begin{equation}
    r_i = \underset{r \in \mathcal{V}}{\arg\max} \, \pi_\theta(r|\mathrm{Q},\mathrm{X}, \mathrm{R}_{<i}),
\end{equation}
where $\mathcal{V}$ denotes the token vocabulary. Depending on $\mathrm{Q}$, $\mathrm{R}$ may take the form of discrete emotion labels, VAD scores, selections from predefined options, or explanations of emotional conclusions, among other formats. Drawing on representative recognition objectives and sources of affective evidence in the reviewed literature, we focus on seven task families, as illustrated in \cref{fig:2-2}. Additional settings, such as emotion recognition from skeletal motion~\cite{lu25eaillm} or body language~\cite{ozaki25bqa}, are not treated as separate task families here due to their comparatively limited exploration in LLM-based research.

\begin{itemize}
\item
\textit{Facial Expression Recognition} (\textbf{FER}) aims to identify observable facial expressions and their associated affective attributes from static facial appearance or temporal changes~\cite{pantic00facial,mollah19affect}. $\mathrm{X}$ comprises face-centered images ($\mathrm{I}$) or video sequences ($\mathrm{V}$), with audio or contextual text providing auxiliary information sometimes.

\item
\textit{Audio Emotion Recognition} (\textbf{AER}) aims to recognize emotion expressed in or evoked by speech, music, and %other vocal expressions
%BS: changed to
sound~\cite{schull03ser,aljana17deam}. $\mathrm{X}$ primarily comprises audio ($\mathrm{A}$), optionally accompanied by transcripts or contextual text.

\item
\textit{Visual Content Emotion Recognition} (\textbf{VCER}) aims to interpret affect associated with generic visual content, encompassing emotions evoked in viewers and emotions attributed to depicted individuals in context~\cite{machaj10affect,kosti17emotic}. $\mathrm{X}$ comprises images ($\mathrm{I}$) or videos ($\mathrm{V}$), optionally accompanied by audio for video and textual context or statements.

\item
\textit{Conversational Multimodal Emotion Recognition} (\textbf{CMER}) aims to identify a speaker's emotion or sentiment at the utterance level using multimodal communicative cues and conversational context~\cite{busso08iemocap,poria19meld}. $\mathrm{X}$ typically combines audio ($\mathrm{A}$), video ($\mathrm{V}$), and transcripts ($\mathrm{T}$). In some cases, visual observations may also be represented by sampled images ($\mathrm{I}$).

\item
\textit{Image-Text Emotion Recognition} (\textbf{ITER}) aims to determine the emotion or sentiment polarity conveyed through paired images and text, either at the post level or toward a specified entity~\cite{niu16sentiment,yu19p751,yang21tumemo}. $\mathrm{X}=\{\mathrm{I},\mathrm{T}\}$ typically comprises media posts, with the target entity specified on demand.

\item
\textit{Multimodal Social-Meaning Understanding} (\textbf{MSU}) extends beyond emotion recognition to interpret higher-order, affect-related social meanings, including intent, sarcasm, humor, and mental states~\cite{zadeh19sociq,castro19mustar,zhang22mintrec}. Representative input formats include audiovisual content ($\mathrm{X}=\{\mathrm{A},\mathrm{V}\}$) and image-text pairs ($\mathrm{X}=\{\mathrm{I},\mathrm{T}\}$).

\item
\textit{Physiological Emotion Recognition} (\textbf{PER}) aims to estimate an individual's experienced affective state from physiological measurements~\cite{picard01phys,koelst12deap}. $\mathrm{X}=\{\mathrm{P}\}$ comprises one or more physiological signal types, including Electroencephalography (EEG), a key input to affective Brain-Computer Interfaces (BCIs), and complementary peripheral signals such as Electrocardiography (ECG) and Electrodermal Activity (EDA). Recordings are often obtained during emotion-elicitation procedures, with affective targets defined through participant reports or stimulus conditions.
\end{itemize}

%BS: Feels like a random choice of buzz-words... Where is simple sentiment analysis, where are micro-gestures, etc.? Perhaps motivate the choice in a sentence or at least state that more exist and this is a typical selection - where is also BCI, etc.

\begin{figure*}[t]
    \centering
    %\vskip -0.04in
    \includegraphics[width=1\linewidth]{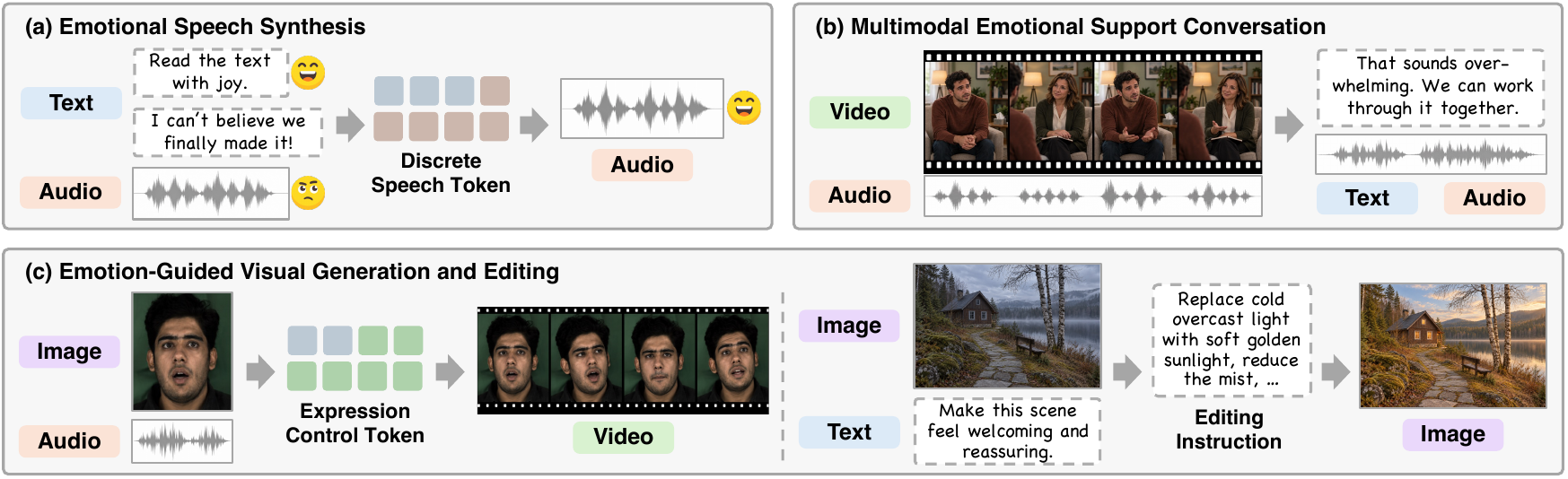}
    \vskip -0.1in
    \caption{Illustration of the three MEG-with-LLMs subtasks, with representative examples drawn from the corresponding datasets~\cite{wang20mead,chu25mesc} or synthesized.}
    \label{fig:2-3}
    \vskip -0.1in
\end{figure*}

\textbf{MEG-with-LLMs}.
MEG aims to generate content that conveys or evokes intended emotions to fulfill affective communicative purposes. Within this framework, LLMs interpret affective requirements and context, formulate appropriate responses, and guide their realization across modalities. Unlike MER-with-LLMs, which requires non-text input, we define MEG-with-LLMs to encompass generation tasks involving at least one non-text modality in either the input or the output.

Following the notation above, let $\mathrm{X}$ denote the input and $\mathrm{Q}$ a textual query. As in MER-with-LLMs, the LLM $\pi_{\theta}$ first processes these inputs to produce an output $\mathrm{R}$:
\begin{equation}
    \mathrm{R}
    =
    \left[
        \underset{r \in \mathcal{V}}{\arg\max}\,
        \pi_{\theta}\!\left(r \mid \mathrm{Q}, \mathrm{X}, \mathrm{R}_{<i}\right)
    \right]_{i=1}^{n}.
\end{equation}
$\mathrm{R}$ may contain response text, affective plans, control representations, or modality-specific tokens. Unlike recognition, where the textual response typically constitutes the final output, generation may require further realization in a target modality. Therefore, a realization module $g_{\phi}$, implemented as a decoder or conditioned generator, converts $\mathrm{R}$ into the final output, optionally conditioned on $\mathrm{X}$:
\begin{equation}
    \mathrm{Z} = g_{\phi}(\mathrm{R},\mathrm{X}).
\end{equation}
The output $\mathrm{Z}$ may comprise text ($\mathrm{T}$), audio ($\mathrm{A}$), images ($\mathrm{I}$), video ($\mathrm{V}$), or their combinations. According to the generation objective and output modality, we organize MEG-with-LLMs into three main subtasks, as illustrated in \cref{fig:2-3}.

%BS: the below again feels random. Why would you put a highly specialised task such as MESC before the traditional and for MESC required ESS? Please motivate these choices and again make clear that these are just examples. I highly recommend sorting by "classic basis" such as ESS first :) 

\begin{itemize}
\item
\textit{Emotional Speech Synthesis} (\textbf{ESS}) aims to synthesize speech that conveys intended emotional characteristics while preserving the supplied linguistic content~\cite{schrod01ess}. $\mathrm{X}$ typically contains text to be synthesized ($\mathrm{T}$), optionally accompanied by reference audio ($\mathrm{A}$) specifying speaker identity or delivery style, while $\mathrm{Z}$ is generated speech ($\mathrm{A}$). The LLM produces speaking-style descriptions or discrete speech tokens, while $g_{\phi}$ realizes the waveform through a conditioned speech synthesizer or an audio-token decoder, such as a neural codec decoder~\cite{ji24textrolspe} or a flow-matching model followed by a vocoder~\cite{ren26ovtts}.

\item
\textit{Multimodal Emotional Support Conversation} (\textbf{MESC}) aims to generate supportive responses that address a user's emotional needs and help alleviate distress, drawing on multimodal evidence within a conversation~\cite{liu21esconv,chu25mesc}. $\mathrm{X}$ comprises spoken or audiovisual exchanges ($\mathrm{A}$ or $\mathrm{V},\mathrm{A}$), optionally accompanied by transcripts, dialogue history, or other textual context ($\mathrm{T}$). In early systems, $\mathrm{Z}$ mainly takes the form of supportive text ($\mathrm{T}$), which is later extended to coordinated vocal expression ($\mathrm{A}$).

\item
\textit{Emotion-Guided Visual Generation and Editing} (\textbf{EVGE}) aims to generate or modify visual content to convey intended emotional expressions or evoke particular affective responses in viewers~\cite{wang20mead,yang24emogen}. $\mathrm{X}$ may comprise textual descriptions ($\mathrm{T}$), source or reference images ($\mathrm{I}$) and videos ($\mathrm{V}$), and driving audio ($\mathrm{A}$) for speech-linked animation, while $\mathrm{Z}$ is an image ($\mathrm{I}$) or video ($\mathrm{V}$). The LLM specifies affect-relevant visual attributes, editing instructions, or expression controls rather than directly rendering pixels. These specifications guide $g_{\phi}$, which may take the form of a conditioned image-editing model~\cite{ye25moodifier} or an expression-controlled talking-head generator~\cite{lyu26auhead}. For editing, the intended affective changes are additionally constrained to preserve task-relevant identity, structure, or semantics.
\end{itemize}

\section{Challenges and Taxonomy}
\label{sec:3}

\begin{figure*}[t]
    \centering
    %\vskip -0.04in
    \includegraphics[width=1\linewidth]{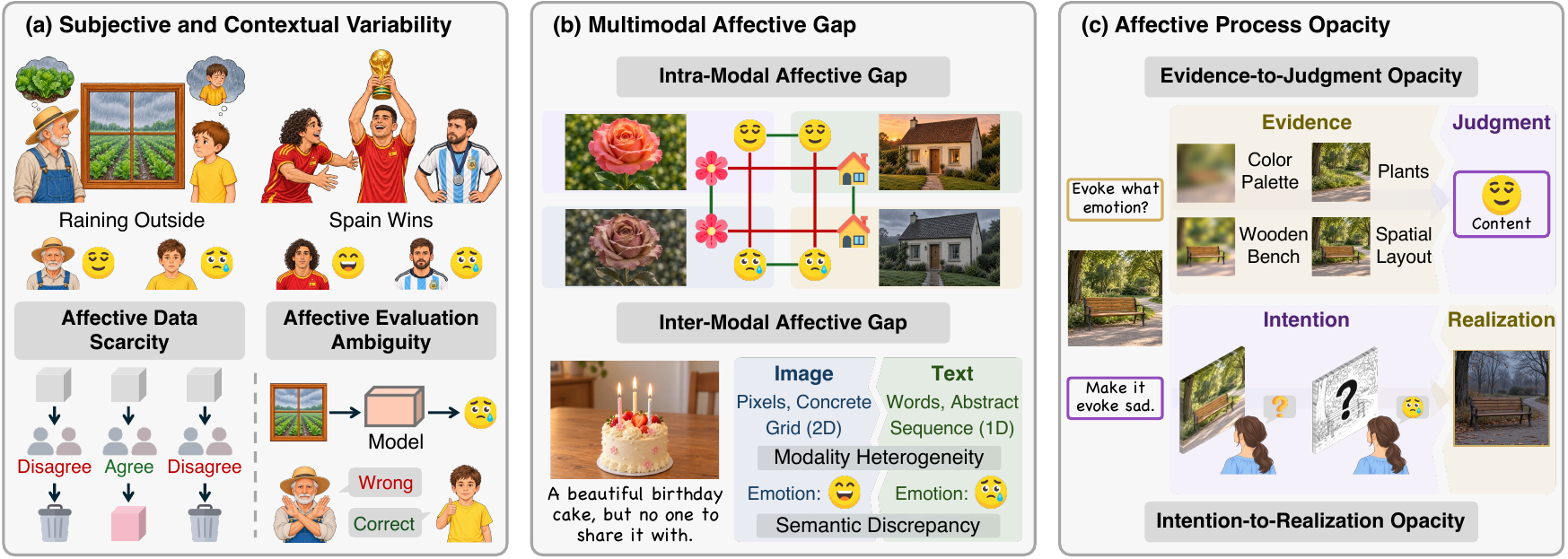}
    \vskip -0.1in
    \caption{Illustration of the three core challenges in MER-with-LLMs and MEG-with-LLMs. \textit{Subjective and Contextual Variability} (a) refers to the fact that the same scene can evoke different emotional responses and admit multiple plausible affective interpretations, contributing to \textit{Affective Data Scarcity} and \textit{Affective Evaluation Ambiguity}. \textit{Multimodal Affective Gap} (b) encompasses the disconnect between observable content features and affective meanings within individual modalities, as well as heterogeneity and semantic discrepancies across modalities. \textit{Affective Process Opacity} (c) concerns the limited transparency and interpretability of the processes linking evidence to judgment in recognition and intention to realization in generation.}
    \label{fig:3-1}
    \vskip -0.1in
\end{figure*}

\subsection{Challenges in Affective Computing with LLMs}

LLMs have opened new paradigms for MER and MEG by introducing instruction-following capabilities and extensive world knowledge, yet these advances do not eliminate the underlying challenges of affective computing. As illustrated in \cref{fig:3-1}, we organize these challenges into \textit{Subjective and Contextual Variability}, \textit{Multimodal Affective Gap}, and \textit{Affective Process Opacity}, which collectively capture the principal motivations and research foci under these paradigms.

\textbf{Subjective and Contextual Variability}.
Emotional responses and interpretations depend on individual goals, experiences, and situational appraisals, rather than being uniquely determined by observable content~\cite{scherer05emotions}. In \cref{fig:3-1} (a), rain may please a farmer but disappoint a child, while the same football victory evokes contrasting reactions among opposing supporters. This variability contributes to two challenges. First, \textit{Affective Data Scarcity} is exacerbated by the cost of obtaining reliable annotations: multiple judgments and majority voting are often required, increasing annotation costs and reducing the samples retained through consensus-based filtering. For example, FI~\cite{you16fi} retains only 25.9\% of candidate images after majority-vote verification. Second, \textit{Affective Evaluation Ambiguity} makes it difficult to establish evaluation frameworks that adequately accommodate diverse yet defensible affective judgments. Although a majority label provides an operational reference, consequential evaluation still risks overlooking reasonable minority~\cite{davani22disagr} or open-vocabulary responses~\cite{lian23emer}.

\textbf{Multimodal Affective Gap}.
Earlier research characterized the affective gap as the mismatch between computational features and the affective states evoked in viewers~\cite{zhao22aica}. Under the multimodal setting, this formulation corresponds to the \textit{Intra-Modal Affective Gap}, where recognizing observable content within a modality does not necessarily reveal its emotional meaning. In \cref{fig:3-1} (b), blooming and withered roses share an object category but convey contrasting affect, whereas visually dissimilar flowers and houses may evoke similar feelings. Beyond this intra-modality challenge, coordinating multiple signals introduces an \textit{Inter-Modal Affective Gap}, which involves modality heterogeneity~\cite{wei23mvcn} and semantic discrepancy~\cite{wu26code}. Modality heterogeneity encompasses differences in representational structure and information granularity. For example, images encode concretely in spatial grids, whereas text conveys meaning through sequences of abstract symbols, complicating cross-modal alignment and fusion. Semantic discrepancy arises when modalities express different or contradictory sentiments. An image may appear positive in isolation, yet its interpretation can shift toward sadness when combined with the accompanying text. Such divergence significantly complicates the overall affective judgment.

\textbf{Affective Process Opacity}.
Affective applications often require not only predictions or generated outputs but also explanations of their underlying rationale. However, articulating how affective judgments are reached and intentions are realized remains challenging in two respects. First, \textit{Evidence-to-Judgment Opacity} concerns the difficulty of explaining how perceived cues support emotional conclusions~\cite{lian23emer}. In \cref{fig:3-1} (c), predicting contentment alone does not explain the contributions of the park's color palette, vegetation, and spatial layout. On the generation side, \textit{Intention-to-Realization Opacity} concerns the difficulty of translating abstract affective goals into concrete generation decisions~\cite{ye25moodifier}. For the same scene, an instruction to evoke sadness leaves unclear which visual attributes should change, why those changes would convey sadness, and which content should remain intact.

\input{method/taxonomy}

% \begin{figure*}[t]
%     \centering
%     %\vskip -0.04in
%     \includegraphics[width=1\linewidth]{figs/fig3-2.pdf}
%     \vskip -0.1in
%     \caption{Taxonomy of datasets and methods, and their relationships to the challenges. Datasets directly support efforts to address \textit{Subjective and Contextual Variability} and underpin both recognition and generation branches, whose methodological categories are organized around the corresponding challenges.}
%     \label{fig:3-2}
%     \vskip -0.1in
% \end{figure*}

Building on the challenges outlined above, we organize the field into affective datasets, MER-with-LLMs methods, and MEG-with-LLMs methods, as illustrated in \cref{fig:3-2}. Datasets underpin both paradigms and help address \textit{Subjective and Contextual Variability} through supervision and evaluation. \textit{Conventional Affective Datasets} lay the empirical foundation for subsequent research, \textit{Affective Model-Tuning Datasets} support model training and adaptation, and \textit{Affective Evaluation Benchmarks} assess affective capabilities.

From a challenge-oriented perspective, we organize MER-with-LLMs methods into three categories. \textit{Data-Efficient Affective Elicitation} activates and adapts existing capabilities with limited supervision, addressing the data scarcity associated with \textit{Subjective and Contextual Variability}. \textit{Multimodal Affective Representation} shapes how affective cues are encoded within and integrated across modalities to bridge the \textit{Multimodal Affective Gap}. \textit{Affective Reasoning and Explanation} extends prediction to explicit reasoning and evidence-supported interpretation, addressing \textit{Affective Process Opacity}. Together, these categories reflect progressively deeper intervention, from configuring model use and reshaping representations to enriching the recognition objective itself.

MEG-with-LLMs methods follow a complementary progression through planning, realization, and alignment. \textit{Affective Generation Planning} specifies affective intentions and generation decisions, addressing \textit{Affective Process Opacity}. \textit{Affective Control and Realization} translates these specifications into expressive features and outputs, bridging the \textit{Multimodal Affective Gap} between intended affect and its realization. \textit{Human-Oriented Affective Alignment} optimizes generated behavior against human preferences and expectations, addressing \textit{Subjective and Contextual Variability}. These categories connect what to express, how to express it, and whether the result meets human affective needs. As methods increasingly integrate multiple capabilities, we allow them to appear in multiple categories across both paradigms, discussing their relevant contributions in each.

\section{Datasets}
\label{sec:4}

\begin{figure*}
    \centering
    %\vskip -0.04in
    \includegraphics[width=1\linewidth]{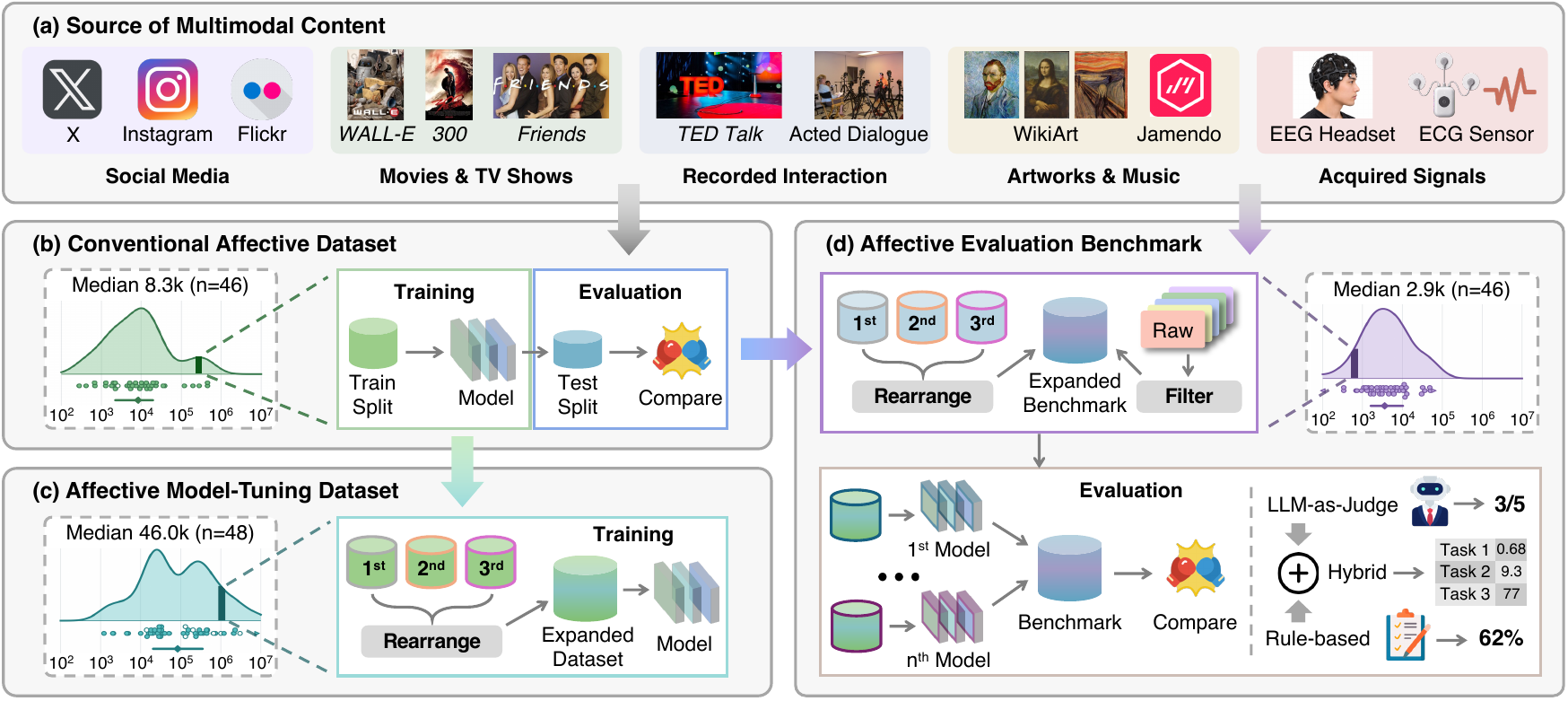}
    \vskip -0.1in
    \caption{Relationships among multimodal content sources and affective
    dataset categories. Diverse content sources (a) provide the basis for \textit{Conventional Affective Datasets} (b), whose training splits are often
    reorganized and augmented to construct \textit{Affective Model-Tuning Datasets} (c) with richer annotations and larger scales. Evaluation benchmarks (d) are constructed by curating and aggregating existing test splits, collecting new content, or combining both approaches.}
    \label{fig:4-1}
    \vskip -0.1in
\end{figure*}

Addressing the challenges outlined above depends not only on methodological advances but also on the data. As models evolve from task-specific predictors to instruction-driven systems, their data requirements likewise expand from affective labels to descriptions, explanations, and interactive responses. 

As illustrated in \cref{fig:4-1}, the data resources supporting this transition draw on five main sources of multimodal content. (1) Social media supplies user-generated images, videos, and image-text posts that capture affect in everyday online communication~\cite{peng15emotion6,you16fi,niu16sentiment}. (2) Movies and TV shows offer expressive audiovisual material embedded in narratives and interpersonal interactions, providing rich observations of contextual emotional behavior~\cite{li17cheavd,poria19meld}. (3) Recorded interactions span scripted and improvised dialogues, acted speech, and public talks, capturing affect across controlled and naturalistic communicative settings~\cite{busso08iemocap,living18ravdes,hasan19urfunny}. (4) Artworks and music convey or evoke emotion through visual composition and acoustic expression, supporting the study of affect beyond interpersonal communication~\cite{machaj10affect,aljana17deam}. (5) Physiological signals offer complementary evidence of experienced affect, typically acquired through dedicated procedures using specialized equipment such as EEG headsets and ECG sensors~\cite{koelst12deap,zheng15seed,katsig18dreame}.

We organize these resources into three categories according to their primary roles. \textit{Conventional Affective Datasets} consolidate multimodal observations into annotated training and test sets. \textit{Affective Model-Tuning Datasets} often reorganize and enrich existing training resources into larger collections of instructions, descriptions, explanations, and responses. \textit{Affective Evaluation Benchmarks} draw on existing test sets, newly collected content, or both to assess model capabilities. The following subsections review each category by primary task, with \cref{tab:4-1,tab:4-2,tab:4-3} summarizing individual resources.

\begin{table*}[t]
\centering
\caption{Conventional Affective Datasets with their key attributes. Datasets are grouped by their primary tasks. MESC$^\dagger$ denotes dialogue benchmarks that support subsequent MESC methods. In the Modality column, I, A, V, P, and T denote image, audio, video, physiological signals, and text, respectively. QA denotes Question-Answer pairs.}
\label{tab:4-1}
\vskip -0.1in
\begin{adjustbox}{max width=\textwidth}
\footnotesize
\renewcommand{\arraystretch}{1.1}
\setlength{\tabcolsep}{5pt}
\begin{tabular}{@{}lccccccc@{}}
\toprule
\textbf{Dataset} &
\textbf{Year} &
\textbf{Task} &
\textbf{Modality} &
\textbf{Scale} &
\textbf{Target Space} &
\textbf{Annotation} &
\textbf{Source} \\
\midrule

RAF-DB \cite{li17rafdb} &
2017 & FER & I &
15,339 Images &
Categorical-7 &
Human ($\sim$40/image) &
Web Face Images \\

AffectNet \cite{mollah19affect} &
2019 & FER & I &
318,969 Images &
Categorical-8 + VA &
Human (1-2/image) &
Image Search Engines \\

DFEW \cite{jiang20dfew} &
2020 & FER & V &
12,059 Clips &
Categorical-7 &
Human (10/clip) &
Movies \\

MAFW \cite{liu22mafw} &
2022 & FER & A,V &
10,045 Clips &
Single + Compound &
Human (11/clip) &
Movies, TV, Web Videos \\

FERV39k \cite{wang22ferv39k} &
2022 & FER & V &
38,935 Clips &
Categorical-7 &
Human (30 total) &
Web Videos and TV \\

\rowcolor{gray!8}
Emo-DB \cite{burkha05emodb} &
2005 & AER & A &
535 Clips &
Categorical-7 &
Human (20 total) &
Performed Speech \\

\rowcolor{gray!8}
CREMA-D \cite{cao14cremad} &
2014 & AER, FER & A,V &
7,442 Clips &
Categorical-6 + Intensity &
Human (2,443 total) &
Performed Speech \\

\rowcolor{gray!8}
DEAM \cite{aljana17deam} &
2017 & AER & A &
1,802 Tracks &
VA &
Human ($\geq$5/track) &
FMA, Jamendo, MedleyDB \\

\rowcolor{gray!8}
RAVDESS \cite{living18ravdes} &
2018 & AER, FER & A,V &
2,452 Clips &
Speech-8 / Song-6 &
Human (247 total) &
Performed Speech and Song \\

\rowcolor{gray!8}
MSP-Podcast \cite{lotfia19msp} &
2019 & AER & A &
151,654 Clips &
Categorical-9 + VAD &
Human ($\geq$5/clip) &
Podcast Recordings \\

Abstract \cite{machaj10affect} &
2010 & VCER & I &
280 Images &
Mikels-8 &
Human ($\sim$14/image) &
Abstract Art \\

Emotion6 \cite{peng15emotion6} &
2015 & VCER & I &
1,980 Images &
Categorical-7 Dist. + VA &
Human (15/image) &
Flickr \\

FI \cite{you16fi} &
2016 & VCER & I &
23,308 Images &
Mikels-8 &
Human (5/image) &
Flickr and Instagram \\

EMOTIC \cite{kosti17emotic} &
2017 & VCER & I &
18,316 Images &
Categorical-26 + VAD &
Human ($\geq$5 total) &
COCO, ADE20K, Web \\

CAER \cite{lee19caer} &
2019 & VCER & A,V &
13,201 Clips &
Categorical-7 &
Human (3/clip) &
TV Shows \\

ArtEmis \cite{achlio21artemi} &
2021 & VCER & I &
454,684 Pairs &
Mikels-8 + Other &
Human ($\geq$5/image) &
WikiArt \\

EmoSet \cite{yang23emoset} &
2023 & VCER & I &
118,102 Images &
Mikels-8 &
Human (10/image) &
Web and Art Images \\

\rowcolor{gray!8}
IEMOCAP \cite{busso08iemocap} &
2008 & CMER & A,V,T &
10,039 Clips &
Categorical-10 + VAD &
Human (2-3/clip) &
Performed Dialogues \\

\rowcolor{gray!8}
CMU-MOSI \cite{zadeh16mosi} &
2016 & CMER & A,V,T &
2,199 Clips &
Sentiment-7 &
Human (5/clip) &
YouTube Monologues \\

\rowcolor{gray!8}
CHEAVD \cite{li17cheavd} &
2017 & CMER & A,V &
$\sim$2,600 Clips &
Categorical-26 &
Human (4/clip) &
Chinese Movies and TV \\

\rowcolor{gray!8}
CMU-MOSEI \cite{zadeh18mosei} &
2018 & CMER & A,V,T &
23,453 Clips &
Sentiment-7 + Basic-6 &
Human (3/clip) &
YouTube Monologues \\

\rowcolor{gray!8}
MELD \cite{poria19meld} &
2019 & CMER & A,V,T &
13,708 Clips &
Categorical-7 + Polarity-3 &
Human (3/clip) &
\textit{Friends} \\

\rowcolor{gray!8}
CH-SIMS v2 \cite{liu22chsims2} &
2022 & CMER & A,V,T &
4,402 Clips &
Sentiment $[-1,1]$ &
Human (7/clip) &
Chinese Movies and TV \\

\rowcolor{gray!8}
M$^{3}$ED \cite{zhao22m3ed} &
2022 & CMER & A,V,T &
24,449 Clips &
Categorical-7 &
Human (3/clip) &
Chinese TV Series \\

\rowcolor{gray!8}
MER2023 \cite{lian23mer2023} &
2023 & CMER & A,V,T &
5,030 Clips &
Categorical-6 + Valence &
Human (4/clip) &
CHEAVD and Web Videos \\

\rowcolor{gray!8}
MER2024 \cite{lian24mer2024} &
2024 & CMER & A,V,T &
7,369 Clips &
Categorical-6 + Open-Vocab. &
Human (4-5/clip) &
MER2023 and Web Videos \\

MVSA-Single \cite{niu16sentiment} &
2016 & ITER & I,T &
5,129 Pairs &
Polarity-3 &
Human (1/pair) &
Twitter Posts \\

MVSA-Multiple \cite{niu16sentiment} &
2016 & ITER & I,T &
19,600 Pairs &
Polarity-3 &
Human (3/pair) &
Twitter Posts \\

Twitter-15 \cite{yu19p751} &
2019 & ITER & I,T &
5,338 Pairs &
Target Polarity-3 &
Human (3/target) &
Twitter Posts \\

Twitter-17 \cite{yu19p751} &
2019 & ITER & I,T &
5,972 Pairs &
Target Polarity-3 &
Human (3/target) &
Twitter Posts \\

TumEmo \cite{yang21tumemo} &
2021 & ITER & I,T &
195,265 Pairs &
Categorical-7 &
Weak (User Tags) &
Tumblr Posts \\

\rowcolor{gray!8}
MUStARD \cite{castro19mustar} &
2019 & MSU & A,V,T &
690 Clips &
Sarcasm-2 &
Human (3/clip) &
TV Series \\

\rowcolor{gray!8}
UR-FUNNY \cite{hasan19urfunny} &
2019 & MSU & A,V,T &
16,514 Clips &
Humor-2 &
Weak (Laughter Tags) &
TED Talks \\

\rowcolor{gray!8}
MIntRec \cite{zhang22mintrec} &
2022 & MSU & A,V,T &
2,224 Clips &
Intent-20 &
Human (5/clip) &
\textit{Superstore} \\

\rowcolor{gray!8}
MSED \cite{jia22msed} &
2022 & MSU, ITER & I,T &
9,190 Pairs &
Emotion; Sentiment; Desire &
Human (5/pair) &
Twitter, Getty, Flickr \\

\rowcolor{gray!8}
Social-IQ 2.0 \cite{wilf23socialiq} &
2023 & MSU & A,V,T &
7,102 QA &
Social Reasoning &
Human &
Web Videos and Movies \\

% \rowcolor{gray!8}
% ECF 2.0 \cite{wang24ecf} &
% 2024 & MSU & A,V,T &
% 1,715 Dialogues &
% Emotion-Cause Pairs &
% Human (3/dialogue) &
% \textit{Friends} \\

DEAP \cite{koelst12deap} &
2012 & PER & P &
1,280 Trials &
VAD + Liking &
Self-Report (32 total) &
Music-Video Elicitation \\

MAHNOB-HCI \cite{soleym12mahnob} &
2012 & PER & P &
532 Trials &
Categorical-9 + VAD &
Self-Report (27 total) &
Film-Clip Elicitation \\

SEED \cite{zheng15seed} &
2015 & PER & P &
675 Trials &
Polarity-3 &
Stimulus-Based &
Film-Clip Elicitation \\

DREAMER \cite{katsig18dreame} &
2018 & PER & P &
414 Trials &
VAD &
Self-Report (23 total) &
Film-Clip Elicitation \\

AMIGOS \cite{mirand21amigos} &
2021 & PER & P &
788 Trials &
VAD + Basic-7 &
Self-Report (40 total) &
Film-Clip Elicitation \\

\rowcolor{gray!8}
DailyDialog \cite{li17dailydial} &
2017 & MESC$^\dagger$ & T &
13K Dialogues &
Emotion-7 + Intent-4 &
Human (3 total) &
Learning Websites \\

\rowcolor{gray!8}
EmpatheticDialogues \cite{rashki19empat} &
2019 & MESC$^\dagger$ & T &
25K Dialogues &
Emotion-32 &
Human (810 total) &
Crowdsourced Dialogues \\

\rowcolor{gray!8}
ESConv \cite{liu21esconv} &
2021 & MESC$^\dagger$ & T &
1,300 Dialogues &
Emotion-7 + Strategy-8 &
Human ($\geq$854 total) &
Crowdsourced Dialogues \\

ESD \cite{zhou21esd} &
2021 & ESS, AER & A,T &
35,000 Clips &
Categorical-5 &
Acted (20 speakers) &
Parallel Speech \\

\rowcolor{gray!8}
MEAD \cite{wang20mead} &
2020 & EVGE & A,V &
281,400 Clips &
Categorical-8 + Intensity-3 &
Acted (60 actors) &
Talking-Head Recordings \\

\bottomrule
\end{tabular}
\end{adjustbox}
\vskip -0.1in
\end{table*}

\subsection{Conventional Affective Datasets} \label{sec:4_1}

\textit{Conventional Affective Datasets} establish much of the empirical foundation inherited by LLM-based research. We focus on representative resources that inform subsequent model tuning and evaluation. Their construction remains largely dependent on labor-intensive human annotation, with discrete categories predominant alongside dimensional, distributional, and multi-label schemes. Although shared affective content permits cross-task reuse, most resources originate from a specific task objective, which guides the organization of \cref{tab:4-1} and the discussion below.

\textbf{FER Datasets.} FER resources trace a progression from static facial appearance to expressions unfolding in unconstrained environments. AffectNet~\cite{mollah19affect} combines categorical and VA annotations, while RAF-DB~\cite{li17rafdb} uses extensive crowdsourcing to capture expressions across diverse subjects and imaging conditions. Video datasets recover the temporal information absent from individual frames. DFEW~\cite{jiang20dfew} draws expression clips from movies, and FERV39k~\cite{wang22ferv39k} expands coverage to 38,935 clips across 22 scenes. MAFW~\cite{liu22mafw} adds audiovisual evidence and compound expression labels. This progression broadens from facial configurations to their temporal dynamics, contextual variation, and expressive complexity.

\textbf{AER Datasets.} Acoustic resources capture affect through both vocal delivery and musical structure. Emo-DB~\cite{burkha05emodb} and CREMA-D~\cite{cao14cremad} emphasize acted speech under controlled conditions. RAVDESS~\cite{living18ravdes} extends this setting to speech and song, with standardized verbal content, controlled intensities, and corresponding visual recordings. MSP-Podcast~\cite{lotfia19msp} complements these performances with naturalistic speech, exposing models to less constrained emotional expression. Music brings a different temporal emphasis, as affect develops through rhythm, melody, and timbre. DEAM~\cite{aljana17deam} captures this evolution through time-varying valence and arousal annotations. Together, these resources span categorical vocal expressions and continuously varying musical affect.

\textbf{VCER Datasets.} Moving beyond facial expression, VCER resources examine affect associated with broader visual content. Abstract and ArtPhoto~\cite{machaj10affect} explore abstract compositions and artistic photographs, while Emotion6~\cite{peng15emotion6} preserves variation in viewers' reactions through probability distributions. FI~\cite{you16fi} and WEBEmo~\cite{panda18webemo} broaden the visual domain through social-media and web imagery. Within complex scenes, EMOTIC~\cite{kosti17emotic} and CAER~\cite{lee19caer} address a complementary target: the emotions of depicted individuals as inferred from their behavior and surroundings, rather than those evoked in viewers. Annotation also extends beyond emotional labels. ArtEmis~\cite{achlio21artemi} pairs responses to artworks with natural-language explanations, an approach extended to real-world images by Affection~\cite{achlio23affect}. EmoSet~\cite{yang23emoset} supplies visual attributes spanning brightness, colorfulness, scenes, objects, expressions, and actions. These developments connect visual affect to both its perceptual basis and the interpretations through which it acquires meaning.

\textbf{CMER Datasets.} CMER resources place affect within utterances and interactions, bringing linguistic, acoustic, and visual evidence into a common setting. IEMOCAP~\cite{busso08iemocap} provides scripted and improvised actor dialogues with categorical and dimensional annotations. CMU-MOSI~\cite{zadeh16mosi} and CMU-MOSEI~\cite{zadeh18mosei} extend coverage to online opinion monologues, supplying sentiment intensity labels and, in MOSEI, emotion categories. MELD~\cite{poria19meld} preserves the dependencies of multiparty conversation through dialogues from \textit{Friends}. Chinese resources contribute complementary settings. CHEAVD~\cite{li17cheavd} captures diverse emotional expressions, CH-SIMS v2~\cite{liu22chsims2} emphasizes nonverbal sentiment cues, and M$^{3}$ED~\cite{zhao22m3ed} covers dyadic interactions across television series and relationship contexts. More recent challenges broaden not only the data but also the evaluation objectives. MER2023~\cite{lian23mer2023} examines joint categorical-dimensional prediction, noise robustness, and limited supervision, while MER2024~\cite{lian24mer2024} introduces open-vocabulary recognition, loosening the constraints on both observed conditions and emotional labels.

\textbf{ITER Datasets.} Image-text resources capture affect in social-media communication, where the two modalities may reinforce or contradict one another. MVSA-Single and MVSA-Multiple~\cite{niu16sentiment} provide separate sentiment judgments for Twitter images and text, differing in the number of annotators. Whole-post labels, however, cannot distinguish attitudes toward different entities within the same post. Twitter-15 and Twitter-17~\cite{yu19p751} address this limitation through target-oriented sentiment annotations. TumEmo~\cite{yang21tumemo} instead broadens the affective vocabulary, using Tumblr posts and user-associated tags to move beyond polarity toward multiple emotion categories.

\textbf{MSU Datasets.} Understanding affect often requires interpreting what people imply, intend, or desire. MUStARD~\cite{castro19mustar} examines sarcasm in television dialogue, while the Twitter dataset accompanying HFM~\cite{cai19hfm} captures incongruity in image-text communication. UR-FUNNY~\cite{hasan19urfunny} investigates humor through audience-laughter markers in TED talks. Beyond nonliteral expression, MIntRec~\cite{zhang22mintrec} annotates communicative intentions in television dialogue, and MSED~\cite{jia22msed} associates desire with emotion and sentiment in image-text posts. Social-IQ 2.0~\cite{wilf23socialiq} broadens the scope through audiovisual question answering, requiring contextual understanding of interactions rather than identification of a single social category.

\textbf{PER Datasets.} Physiological resources complement outward expression with measurements of the bodily responses accompanying affect. DEAP~\cite{koelst12deap} and MAHNOB-HCI~\cite{soleym12mahnob} collectively cover EEG and peripheral signals such as ECG, EDA, respiration, and skin temperature, although dedicated instrumentation and elicitation procedures constrain collection scale. DEAP pairs physiological recordings with self-reported responses to music videos, while SEED~\cite{zheng15seed} uses repeated film-elicited EEG sessions to capture individual and temporal variation. DREAMER~\cite{katsig18dreame} combines low-cost wearable EEG and ECG acquisition with dimensional self-reports. AMIGOS~\cite{mirand21amigos} extends the experimental setting to individual and group viewing, incorporating social conditions into the study of physiological affect. These measurements do not constitute emotional ground truth, but they provide a complementary observational basis for context-sensitive affective modeling.

\textbf{MESC, ESS, and EVGE Datasets.} Generation requires additional responses and performances beyond recognition labels, which provide an empirical basis for emotionally expressive language and media generation with autoregressive~\cite{brown20gpt3} and diffusion models~\cite{ho20ddpm}. For MESC, DailyDialog~\cite{li17dailydial} captures everyday emotional exchanges, EmpatheticDialogues~\cite{rashki19empat} emphasizes responses to emotionally grounded situations, and ESConv~\cite{liu21esconv} introduces explicit support strategies within help-seeker-supporter conversations. Complementing response content, ESD~\cite{zhou21esd} captures expressive delivery through parallel emotional speech for voice conversion, holding linguistic content constant across emotional conditions. MEAD~\cite{wang20mead} provides a visual counterpart through talking-face recordings with controlled expression categories and intensities. Together, these resources anchor affective generation in conversational behavior and its acoustic and visual realization.

\begin{table*}[t]
\centering
\caption{Affective Model-Tuning Datasets with their key attributes. Datasets are grouped by their primary tasks. Training Stage includes Pre-training for initial representation alignment or generative learning, supervised fine-tuning (SFT) for task adaptation, and reinforcement learning (RL) for reward-based optimization. In the Construction column, Hybrid denotes the involvement of both humans and models, such as model-generated annotations with human validation.}
\label{tab:4-2}
\vskip -0.1in
\begin{adjustbox}{max width=\textwidth}
\footnotesize
\renewcommand{\arraystretch}{1.1}
\setlength{\tabcolsep}{2pt}
\begin{tabular}{@{}lcccccccc@{}}
\toprule
\textbf{Dataset} &
\textbf{Year} &
\textbf{Task} &
\textbf{Modality} &
\textbf{Scale} &
\textbf{Training Stage} &
\textbf{Supervision} &
\textbf{Construction} &
\textbf{Source} \\
\midrule

FABA-Instruct \cite{li24faba} & 2024 & FER & I & 19,877 Images & SFT & Rationale & Hybrid & AffectNet \\

FEID \cite{wang25ekevllm} & 2025 & FER & I & 26,238 Images & SFT & Label & Hybrid & Six Face Datasets \\

UniFER \cite{zhang25unifer} & 2025 & FER & I & 229K/359K Samples & SFT + RL & Rationale / Label & Hybrid & RAF-DB, AffectNet, \textit{etc.} \\

MMAFFIn \cite{liu25mmaffben} & 2025 & FER, VCER, CMER & I,V,T & 272,656 Samples & SFT & Label & Human & 19 Dataset Tracks \\

MEU-Instruct \cite{zhao25mellm} & 2025 & FER & I & 4,793 Samples & SFT & Rationale & Hybrid & DFME \\

R1-Omni-RL \cite{zhao25r1omni} & 2025 & FER & A,V & 15,306 Clips & RL & Label & Human & DFEW, MAFW \\

Emo-CFG \cite{zhang25videmo} & 2025 & FER, VCER & V,T & 2.1M Samples & SFT + RL & Rationale & Hybrid & Seventeen Video Datasets \\

FaceInstruct-1M \cite{chaube26facell} & 2026 & FER & I,V & $\sim$1M Pairs & Pre-training + SFT & Rationale & Hybrid & Thirteen Face Datasets \\

FEA-20K \cite{wu26facialr1} & 2026 & FER & I & 17,737 Samples & SFT + RL & Rationale & Hybrid & Seven Face Datasets \\

\rowcolor{gray!8}
EMOSpeech \cite{xu24secap} & 2024 & AER & A,T & 29,326 Utterances & Pre-training + SFT & Description & Human & In-House Speech \\

\rowcolor{gray!8}
PQA$^{*}$ \cite{wang25ctxpara} & 2025 & AER & A,T & 443,815 QA & SFT & Response & Model & MSP-Podcast \\

EmoVIT \cite{xie24emovit} & 2024 & VCER & I & 51,200 Images & SFT & Rationale & Model & EmoSet \\

EmoCause \cite{huang25emodete} & 2025 & VCER, MSU & A,V,T & $\sim$14,000 Samples & SFT & Rationale & Hybrid & LIRIS-ACCEDE, \textit{etc.} \\

Omni-SILA-Train \cite{luo25omnisila} & 2025 & VCER & V,T & 154,000 Clips & SFT & Grounding & Hybrid & TSL-300, CUVA \\

VEC-CoT \cite{wu26emocaliber} & 2025 & VCER & I & 143,446 Samples & SFT & Rationale & Model & EmoSet, WEBEmo, \textit{etc.} \\

EEmoDB-Assess \cite{gao26eemologic} & 2026 & VCER & I & 36,000 QA & RL & Label/Score/Rank & Hybrid & Seven VCER Datasets \\

INSETS-462K \cite{wu26emobserver} & 2026 & VCER & I,T & 462,369 Pairs & RL & Label & Hybrid & FI, EMOTIC, \textit{etc.} \\

EEmoDB-QA \cite{gao26eemologic} & 2026 & VCER & I,T & 1.2M QA & SFT & Rationale & Hybrid & Seven VCER Datasets \\

InsightVQA-Train \cite{wang26insvqa} & 2026 & VCER, MSU & I,T & 725,000 QA & SFT & Rationale & Hybrid & 138,000 Curated Images \\

AICA-Bench-Score \cite{she26aicabench} & 2026 & VCER & I,T & 10,000 Samples & SFT & Score & Human & AICA-Bench Responses \\

\rowcolor{gray!8}
MERR \cite{cheng24emotion} & 2024 & CMER & A,V,T & 28.6K/4.5K Clips & Pre-training + SFT & Desc. / Rationale & Hybrid & MER2023 \\

\rowcolor{gray!8}
R1-Omni-SFT \cite{zhao25r1omni} & 2025 & CMER & A,V & 580 Clips & SFT & Rationale & Human & EMER, HumanOmni \\

\rowcolor{gray!8}
AMT \cite{li25emoverse} & 2025 & CMER, MSU & I,V,T & 32,940 Samples & SFT & Rationale & Hybrid & CMU-MOSEI, MELD, \textit{etc.} \\

\rowcolor{gray!8}
DEEMO-MER \cite{li25deemo} & 2025 & CMER & A,V,T & 2,060 Clips & SFT & Rationale & Hybrid & DEEMO-NFBL \\

\rowcolor{gray!8}
MER-Caption(+) \cite{lian25affectgp} & 2025 & CMER & A,V,T & 115.6K/31.3K Clips & SFT & Description & Hybrid & MER2024 \\

\rowcolor{gray!8}
MMEVerse-Train \cite{peng26emotionl} & 2026 & CMER, FER, AER & A,V,T & $\sim$130,000 Clips & SFT & Rationale & Hybrid & E3, CMU-MOSEI, \textit{etc.} \\

\rowcolor{gray!8}
EmoWorld-130K \cite{huang26oneemo} & 2026 & CMER, MSU, MESC & A,V,T & $\sim$130,000 Samples & SFT + RL & Rationale & Hybrid & DFEW/MERR/MER-Caption+ \\

\rowcolor{gray!8}
EmoCue \cite{zhang26xemogpt} & 2026 & CMER & A,V & 40.7K/31.3K Samples & Pre-training + SFT & Desc. / Rationale & Hybrid & DFEW, MER-Caption+ \\

SoMeData \cite{zhang24somelvl} & 2024 & MSU, ITER & I,T & $\sim$654,000 Pairs & SFT & Rationale & Hybrid & Social-Media Corpora \\

MEC4 \cite{wu25m3f} & 2025 & MSU, CMER & A,V,T & 1,207 Dialogues & SFT & Relations & Human & \textit{Home with Kids} \\

SSRD \cite{kang25ssrgssrd} & 2025 & MSU & A,V,T & 1,202 Clips & SFT & Rationale & Hybrid & MUStARD++ \\

CPQA \cite{wang25ctxpara} & 2025 & MSU, AER & A,T & 32,960 QA & SFT & Response & Model & In-House Speech \\

GRASP \cite{kim26grasp} & 2026 & MSU & V,T & 290,148 Pairs & SFT + RL & Rationale & Hybrid & 46,000 Social Videos \\

\rowcolor{gray!8}
WaveMind-Instruct \cite{zeng25wavemind} & 2025 & PER & P & 28,301 Pairs & SFT & Response & Hybrid & SEED EEG Affect Subset \\

MESC \cite{chu25mesc} & 2025 & MESC, CMER & A,V,T & 28,762 Utterances & SFT & Response & Hybrid & \textit{In Treatment} \\

EChat-200K \cite{geng25osumecha} & 2025 & MESC, ESS & A,T & $\sim$200,000 Pairs & SFT & Rationale + Media & Hybrid & Speech Dialogues \\

LIME-440K \cite{zhao26cogaudio} & 2026 & MESC, AER & A,T & 438,884 Utterances & SFT + RL & Rationale & Hybrid & Synthetic + ECD-TSE/ESD \\

Sympatheia-18K \cite{dindar26sympat} & 2026 & MESC, ESS & A,T & 17,982 Pairs & SFT & Response + Media & Model & Spoken Dialogues \\

\rowcolor{gray!8}
SpeechCraft \cite{jin24speechcra} & 2024 & ESS & A,T & 2,249,579 Utterances & Pre-training + SFT & Description + Media & Model & Zhvoice, GigaSpeech-m, \textit{etc.} \\

\rowcolor{gray!8}
TextrolSpeech \cite{ji24textrolspe} & 2024 & ESS & A,T & 236,220 Utterances & Pre-training & Description + Media & Hybrid & LibriTTS, VCTK, \textit{etc.} \\

\rowcolor{gray!8}
EmoVoice-DB \cite{yang25emovoice} & 2025 & ESS & A,T & 21,050 Utterances & SFT & Description + Media & Model & GPT-4o, GPT-4o-Audio \\

\rowcolor{gray!8}
OV-Speech \cite{ren26ovtts} & 2026 & ESS & A,T & 316,807 Utterances & SFT & Rationale + Media & Model & ContextSpeech \\

\rowcolor{gray!8}
CoT-TTS \cite{xue26i2ctc} & 2026 & ESS & A,T & $\sim$3M Clips & Pre-training + SFT & Rationale + Media & Model & Bilingual Media Speech \\

\rowcolor{gray!8}
MultiEmo-RL \cite{zhou26hybemo} & 2026 & ESS & T & 14,400 Pairs & RL & Description & Model & Synthetic Instructions \\

\rowcolor{gray!8}
HybridEmo-SFT \cite{zhou26hybemo} & 2026 & ESS & A,T & 124,096 Utterances & SFT & Description + Media & Model & Emilia \\

EmotiPrompt \cite{dang25emoticra} & 2025 & EVGE & I,T & 39,843 Images & Pre-training & Description & Hybrid & OASIS, EMOTIC, \textit{etc.} \\

MoodArchive \cite{ye25moodifier} & 2025 & EVGE & I,T & $>$8M Pairs & Pre-training & Description & Hybrid & Unsplash, Pexels, \textit{etc.} \\

L-AVC \cite{luo26epemlavc} & 2026 & EVGE & I,T & 10,000 Triplets & SFT & Description + Media & Hybrid & EmoSet \\

\bottomrule
\end{tabular}
\end{adjustbox}
\vskip -0.1in
\end{table*}

\subsection{Affective Model-Tuning Datasets} \label{sec:4_2}

The transition to LLMs expands both the scale and expressive scope of affective supervision. Instruction following~\cite{ouyang22instru}, explicit reasoning through Chain-of-Thought (CoT) prompting~\cite{wei22cot}, multimodal understanding~\cite{liu23llava}, and generation~\cite{wu24nextgpt} are supported by increasingly diverse training formats, spanning labels, descriptions, explanations, responses, and target media. As illustrated in \cref{fig:4-1}, model-tuning datasets often aggregate and enrich conventional resources through human annotation, model generation, or hybrid pipelines combining automated construction with human guidance and validation. \Cref{tab:4-2} summarizes these resources, which we review below by primary task.

\textbf{FER Datasets.} The enrichment of conventional labels is exemplified by FABA-Instruct~\cite{li24faba}, which augments AffectNet images with expression and Action Unit (AU) descriptions. FEID~\cite{wang25ekevllm} places greater emphasis on category supervision through emotion knowledge and self-verification across facial datasets, while FEA-20K~\cite{wu26facialr1} brings recognition, AU analysis, and explanations into a shared annotation framework. Aggregation broadens this coverage further. FaceInstruct-1M~\cite{chaube26facell} consolidates thirteen facial datasets into approximately one million instruction pairs covering expressions and other facial attributes, and MMAFFIn~\cite{liu25mmaffben} incorporates facial affect into multilingual, multitask supervision.

Facial motion and temporal context add another dimension to these annotations. MEU-Instruct~\cite{zhao25mellm} supplies flow-guided descriptions of DFME micro-expressions, while Emo-CFG~\cite{zhang25videmo} aggregates fine-grained captions, questions, and reasoning annotations for facial and scene-level affect. The organization of supervision also varies with the training objective. UniFER~\cite{zhang25unifer} provides a CoT subset for rationale-based cold-start Supervised Fine-Tuning (SFT) and a category-labeled subset for Reinforcement Learning (RL), specifically Reinforcement Learning with Verifiable Rewards (RLVR). R1-Omni-RL~\cite{zhao25r1omni} similarly uses category-labeled audiovisual clips from DFEW and MAFW for reward-based optimization, complementing reference explanations with verifiable prediction targets.

\textbf{AER Datasets.} Descriptive supervision also brings vocal expression into a language-based format. EMOSpeech~\cite{xu24secap} pairs speech with human-written emotion captions, while PQA$^{*}$~\cite{wang25ctxpara} reformulates MSP-Podcast recordings into model-generated questions and answers about paralinguistic properties. Their distinction lies in the granularity of the account: captions characterize overall affective delivery, whereas questions direct attention to particular vocal attributes. The related CPQA resource situates these attributes within communicative contexts and is discussed under MSU.

\textbf{VCER Datasets.} The range of descriptive and question-based supervision broadens further when affect is associated with entire scenes and their interpretation. EmoVIT~\cite{xie24emovit} reformulates EmoSet into visual emotion instructions, while VEC-CoT~\cite{wu26emocaliber} combines resources to provide reasoning supervision for confidence-aware recognition. In video, descriptions also specify emotional causes and their locations. EmoCause~\cite{huang25emodete} supplies emotional-cause descriptions, and Omni-SILA-Train~\cite{luo25omnisila} integrates sentiment identification, localization, and attribution, giving interpretations an explicit connection to observable evidence.

Alongside this variation in evidence, collections differ in the judgments they elicit. EEmoDB-QA~\cite{gao26eemologic} combines perceptual, ranking, and descriptive questions for SFT, while EEmoDB-Assess~\cite{gao26eemologic} supports reward-based optimization of emotion classification, VAD scoring, and ranking. INSETS-462K~\cite{wu26emobserver} uses statement judgment, pairing emotional and contextual claims with labels indicating whether they are supported. InsightVQA-Train~\cite{wang26insvqa} organizes questions from perception through grounded understanding to higher-level cognition. Human ratings of the resulting responses provide a further form of supervision in AICA-Bench-Score~\cite{she26aicabench}, which trains a dedicated affective evaluator.

\textbf{CMER Datasets.} For audiovisual utterances, descriptive annotation brings together evidence distributed across speech, visual behavior, and language. MERR~\cite{cheng24emotion} augments MER2023 through staged annotation of facial AUs, visual context, and vocal characteristics, which an LLM combines with subtitles. Refinement and expert selection yield coarse- and fine-grained subsets for initial alignment and reasoning-oriented instruction tuning. MER-Caption(+)~\cite{lian25affectgp} follows a similar construction process, consolidating audiovisual cues with textual content and applying model-based filtering. Its descriptions accommodate diverse, co-occurring emotions for open-vocabulary MER.

The contributions of individual modalities remain explicit in EmoCue~\cite{zhang26xemogpt}, which pairs emotional cues with concise captions and detailed explanations. R1-Omni-SFT~\cite{zhao25r1omni} draws explanatory examples from EMER and HumanOmni for initial supervised training before label-based RL. DEEMO-MER~\cite{li25deemo} adds a complementary emphasis on the observations themselves, using de-identified inputs and non-facial behavior to broaden the evidence available for interpretation.

These recognition resources also serve as building blocks for wider task coverage. AMT~\cite{li25emoverse} reorganizes collections such as CMU-MOSEI and MELD around sentiment, emotion, and emotion-related inference, while MMEVerse-Train~\cite{peng26emotionl} spans conversational, facial, and acoustic settings. EmoWorld-130K~\cite{huang26oneemo} brings open-vocabulary MER, basic recognition, social understanding, and supportive response generation into a shared framework through theory-guided reasoning annotation, automated verification, and human revision.

\textbf{MSU Datasets.} This broader coverage brings affect into contact with intentions, nonliteral meanings, and interpersonal relationships. SoMeData~\cite{zhang24somelvl} consolidates social-media corpora into varied instructions, including affect-related image-text tasks. SSRD~\cite{kang25ssrgssrd} focuses on subtype-specific sarcasm and the multimodal incongruity underlying it. Conversational context takes a more structured form in MEC4~\cite{wu25m3f}, whose emotion-cause relations connect emotional expressions with relevant events in Chinese dialogues from \textit{Home with Kids}.

Context is carried not only by conversational events but also by delivery and behavior. CPQA~\cite{wang25ctxpara} complements PQA$^{*}$ with situational questions about vocal characteristics, while GRASP~\cite{kim26grasp} grounds social reasoning in gaze, gestures, and participant-specific events in multiperson video. These collections capture complementary aspects of how affective meaning emerges within interaction.

\textbf{PER Datasets.} Language-based reformulation also accommodates physiological observations. WaveMind-Instruct~\cite{zeng25wavemind} pairs SEED EEG segments with instructions derived from their emotion annotations. Manually curated factual definitions associated with the labels form seed question-answer pairs, which LLM rewriting diversifies into closed-ended, open-ended, and multiple-choice tasks. Source-definition validation, screening for speculative EEG details and neural mechanisms, and instruction deduplication keep this varied linguistic supervision anchored to the original annotations.

\textbf{MESC Datasets.} Alongside resources for interpreting affect, generation-oriented collections pair emotional contexts with appropriate responses. The MESC dataset~\cite{chu25mesc} derives multimodal support conversations from \textit{In Treatment}, associating emotional understanding with supportive dialogue. EChat-200K~\cite{geng25osumecha} incorporates dialogue reasoning and target speech, bringing response content and affective delivery into the same training setting.

Controlled variation helps separate these dimensions. LIME-440K~\cite{zhao26cogaudio} combines synthetic material with existing speech resources to present lexically identical utterances delivered with different emotions, isolating variation in acoustic expression. Sympatheia-18K~\cite{dindar26sympat} varies the intended affect of spoken responses, including different response emotions for emotionally neutral queries. The former emphasizes how emotion is conveyed in an input, while the latter diversifies how it is expressed in a response.

\textbf{ESS Datasets.} The distinction between what is said and how it is delivered also structures supervision for speech synthesis. TextrolSpeech~\cite{ji24textrolspe} and SpeechCraft~\cite{jin24speechcra} pair existing recordings with natural-language descriptions of emotional and acoustic properties, with SpeechCraft providing broader descriptive coverage. EmoVoice-DB~\cite{yang25emovoice} follows a synthetic route, generating expressive speech alongside detailed emotion descriptions. Both construction strategies associate linguistic specifications with acoustic realizations, using recorded or generated performances as targets.

\textbf{EVGE Datasets.} Descriptions play a similar conditioning role in visual generation, where affective intentions are associated with image appearance and its transformation. EmotiPrompt~\cite{dang25emoticra} aligns image content, textual descriptions, and continuous affect, while MoodArchive~\cite{ye25moodifier} expands emotion-attribute associations through hierarchical descriptions of large-scale image collections. L-AVC~\cite{luo26epemlavc} turns the focus toward the transformation itself, supplying image-editing triplets derived from EmoSet that pair affective changes with the preservation of unrelated content. These resources connect learning how visual properties relate to emotion with learning how to modify them under affective instructions.

\begin{table*}[t]
\centering
\caption{Affective Evaluation Benchmarks with their key attributes, grouped by primary task. MCQ, Text, and Rank denote multiple-choice questions, free-form responses, and ordered predictions, respectively. In the Scoring column, R denotes rule-based metrics, M denotes learned evaluators, L denotes LLM-assisted evaluation, and H denotes human assessment. TTS denotes Text-to-Speech techniques.}
\vskip -0.1in
\label{tab:4-3}
\begin{adjustbox}{max width=\textwidth}
\footnotesize
\renewcommand{\arraystretch}{1.1}
\setlength{\tabcolsep}{2pt}
\begin{tabular}{@{}lccccccc@{}}
\toprule
\textbf{Benchmark} &
\textbf{Year} &
\textbf{Task} &
\textbf{Modality} &
\textbf{Scale} &
\textbf{Response Format} &
\textbf{Scoring} &
\textbf{Source Data} \\
\midrule

FEAB \cite{wang25ekevllm} & 2025 & FER & I & 3,035 Images & Label, Score & R & AffectNet, RAF-DB, \textit{etc.} \\

FERBench \cite{zhang25unifer} & 2025 & FER & I & 11,072 Images & Label & R & RAF-DB, AffectNet, \textit{etc.} \\

FaceBench \cite{wang25facebenc} & 2025 & FER & I & 3,988 QA & MCQ & R+L & RAF-DB, RAF-AU \\

FEA-20K \cite{wu26facialr1} & 2026 & FER & I & 1,688 Samples & Label, Text & R & Seven Face Datasets \\

\rowcolor{gray!8}
EMOSpeech \cite{xu24secap} & 2024 & AER & A,T & 600 Utterances & Text & R+M+H & In-House Speech \\

\rowcolor{gray!8}
VoxEmo \cite{zhang26voxemo} & 2026 & AER & A & 35 Corpora & Label, Distribution & R & IEMOCAP, MSP-Podcast, \textit{etc.} \\

EEmo-Bench \cite{gao25eemobench} & 2025 & VCER & I & 6,773 QA & MCQ, Score, Text, Rank & R+L+H & Flickr Images \\

EIBench \cite{lin25eibench} & 2025 & VCER & I,T & 1,665 Samples & Text & R+L & CAER-S, EmoSet \\

Omni-SILA \cite{luo25omnisila} & 2025 & VCER & V & 48,000 Clips & Label, Text, Grounding & R+L+H & TSL-300, CUVA \\

VECBench \cite{wu26emocaliber} & 2025 & VCER & I & 8,235 Images & Label, Score & R & FI, WEBEmo, EmoSet, \textit{etc.} \\

MVEI \cite{wu26mvei} & 2026 & VCER & I,T & 3,086 Pairs & Label & R & EmoSet \\

MultiEmo-Bench \cite{chen26multiemo} & 2026 & VCER & I & 10,344 Images & Distribution, Text & R & EmoSet, FI \\

ArtECulture \cite{chen26artecult} & 2026 & VCER & I,T & 679 Images & Label, Text & R+L+H & ArtELingo, VULCA-BENCH \\

InsightVQA-Bench \cite{wang26insvqa} & 2026 & VCER, MSU & I,T & 30,841 QA & MCQ, Rank & R & Curated Web Images \\

E$^{3}$mo-Bench \cite{gao26e3mobench} & 2026 & VCER, CMER & A,V,T & 12,314 QA & MCQ, Label, Score & R & DFEW, LIRIS-ACCEDE, \textit{etc.} \\

AICA-Bench \cite{she26aicabench} & 2026 & VCER & I,T & 18,124 QA & MCQ, Text & R+M & Nine Affective Datasets \\

\rowcolor{gray!8}
EMER \cite{lian23emer} & 2023 & CMER & A,V,T & 332 Clips & Text & R+L & MER2023 \\

\rowcolor{gray!8}
ECF 2.0 \cite{wang24ecf} & 2024 & CMER & A,V,T & 341 Dialogues & Label, Grounding & R & \textit{Friends} \\

\rowcolor{gray!8}
OV-MERD \cite{lian25ovmer} & 2025 & CMER & A,V,T & 332 Clips & Label & R+L & MER2023 / EMER \\

\rowcolor{gray!8}
CA-MER \cite{han25camer} & 2025 & CMER & A,V,T & 1,500 Clips & Text & R+L & MER2023 \\

\rowcolor{gray!8}
EmoBench-M \cite{hu25emobenchm} & 2025 & CMER, AER, MSU & A,V,T & 5,646 Samples & MCQ, Text & R+L & MELD, CH-SIMS, RAVDESS, \textit{etc.} \\

\rowcolor{gray!8}
EmotionHallucer \cite{xing26emotionh} & 2026 & CMER, AER, VCER, MSU & I,A,V,T & 2,742 QA & Label & R & RAVDESS, MER2023, Social-IQ, \textit{etc.} \\

\rowcolor{gray!8}
MME-Emotion \cite{zhang26mmeemot} & 2026 & CMER, FER, MSU & A,V,T & 6,500 QA & Label, Text & R+L & IEMOCAP, MELD, DFEW, \textit{etc.} \\

\rowcolor{gray!8}
EmoTrans \cite{hu26emotrans} & 2026 & CMER & A,V,T & 3,274 QA & Label, Text & R+M+L & OpenHumanVid, CH-SIMS v2 \\

\rowcolor{gray!8}
EmoPrefer-Bench \cite{lian26emoprefe} & 2026 & CMER & A,V,T & 1,368 Samples & Label & R & MERR-Fine, MER-Caption+ \\

\rowcolor{gray!8}
EmoS \cite{guo26emos} & 2026 & CMER & A,V,T & 9,403 Utterances & Label, Score & R & MELD, CH-SIMS v2 \\

\rowcolor{gray!8}
MMEVerse-Bench \cite{peng26emotionl} & 2026 & CMER, FER, VCER, MSU & A,V,T & 36,133 Samples & Label & R+L & IEMOCAP, MELD, DFEW, \textit{etc.} \\

\rowcolor{gray!8}
EmoCue-Eval \cite{zhang26xemogpt} & 2026 & CMER & A,V & 400 Samples & Text & R+M+L & Curated Emotion Videos \\

MM-BigBench \cite{yang23mmbigben} & 2023 & ITER, CMER, MSU & I,T & 28,745 Samples & MCQ, Label & R & MVSA, Twitter15/17, MOSI, \textit{etc.} \\

\rowcolor{gray!8}
Hummus \cite{tong25hummus} & 2025 & MSU & I,T & 940 Pairs & Label, Text, Grounding & R & \textit{New Yorker} Caption Contest \\

\rowcolor{gray!8}
MoMentS \cite{villac25moment} & 2025 & MSU & A,V,T & 2,344 QA & MCQ & R & SF20K / Omeleto Films \\

\rowcolor{gray!8}
IntentBench \cite{yang25humanomn} & 2025 & MSU & A,V,T & 2,689 QA & MCQ & R & Social-IQ 2.0, EMER, MDPE \\

\rowcolor{gray!8}
CPQA-Eval \cite{wang25ctxpara} & 2025 & MSU, AER & A,T & 3,857 QA & Text & L & In-House Speech \\

\rowcolor{gray!8}
GridToM \cite{li25gridtom} & 2025 & MSU & V,T & 1,296 Pairs & MCQ & R & MultiGrid Simulations \\

\rowcolor{gray!8}
MTMEUR \cite{hu25mtmeur} & 2025 & MSU, CMER & V,T & 5,101 QA & MCQ & R & Pexels, Mixkit \\

\rowcolor{gray!8}
CEDAR \cite{dai26cedar} & 2026 & MSU, VCER & I,T & 2,800 Samples & Label & R & Web Images + Synthetic Scenarios \\

\rowcolor{gray!8}
HumanSense \cite{qin26humansns} & 2026 & MSU, FER & A,V,T & 3,882 QA & MCQ, Score & R & Public Datasets + YouTube \\

\rowcolor{gray!8}
SpeechEQ \cite{wu26speecheq} & 2026 & MSU & A,T & 2,265 Dialogues & MCQ & R & GPT-4o Scripts + TTS \\

\rowcolor{gray!8}
EmoSBench \cite{wang26emos} & 2026 & MSU, AER & A,T & $\sim$4.5K Samples & Score, Text & R & EmoDialogue, YouTube \\

\rowcolor{gray!8}
HumDial \cite{zhao26i2hc} & 2026 & MSU, MESC, ESS & A & 2,332 Utterances & Text, Speech & L+H & Actor-Recorded Dialogues \\

\rowcolor{gray!8}
HumDial-EIBench \cite{wang26humdiale} & 2026 & MSU, MESC, ESS & A & 1,077 Samples & MCQ, Text, Speech & R+L+H & HumDial + Recorded Dialogues \\

WaveMind-Bench \cite{zeng25wavemind} & 2025 & PER & P & 6,000 QA & MCQ & R & SEED \\

\rowcolor{gray!8}
AEQ-Bench \cite{luo26aeqbench} & 2026 & MESC, ESS & A,T & 1,885 Samples & Label, Text, Speech & R+L+H & MELD, GigaSpeech, EmoV-DB \\

E-VOC \cite{lin26instructi} & 2026 & ESS & T & 3,600 Utterances & Speech & H & Gemini Scripts + TTS \\

CoT-TTS \cite{xue26i2ctc} & 2026 & ESS & A,T & 1,200 Samples & Text, Speech & R+M+L+H & Film, TV and Radio Drama \\

MultiEmo-Test \cite{zhou26hybemo} & 2026 & ESS & A,T & 720 Conditions & Speech & R+M+L+H & Synthetic + Seed-TTS \\

\bottomrule
\end{tabular}
\end{adjustbox}
\vskip -0.1in
\end{table*}

\subsection{Affective Evaluation Benchmarks} \label{sec:4_3}

\textit{Affective Evaluation Benchmarks} complement model-tuning datasets by providing a common basis for assessing capabilities. They draw on existing test resources, newly collected content, or both, pairing them with task-specific response formats and scoring protocols to evaluate recognition, reasoning, and generation (\cref{fig:4-1}). \Cref{tab:4-3} summarizes their main characteristics, and we review them below by primary task.

\textbf{FER Benchmarks.} Established expression datasets remain a foundation for evaluating facial affect recognition. FEAB~\cite{wang25ekevllm} and FERBench~\cite{zhang25unifer} build on resources such as AffectNet and RAF-DB, with the former covering categorical and numerical judgments and the latter focusing on label-based recognition. Other benchmarks extend this assessment to the facial evidence underlying a prediction. The expression and AU components of FaceBench~\cite{wang25facebenc} form part of a broader multiple-choice evaluation of facial perception, while the held-out portion of FEA-20K~\cite{wu26facialr1} pairs emotional predictions with AU analysis and explanations. Together, these resources extend evaluation from identifying an expression to articulating its observable basis.

\textbf{AER Benchmarks.} A similar shift toward descriptive assessment appears in acoustic evaluation. The EMOSpeech test set~\cite{xu24secap} evaluates emotion captions against human references using computational metrics, learned similarity measures, and human judgments, emphasizing how accurately a description conveys the affective qualities of a recording. VoxEmo~\cite{zhang26voxemo} instead focuses on robustness across conditions, bringing together 35 corpora across 15 languages and examining sensitivity to recording settings and prompts. Its distribution-aware evaluation also accounts for annotator disagreement, recognizing that reliable acoustic understanding involves both generalization across domains and sensitivity to differences in human interpretation.

\textbf{VCER Benchmarks.} Differences in interpretation are particularly salient when evaluating emotions evoked by visual content. MultiEmo-Bench~\cite{chen26multiemo} captures multiple emotional judgments through distributions and descriptions, whereas VECBench~\cite{wu26emocaliber} examines whether predictive confidence corresponds to recognition reliability. These benchmarks distinguish variation in human responses from uncertainty in model predictions. EEmo-Bench~\cite{gao25eemobench} broadens the assessment through categorical and dimensional judgments, emotion rankings, and descriptive responses, capturing aspects of visual affect that a single response format may not fully represent.

Context and perspective further shape these judgments. MVEI~\cite{wu26mvei} evaluates emotional and contextual statements, while ArtECulture~\cite{chen26artecult} incorporates cultural perspectives into artwork judgments and explanations. EIBench~\cite{lin25eibench} examines the visual and contextual triggers underlying an interpretation, and InsightVQA-Bench~\cite{wang26insvqa} extends assessment from perception to grounded understanding and higher-level cognition. AICA-Bench~\cite{she26aicabench} also includes emotion-conditioned text generation, using a learned evaluator to assess open-ended responses. Across these resources, emotional judgments are increasingly assessed alongside the context, evidence, and explanations that support them.

Video adds questions about where affective evidence occurs and whose emotions are being evaluated. Omni-SILA~\cite{luo25omnisila} combines sentiment identification, localization, and attribution, explicitly assessing where sentiment is expressed and to whom it is attributed. E$^{3}$mo-Bench~\cite{gao26e3mobench} distinguishes emotions expressed within multimodal content from those evoked in viewers. These complementary emphases clarify both the evidence to which a judgment refers and the perspective from which affect is interpreted.

\textbf{CMER Benchmarks.} The relationship between interpretation and evidence also guides evaluation across linguistic, acoustic, and visual signals. EMER~\cite{lian23emer} and OV-MERD~\cite{lian25ovmer} use MER2023 clips to assess textual emotional interpretation and open-vocabulary recognition. They demonstrate how different annotations and task formulations can broaden evaluation without changing the underlying media. ECF 2.0~\cite{wang24ecf} and EmoCue-Eval~\cite{zhang26xemogpt} focus more directly on supporting evidence, assessing conversational emotion-cause relations and the identification of expert-annotated emotional cues.

The reliability of an interpretation, however, depends on more than the detail of its explanation. CA-MER~\cite{han25camer} examines conflicting affective cues across modalities, while EmotionHallucer~\cite{xing26emotionh} probes unsupported emotional inferences through structured questions. EmoPrefer-Bench~\cite{lian26emoprefe} introduces a human-centered criterion by testing whether models can identify preferred emotional descriptions. These resources assess distinct aspects of interpretive quality: handling conflicting evidence, avoiding unsupported conclusions, and recognizing descriptions that align with human preferences.

Evaluation also extends beyond individual judgments to emotions unfolding over time. EmoTrans~\cite{hu26emotrans} examines the understanding, explanation, and prediction of emotion transitions, while EmoS~\cite{guo26emos} addresses streaming emotional understanding. Broader suites complement these targeted assessments by bringing multiple affective tasks into a common framework. EmoBench-M~\cite{hu25emobenchm} connects basic recognition with conversational and socially complex understanding, MME-Emotion~\cite{zhang26mmeemot} covers recognition and reasoning across affective settings, and MMEVerse-Bench~\cite{peng26emotionl} spans conversational, facial, visual, and social tasks. Their broader coverage helps establish whether capabilities demonstrated in specific settings extend to other forms of affective understanding.

\textbf{ITER Benchmarks.} This approach to unified evaluation also applies to tasks that share an image-text input format but differ in their prediction targets. The affect-related subset of MM-BigBench~\cite{yang23mmbigben} brings together resources such as MVSA and Twitter-15/17 alongside related sentiment and social-understanding tasks. It distinguishes whole-post sentiment from entity-specific judgments while varying task instructions to examine sensitivity to prompt formulation. The resulting framework supports comparisons across tasks and prompting conditions while preserving differences in what each prediction concerns.

\textbf{MSU Benchmarks.} Beyond sentiment toward a post or entity, multimodal content can convey implicit intentions, humor, and social meaning. Hummus~\cite{tong25hummus} and IntentBench~\cite{yang25humanomn} examine these dimensions in complementary settings. Hummus assesses the interpretation and grounding of humorous multimodal metaphors in \textit{New Yorker} cartoons, whereas IntentBench uses questions derived from established audiovisual resources to evaluate communicative intentions. Both extend assessment beyond observable expressions to the meanings conveyed through them.

Interpreting such meanings often requires understanding participants' circumstances and perspectives. MoMentS~\cite{villac25moment} captures the narrative complexity of social interactions in short films, while GridToM~\cite{li25gridtom} uses controlled simulations to isolate differences in what participants observe. These settings offer complementary approaches to evaluating mental-state reasoning under naturalistic and controlled conditions. MTMEUR~\cite{hu25mtmeur} and CEDAR~\cite{dai26cedar} introduce situational and cultural perspectives, while HumanSense~\cite{qin26humansns} connects perception and contextual understanding with appropriate behavioral choices. Assessment thus extends from interpreting a situation to determining a suitable response.

In spoken interaction, appropriate responses also depend on vocal delivery and dialogue context. CPQA-Eval~\cite{wang25ctxpara} assesses contextual paralinguistic interpretation through open-ended responses, whereas SpeechEQ~\cite{wu26speecheq} evaluates socially appropriate response selection in theory-grounded, multi-turn scenarios. EmoSBench~\cite{wang26emos} adopts a broader view, covering the abilities to perceive, understand, use, and manage emotion. HumDial~\cite{zhao26i2hc} extends evaluation to response generation in actor-recorded spoken interactions, and HumDial-EIBench~\cite{wang26humdiale} combines emotional tracking and causal reasoning with empathetic speech generation. These benchmarks increasingly assess social understanding together with the ability to respond appropriately during an interaction.

\textbf{PER Benchmarks.} Physiological recordings offer a complementary source of affective evidence beyond observable behavior and communicative content. The affect-related subset of WaveMind-Bench~\cite{zeng25wavemind} converts held-out SEED EEG recordings into multiple-choice questions, using alternative emotion labels as distractors. Varying the candidate sets tests recognition under different question formulations, while comparisons between real and random EEG inputs examine whether predictions depend on the physiological evidence. This design assesses not only answer accuracy but also the contribution of the underlying signal.

\textbf{MESC Benchmarks.} Moving from emotion recognition to empathetic interaction, AEQ-Bench~\cite{luo26aeqbench} pairs speech from MELD, GigaSpeech, and EmoV-DB with model-generated, human-validated textual contexts. By varying the contexts of MELD and GigaSpeech utterances and the emotional tones of identical verbal content from EmoV-DB, it tests whether responses adapt to situational meaning and vocal cues, assessing response differentiation, coherence, supportiveness, and vocal delivery. It further compares model judgments of spoken responses with human assessments, linking empathetic generation to the reliability of automatic evaluation.

\textbf{ESS Benchmarks.} Beyond empathetic delivery, expressive speech benchmarks examine how a wider range of affective intentions are specified and realized. E-VOC~\cite{lin26instructi} uses human assessment to evaluate how well generated speech conveys explicit expressive instructions. CoT-TTS~\cite{xue26i2ctc} additionally assesses contextual reasoning about an appropriate speaking manner alongside the resulting speech, while MultiEmo-Test~\cite{zhou26hybemo} distinguishes sequential emotional trajectories from simultaneous emotional blends. These settings extend evaluation from instruction fidelity to context-sensitive delivery and complex emotional expression, with human and automated assessments covering affective realization, contextual appropriateness, and acoustic quality.

\begin{figure*}
    \centering
    %\vskip -0.04in
    \includegraphics[width=1\linewidth]{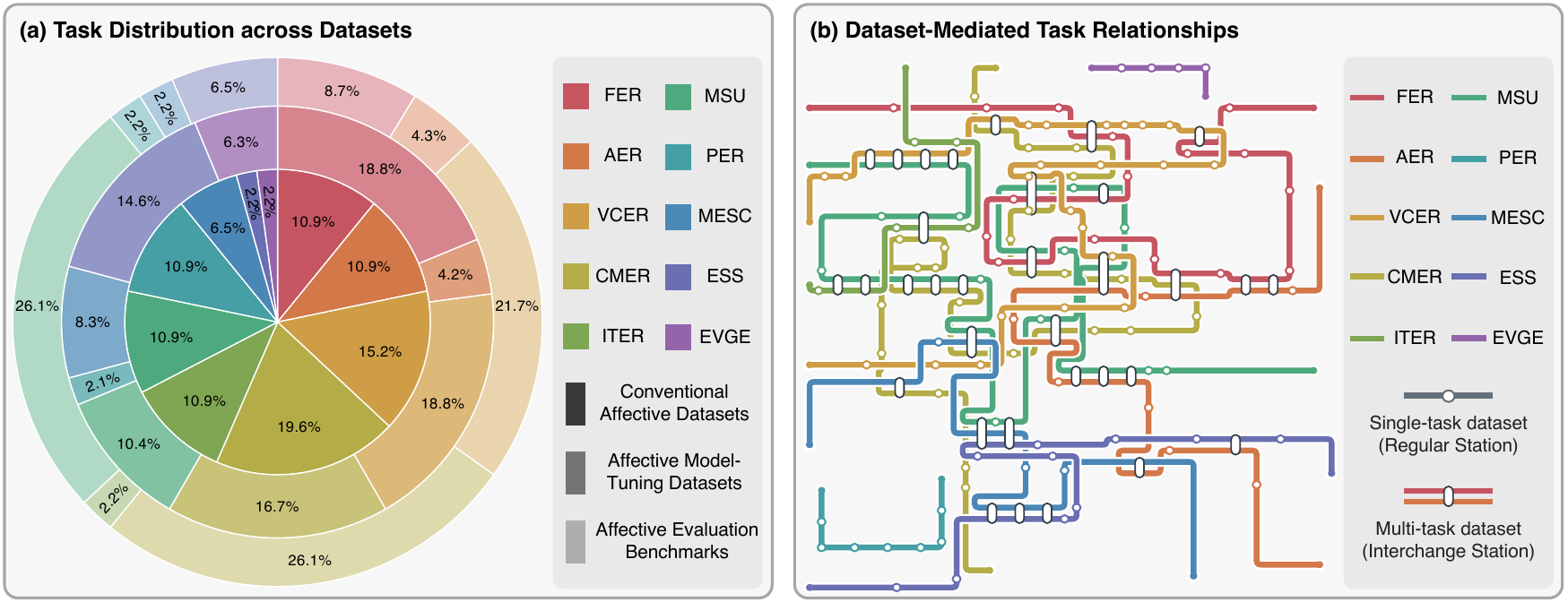}
    \vskip -0.1in
    \caption{Visualization of task distribution and relationships from the dataset perspective. Panel (a) shows dataset proportions by primary task for conventional datasets, model-tuning datasets, and evaluation benchmarks from the innermost to the outermost ring. Primary task denotes the first task listed for each dataset in \cref{tab:4-1,tab:4-2,tab:4-3}. Panel (b) represents datasets as metro stations and tasks as lines, with interchanges indicating where multiple tasks share the same dataset.}
    \label{fig:4-2}
    \vskip -0.1in
\end{figure*}

\subsection{Discussion}

Together, resources developed across diverse tasks provide a shared foundation for affective model development and evaluation. The synthesis in \cref{fig:4-2} offers a broader view of the field's current priorities and gaps.

\subsubsection{\textbf{Quantitative Analysis}}
The task distributions in \cref{fig:4-2} (a) reveal different emphases across the three dataset categories. Conventional resources cover a broad range of recognition tasks, whereas approximately 74\% of the evaluation benchmarks focus on CMER, MSU, and VCER, concentrating assessment on contextual and interpretive capabilities. In contrast, ESS accounts for a smaller share of benchmarks than model-tuning datasets, and EVGE has no dedicated benchmark in the distribution. This imbalance suggests that the growing provision of generative supervision needs stronger evaluation support, particularly for assessing the quality and controllability of affective outputs.

\subsubsection{\textbf{Qualitative Analysis}}
Beyond individual tasks, the metro-style map in \cref{fig:4-2} (b) illustrates their connections through shared datasets. CMER and MSU exhibit the greatest overlap, linking emotional interpretation with broader social understanding. Shared resources for MSU, MESC, and ESS further connect social interpretation with supportive responses and expressive speech. PER and EVGE, by contrast, remain separate lines without shared stations. Overall, existing resources provide an abundant foundation for LLM-based affective intelligence, although important gaps remain. Because individual tasks essentially capture complementary aspects of the same real-world situations, the next stage of dataset development calls for a more unified perspective, bringing modalities and tasks together within richer contexts that reflect the complexity of human affective experience.

\section{MER-with-LLMs}
\label{sec:5}

Building on these data resources, this section introduces MER-with-LLMs methods, which typically use LLMs to infer affective states and meanings from multimodal evidence. Such understanding is fundamental to affective intelligence and can inform contextually appropriate affective generation. Following our challenge-oriented taxonomy, we organize these methods into three categories aligned with the core challenges: \textit{Data-Efficient Affective Elicitation}, \textit{Multimodal Affective Representation}, and \textit{Affective Reasoning and Explanation}.

\subsection{Data-Efficient Affective Elicitation}
\label{sec:5_1}

\begin{figure*}
    \centering
    \includegraphics[width=1\linewidth]{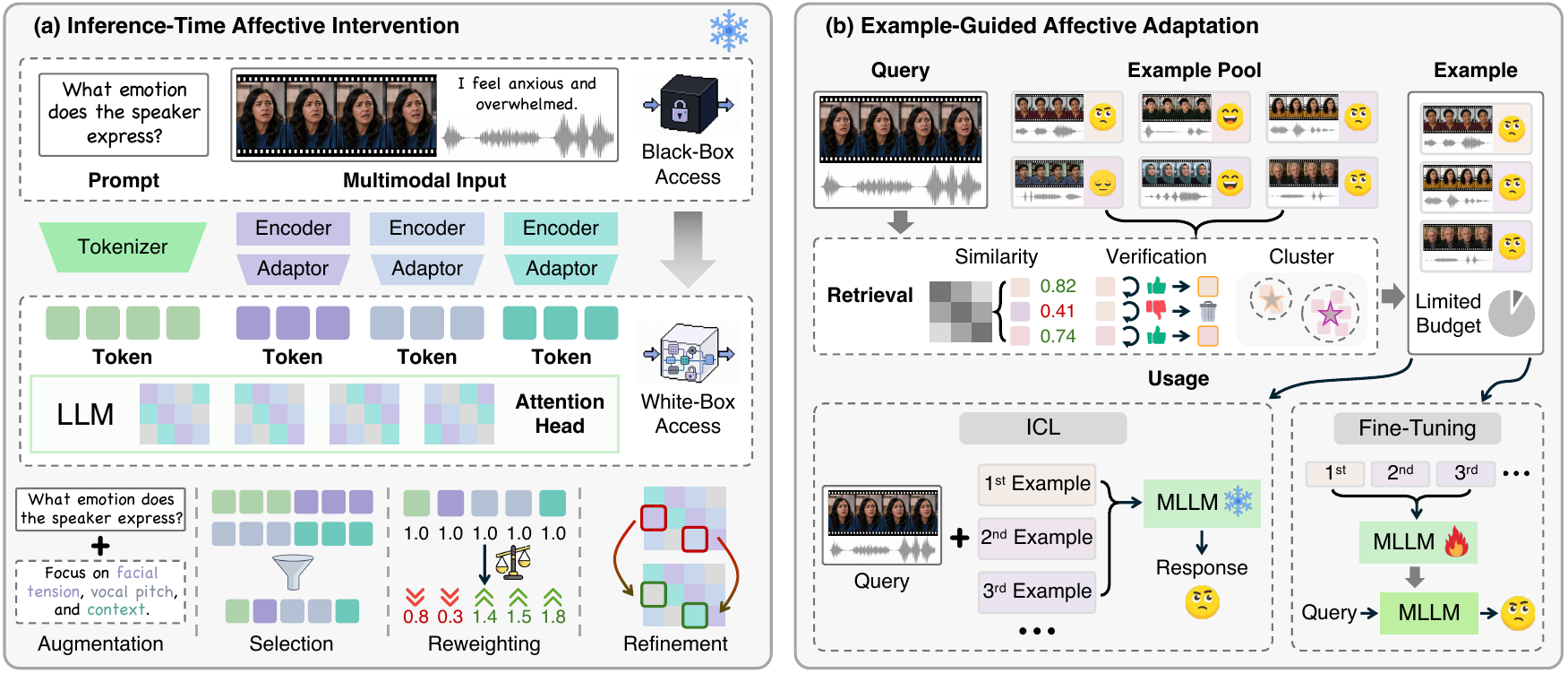}
    \vskip -0.1in
    \caption{Illustration of the two branches of \textit{Data-Efficient Affective Elicitation}. \textit{Inference-Time Affective Intervention} (a) modifies prompts and multimodal inputs under black-box access, or tokens and attention heads under white-box access, through operations such as augmentation and selection. \textit{Example-Guided Affective Adaptation} (b) retrieves examples within a limited budget and organizes them as In-Context Learning (ICL) demonstrations or fine-tuning data.}
    \label{fig:5-2}
    \vskip -0.1in
\end{figure*}

\textit{Subjective and Contextual Variability} gives rise to both \textit{Affective Evaluation Ambiguity} and \textit{Affective Data Scarcity}. While the former informs the design of datasets and evaluation protocols, the latter constrains model training and motivates methods collectively termed \textit{Data-Efficient Affective Elicitation}. These methods draw on affective knowledge encoded in pretrained models while requiring little or no additional task-specific data. We distinguish two approaches by their use of reference examples: \textit{Inference-Time Affective Intervention} elicits existing capabilities during inference, whereas \textit{Example-Guided Affective Adaptation} uses a small set of examples to guide or adapt model behavior. \Cref{fig:5-2} illustrates their workflows, which we elaborate on in the following subsections.

\begin{table*}[t]
\centering
\caption{Inference-Time Affective Intervention methods with their key attributes. Model Access indicates the access required, and Intervention Site identifies the principal intervention location.}
\label{tab:5-1}
\vskip -0.1in
\begin{adjustbox}{max width=\textwidth}
\footnotesize
\renewcommand{\arraystretch}{1.1}
\setlength{\tabcolsep}{5pt}
\begin{tabular}{@{}lccccccc@{}}
\toprule
\textbf{Method} &
\textbf{Year} &
\textbf{Task} &
\textbf{Modality} &
\textbf{Model Access} &
\textbf{Intervention Site} &
\textbf{Operation} &
\textbf{Key Design} \\
\midrule
% OmniVox~\cite{murzak25omnivo}, SpeechCueLLM~\cite{wu25beyond}, NarraCap~\cite{etesam24contex}, ELENA~\cite{saim25anatomy}, Contextual PQA~\cite{wang25ctxpara}, Revise-Reason-Recognize~\cite{li25revise}, SoV~\cite{zhang24visual}, SoVTP~\cite{wang25visualte}, MultiEEG-GPT~\cite{hu24multieeggp}, MERMAID~\cite{yang25mermaid}, SEPM~\cite{fang25catch}, WETR~\cite{liu26wetr}, GridToM~\cite{li25gridtom}
% Black-box / Prompt / Augmentation

OmniVox~\cite{murzak25omnivo} & 2025 & AER & A,T &
Black-box & Prompt & Augmentation &
Acoustic and Conversational Prompting \\

SpeechCueLLM~\cite{wu25beyond} & 2025 & AER & A,T &
Black-box & Prompt & Augmentation &
Vocal-Cue Verbalization in Prompts \\

NarraCap~\cite{etesam24contex} & 2024 & VCER & I &
Black-box & Prompt & Augmentation &
Visual-Cue Narrative Captions \\

ELENA~\cite{saim25anatomy} & 2025 & VCER & I &
Black-box & Prompt & Augmentation &
Body-Part Emotion Prompting \\

Contextual PQA~\cite{wang25ctxpara} & 2025 & AER, MSU & A,T &
Black-box & Prompt & Augmentation &
Timestamped Emotion Metadata \\

% Black-box / Prompt / Refinement

Revise-Reason-Recognize~\cite{li25revise} & 2025 & AER & A,T &
Black-box & Prompt & Refinement &
Transcripts Revision and Reasoning \\

% Black-box / Multimodal Input / Augmentation

\rowcolor{gray!8}
SoV~\cite{zhang24visual} & 2024 & FER & I &
Black-box & Multimodal Input & Augmentation &
Numbered Boxes and Facial Landmarks \\

\rowcolor{gray!8}
SoVTP~\cite{wang25visualte} & 2025 & FER & V &
Black-box & Multimodal Input & Augmentation &
Hierarchical Visual-Textual Cues \\

\rowcolor{gray!8}
MultiEEG-GPT~\cite{hu24multieeggp} & 2024 & PER & I,A,P,T &
Black-box & Multimodal Input & Augmentation &
EEG Maps with Multimodal Cues \\

% Black-box / Multimodal Input / Refinement

\rowcolor{gray!8}
MERMAID~\cite{yang25mermaid} & 2025 & VCER & I &
Black-box & Multimodal Input & Refinement &
Visual Augmentation and Reflection \\

% White-box / Token / Selection

SEPM~\cite{fang25catch} & 2025 & VCER & I &
White-box & Token & Selection &
Attention-Guided Token Pruning \\

% White-box / Token / Selection + Reweighting

WETR~\cite{liu26wetr} & 2026 & FER, CMER & V,T &
White-box & Token & Reweighting &
Noise Filtering and Emotion Reweighting \\

% White-box / Attention Head / Refinement

\rowcolor{gray!8}
GridToM~\cite{li25gridtom} & 2025 & MSU & I,T &
White-box & Attention Head & Refinement &
Probe-Guided Activation Steering \\

\bottomrule
\end{tabular}
\end{adjustbox}
\vskip -0.1in
\end{table*}

\subsubsection{\textbf{Inference-Time Affective Intervention}} \label{sec:5_1_1}
\textit{Inference-Time Affective Intervention} modifies model inputs or internal activations to elicit pretrained affective knowledge and strengthen the influence of relevant evidence while keeping backbones frozen. As summarized in \cref{tab:5-1}, we organize these methods by model access, then by intervention site and operation, with mechanisms illustrated in \cref{fig:5-2} (a). The following discussion distinguishes these methods by intervention site.

\textbf{Prompt-level Intervention.}
Textual prompts make affective evidence easier to interpret by verbalizing non-textual cues, supplying context, and structuring reasoning. Acoustic approaches connect linguistic content with vocal expression: SpeechCueLLM~\cite{wu25beyond} converts prosodic features into textual descriptions, OmniVox~\cite{murzak25omnivo} combines acoustic and conversational prompting, and Contextual PQA~\cite{wang25ctxpara} adds temporal specificity through timestamped emotion metadata. The same principle extends to visual evidence, with NarraCap~\cite{etesam24contex} describing people and their surroundings and ELENA~\cite{saim25anatomy} directing interpretation toward bodily expressions. Beyond enriching the available context, Revise-Reason-Recognize~\cite{li25revise} refines noisy transcripts before reasoning and prediction to limit error propagation into affective judgments.

\textbf{Multimodal-Input Intervention.}
Complementing this linguistic guidance, interventions on multimodal inputs make affective cues more explicit in the observations themselves. SoV~\cite{zhang24visual} uses spatial annotations to highlight relevant individuals and facial details while preserving context, and SoVTP~\cite{wang25visualte} extends this guidance to video through coordinated visual and textual cues. Beyond highlighting visible behavior, MultiEEG-GPT~\cite{hu24multieeggp} introduces complementary physiological evidence by combining EEG topographic maps with facial or audio-derived information for zero-shot inference. MERMAID~\cite{yang25mermaid} further expands the evidence available for comparison through emotion-conditioned visual variants. Reflection and cross-modal verification then connect these generated alternatives with the original observations to iteratively refine predictions.

\textbf{Token-level Intervention.}
Making evidence available, however, does not ensure that it contributes effectively to prediction, as redundant tokens can dilute informative cues. Moving from black-box input manipulation to white-box access allows this imbalance to be addressed within the model. SEPM~\cite{fang25catch} couples confidence-guided coarse-to-fine inference with attention-guided visual token pruning, narrowing the emotion search space while reducing visual distraction. WETR~\cite{liu26wetr} complements selection with reweighting: the former suppresses noisy video-text evidence, while the latter strengthens complementary emotional cues. These approaches thus regulate both the inclusion and relative influence of evidence during inference.

\textbf{Attention-Head Intervention.}
Beyond regulating token contributions, internal intervention can steer the representations involved in perspective-sensitive reasoning. GridToM~\cite{li25gridtom} uses supervised linear probes to identify belief-sensitive attention heads and derive activation-steering directions that improve Theory-of-Mind reasoning. Across these intervention sites, the emphasis shifts from making affective evidence explicit in external inputs to regulating its contribution and interpretation within the model.

\begin{table*}[t]
\centering
\caption{Example-Guided Affective Adaptation methods with their key attributes. Methods are grouped by example usage, with selection mechanisms, presentation formats, and example budgets summarized for In-Context Learning (ICL) and fine-tuning.}
\vskip -0.1in
\label{tab:5-2}
\begin{adjustbox}{max width=\textwidth}
\footnotesize
\renewcommand{\arraystretch}{1.1}
\setlength{\tabcolsep}{5pt}
\begin{tabular}{@{}lcccccccc@{}}
\toprule
\textbf{Method} &
\textbf{Year} &
\textbf{Task} &
\textbf{Modality} &
\textbf{Example Budget} &
\textbf{Usage} &
\textbf{Retrieval} &
\textbf{Presentation} &
\textbf{Key Design} \\
\midrule
% EmoGist\cite{seoh25emogist}{,} JMI\cite{arefa24jmi}{,} EA-ICL\cite{yang24entityaw}{,} ICL-Config\cite{wu25iclconfigu}{,} VERO\cite{sun25vero}{,} 
% ICL - white

EmoGist \cite{seoh25emogist} &
2025 &
VCER &
I &
4 images/description &
ICL &
Cluster &
Label Description &
Centroid-Retrieved Label Descriptions \\

JMI \cite{arefa24jmi} &
2024 &
CMER &
V,T &
1 case/stage &
ICL &
Similarity &
Input-Rationale-Output &
Retrieved Rationale Demonstrations \\

EA-ICL \cite{yang24entityaw} &
2024 &
ITER &
I,T &
10 shots &
ICL &
Similarity &
Input-Output Pair &
Entity-Aware Contrastive Retrieval \\

ICL-Config \cite{wu25iclconfigu} &
2025 &
ITER &
I,T &
4/8/16 shots &
ICL &
Similarity &
Input-Output Pair &
Joint Demonstration Configuration \\

% Fine-Tuning - gray

\rowcolor{gray!8}
VERO \cite{sun25vero} &
2025 &
ITER &
I,T &
1\%/7\% of train &
Fine-Tuning &
Verification &
Input-Label Pair &
Confidence-Ranked Hard Examples \\

\bottomrule
\end{tabular}
\end{adjustbox}
\vskip -0.1in
\end{table*}

\subsubsection{\textbf{Example-Guided Affective Adaptation}} \label{sec:5_1_2}
Complementing direct inference-time intervention, example-guided adaptation draws on selected examples to guide affective interpretation through In-Context Learning (ICL) or limited-data fine-tuning. Its effectiveness depends on both example selection and presentation, which determine the guidance available under a constrained budget. \Cref{tab:5-2} and \cref{fig:5-2} (b) summarize representative methods.

\textbf{ICL.}
For frozen models, demonstrations provide contextual guidance that general instructions alone may not convey. EA-ICL~\cite{yang24entityaw} uses entity-aware contrastive retrieval to select examples relevant to the target entity and sentiment, while ICL-Config~\cite{wu25iclconfigu} broadens this perspective by jointly considering multimodal similarity, modality presentation, and sentiment distribution. Beyond selecting relevant instances, the examples can communicate different levels of affective knowledge. JMI~\cite{arefa24jmi} supplements input-output demonstrations with rationales, exposing reasoning patterns alongside predictions. EmoGist~\cite{seoh25emogist} instead distills clustered visual examples into label descriptions retrieved for each test image, replacing individual demonstrations with compact, example-derived guidance. 

\textbf{Limited-Data Fine-Tuning.}
When examples support parameter adaptation, selection shifts from constructing informative contexts to allocating a limited training budget. VERO~\cite{sun25vero} combines self-verification and confidence estimates to prioritize informative hard examples, supplemented by easier instances. 

Together, these elicitation methods offer complementary routes to adapting affective inference to available data and model access, from external guidance to selective internal intervention and limited parameter updates.

\subsection{Multimodal Affective Representation}
\label{sec:5_2}

\begin{figure*}
    \centering
    \includegraphics[width=1\linewidth]{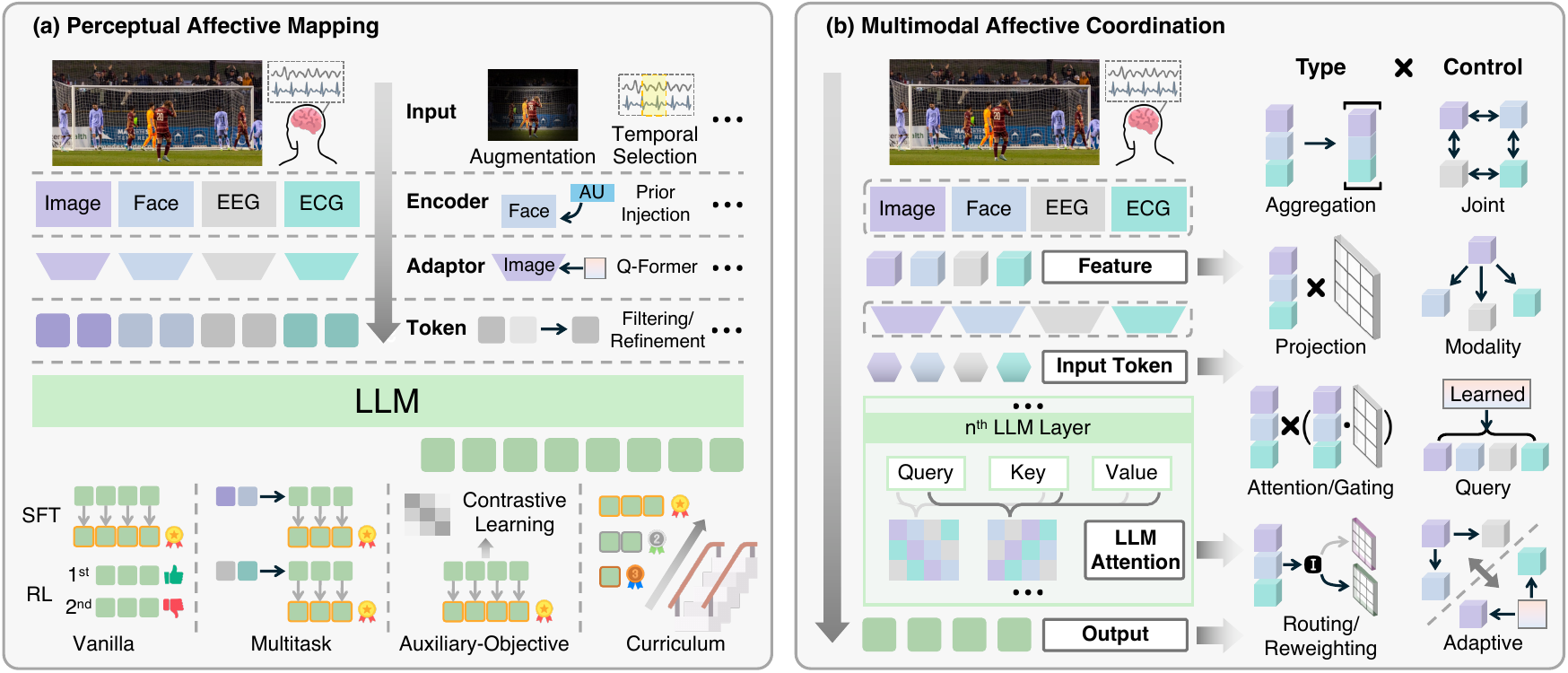}
    \vskip -0.1in
    \caption{Illustration of the two \textit{Multimodal Affective Representation} branches. \textit{Perceptual Affective Mapping} (a) captures intra-modal emotional cues through direct operations at different stages of the encoding stream or implicitly through learning objectives that shape emotion-oriented representations. \textit{Multimodal Affective Coordination} (b) facilitates cross-modal interactions at different model stages, including the feature, input-token, LLM-attention, and output levels.}
    \label{fig:5-3}
    \vskip -0.1in
\end{figure*}

Beyond unveiling existing capabilities, affective understanding essentially depends on how emotional evidence is represented. The \textit{Multimodal Affective Gap} arises both within individual modalities, where observable features do not directly reveal emotional meaning, and across modalities, where heterogeneous representations and divergent affective cues complicate joint interpretation. \textit{Multimodal Affective Representation} connects these two levels through \textit{Perceptual Affective Mapping}, which translates modality-specific observations into emotion-sensitive representations, and \textit{Multimodal Affective Coordination}, which coordinates these representations to support coherent interpretation. Together, they connect intra-modal encoding and inter-modal fusion, as illustrated in \cref{fig:5-3} and elaborated below.

\begin{table*}[t]
\centering
\caption{Perceptual Affective Mapping methods with their key attributes. Target indicates the mapping location, and ``-'' in Training Strategy denotes training-free methods. Abbreviations: CE (Cross-Entropy), BCE (Binary Cross-Entropy).}
\vskip -0.1in
\label{tab:5-3}
\begin{adjustbox}{max width=\textwidth}
\footnotesize
\renewcommand{\arraystretch}{1.1}
\setlength{\tabcolsep}{2pt}
\begin{tabular}{@{}lccccccc@{}}
\toprule
\textbf{Method} &
\textbf{Year} &
\textbf{Task} &
\textbf{Modality} &
\textbf{Target} &
\textbf{Perceptual Design} &
\textbf{Training Strategy} &
\textbf{Training Design} \\
\midrule

% Input - white
% ActFER\cite{liu26actfer}{,} EEG Emotion Copilot\cite{chen25eegcopil}{,} MELLM\cite{zhao25mellm}{,} ES4R\cite{gao26es4r}{,} StimuVAR\cite{guo25stimuvar}{,} MERMAID\cite{yang25mermaid}{,} PH-EMO\cite{liu26phemo}{,} EmoLA\cite{li24faba}{,} EMO-LLaMA\cite{xing26emollama}{,} MicroEmo\cite{zhang24microem}{,} EmoChat\cite{xu25emochat}{,} CG SLLM\cite{morais25elcces}{,} SpeechER\cite{zhang25speechr}{,} FaceInsight\cite{li25faceinsigh}{,} DEFT-LLM\cite{zhang25deftllm}{,} ProFocus\cite{zhang26profoc}{,} ICM\cite{luo25omnisila}{,} Omni-Emotion\cite{yang25omniemot}{,} Emotion-LLaMA\cite{cheng24emotion}{,} Emotion-LLaMAv2\cite{peng26emotionl}{,} Nano-EmoX\cite{huang26nanoemo}{,} SEKE\cite{wang25ekevllm}{,} EAA\cite{du25eaa}{,} MSE-Adapter\cite{yang25mseadapt}{,} AcoustEmo\cite{zhang26acouste}{,} E$^3$ Emotion-LLaMa\cite{lin24e3}{,} Modality Reprogramming\cite{babu26reprogra}{,} SeaLLMs-Audio\cite{liu25seallmsau}{,} UniMind\cite{lu25unimind}{,} SECap\cite{xu24secap}{,} PL-Distill\cite{yang26pldistil}{,} EmoQ\cite{yang25emoq}{,} Light-MER\cite{zheng26lightme}{,} HumanOmni\cite{zhao25humanomn}{,} Emotion-Qwen\cite{huang25emoqwen}{,} BLSP-Emo\cite{wang24blspemo}{,} WaveMind\cite{zeng25wavemind}{,} E$^2$-LLM\cite{ma26e2llm}{,} Prototype-as-Prompt\cite{zhao26prototyp}{,} NeuroLM\cite{jiang25neurolm}{,} WETR\cite{liu26wetr}{,} SEPM\cite{fang25catch}
ActFER \cite{liu26actfer} & 2026 & FER & I &
Input & Region-of-Interest Tool Calls &
Vanilla & AU-Grounded Tool-Use RL \\

EEG Emotion Copilot \cite{chen25eegcopil} & 2025 & PER & P,T &
Input & Wavelet-Compressed EEG Prompts &
Vanilla & Pruned EEG-to-Dialogue SFT \\

MELLM \cite{zhao25mellm} & 2025 & FER & I,T &
Input & Motion-Enhanced Flow Maps &
Multitask & Recognition-Understanding SFT \\

ES4R \cite{gao26es4r} & 2026 & MESC, ESS & A,T &
Input & Dual-Level Affective Attention &
Auxiliary-Objective & Cross-Path CE + KL Distillation \\

StimuVAR \cite{guo25stimuvar} & 2025 & VCER, MSU & V &
Input & Frame-Tube Cue Selection &
Curriculum & Emotion Alignment then Reasoning SFT \\

MERMAID \cite{yang25mermaid} & 2025 & VCER & I &
Input & Emotion-Guided Image Augmentation &
- & - \\

PH-EMO \cite{liu26phemo} & 2026 & PER & A,V,P &
Input & Modality-Specific Perception Agents &
- & - \\

% Encoder - gray

\rowcolor{gray!8}
EmoLA \cite{li24faba} & 2024 & FER & I &
Encoder & Facial Prior Expert &
Vanilla & - \\

\rowcolor{gray!8}
EMO-LLaMA \cite{xing26emollama} & 2026 & FER & I,V &
Encoder & FaceXFormer Prior Injection &
Vanilla & - \\

\rowcolor{gray!8}
MicroEmo \cite{zhang24microem} & 2024 & CMER & A,V,T &
Encoder & Global-Local Micro-Dynamics &
Vanilla & - \\

\rowcolor{gray!8}
EmoChat \cite{xu25emochat} & 2025 & CMER & A,V,T &
Encoder & AU-Aware Encoding &
Vanilla & - \\

\rowcolor{gray!8}
CG SLLM \cite{morais25elcces} & 2025 & AER & A &
Encoder & Connectionist Temporal Features &
Multitask & Shared Speech-Decoding Interface \\

\rowcolor{gray!8}
SpeechER \cite{zhang25speechr} & 2025 & AER & A &
Encoder & Semantic-Emotion Dual Encoders &
Multitask & Semantic-Emotion Task Alternation \\

\rowcolor{gray!8}
FaceInsight \cite{li25faceinsigh} & 2025 & FER & I &
Encoder & Segmentation-Guided Encoding &
Auxiliary-Objective & Attribute BCE + Logical Constraints \\

\rowcolor{gray!8}
DEFT-LLM \cite{zhang25deftllm} & 2025 & FER & V &
Encoder & Disentangled Motion Experts &
Auxiliary-Objective & Generation + Emotion/AU Losses \\

\rowcolor{gray!8}
ProFocus \cite{zhang26profoc} & 2026 & VCER & I &
Encoder & Hierarchical Linguistic Priors &
Auxiliary-Objective & Contrastive + VAD Consistency Losses \\

\rowcolor{gray!8}
ICM \cite{luo25omnisila} & 2025 & VCER & V &
Encoder & Omni-Scene Expert Encoding &
Auxiliary-Objective & Router Balancing + Language Modeling \\

\rowcolor{gray!8}
Omni-Emotion \cite{yang25omniemot} & 2025 & FER, CMER & A,V &
Encoder & Dedicated Facial Encoder &
Curriculum & Audio/Face Alignment then SFT \\

\rowcolor{gray!8}
Emotion-LLaMA \cite{cheng24emotion} & 2024 & FER, CMER & A,V,T &
Encoder & Global-Local-Temporal Encoders &
Curriculum & Coarse-to-Fine Recognition/Reasoning \\

\rowcolor{gray!8}
Emotion-LLaMAv2 \cite{peng26emotionl} & 2026 & FER, AER, CMER & A,V,T &
Encoder & Spatiotemporal Multiview Encoding &
Curriculum & Perception-First Instruction Tuning \\

\rowcolor{gray!8}
Nano-EmoX \cite{huang26nanoemo} & 2026 & CMER, MSU, MESC & A,V,T &
Encoder & Temporal Facial Expert &
Curriculum & Alignment, Fusion, and Multitask SFT \\

% Adapter - white

SEKE \cite{wang25ekevllm} & 2025 & FER & I &
Adapter & Visual-Language Projection &
Vanilla & Self-Verified Facial-Emotion Labels \\

EAA \cite{du25eaa} & 2025 & AER & A,T &
Adapter & Dual Cross-Attention Connector &
Vanilla & Context-Aware Instruction Tuning \\

MSE-Adapter \cite{yang25mseadapt} & 2025 & CMER & A,V,T &
Adapter & Text-Guided Pseudo Tokens &
Vanilla & - \\

AcoustEmo \cite{zhang26acouste} & 2026 & CMER & A,V,T &
Adapter & Utterance-Aware Acoustic Queries &
Vanilla & - \\

E$^3$ Emotion-LLaMa \cite{lin24e3} & 2024 & MSU & A,V,T &
Adapter & AudioMAE-to-LLM Projection &
Vanilla & Egocentric Affect Instruction Tuning \\

Modality Reprogramming \cite{babu26reprogra} & 2026 & PER & P &
Adapter & EEG-to-Token Reprogramming &
Vanilla & Frozen-LLM Patch/Projector Training \\

SeaLLMs-Audio \cite{liu25seallmsau} & 2025 & AER & A &
Adapter & Regional Audio-Language Connector &
Multitask & Cross-Task Audio Instruction Mixture \\

UniMind \cite{lu25unimind} & 2025 & PER & P &
Adapter & Task-Aware Neural Connector &
Multitask & Task-Aware Multitask SFT \\

SECap \cite{xu24secap} & 2024 & AER & A &
Adapter & Emotion-Disentangled Queries &
Auxiliary-Objective & Mutual Information + Contrastive \\

PL-Distill \cite{yang26pldistil} & 2026 & AER & A &
Adapter & Attention-Weighted Alignment &
Auxiliary-Objective & Projector/Logit Distillation \\

EmoQ \cite{yang25emoq} & 2025 & AER & A,T &
Adapter & Speech-Aware Q-Former &
Auxiliary-Objective & Supervised Contrastive + Focal Losses \\

Light-MER \cite{zheng26lightme} & 2026 & CMER & A,V,T &
Adapter & Temporal Pooling and Projection &
Auxiliary-Objective & Hidden-State Distillation + RL \\

HumanOmni \cite{zhao25humanomn} & 2025 & FER, CMER & A,V &
Adapter & Face-Body-Interaction Projectors &
Curriculum & Unimodal Training then Integration \\

Emotion-Qwen \cite{huang25emoqwen} & 2025 & FER, VCER, CMER & I,V &
Adapter & Emotion-General Compressor &
Curriculum & Expert Warm-Up then SFT \\

BLSP-Emo \cite{wang24blspemo} & 2024 & AER, MESC & A &
Adapter & Speech Modality Adapter &
Curriculum & Semantic-to-Emotion Stages \\

WaveMind \cite{zeng25wavemind} & 2025 & PER & P &
Adapter & Semantic EEG-Token Alignment &
Curriculum & Alignment, Cold-Start, and SFT \\

E$^2$-LLM \cite{ma26e2llm} & 2026 & PER & P,T &
Adapter & EEGPT-to-Qwen3 Projection &
Curriculum & Pretraining, Alignment, and SFT \\

% Token - gray

\rowcolor{gray!8}
Prototype-as-Prompt \cite{zhao26prototyp} & 2026 & CMER & A,V,T &
Token & Sentiment Prototype Mapping &
Auxiliary-Objective & Alignment + Diversity Constraints \\

\rowcolor{gray!8}
NeuroLM \cite{jiang25neurolm} & 2025 & PER & P &
Token & Text-Aligned Neural Tokenizer &
Curriculum & Tokenization, Pretraining, and SFT \\

\rowcolor{gray!8}
WETR \cite{liu26wetr} & 2026 & FER, CMER & V,T &
Token & Token Selection and Reweighting &
- & - \\

\rowcolor{gray!8}
SEPM \cite{fang25catch} & 2025 & VCER & I &
Token & Attention-Guided Token Pruning &
- & - \\

\bottomrule
\end{tabular}
\end{adjustbox}
\vskip -0.1in
\end{table*}

\subsubsection{\textbf{Perceptual Affective Mapping}} \label{sec:5_2_1}
\textit{Perceptual Affective Mapping} focuses on enhancing the affective representational capacity of modality-specific embeddings. \Cref{tab:5-3} summarizes two complementary dimensions: the target of perceptual intervention and the training strategy used to shape the resulting representations. Their mechanisms are illustrated in \cref{fig:5-3} (a). We first trace the perceptual pipeline and then examine the associated learning strategies.

\textbf{Input-level Mapping.}
Affective evidence is often sparse, spatially localized, or temporally brief, motivating interventions that enhance its salience before encoding. StimuVAR~\cite{guo25stimuvar} selects emotion-relevant temporal regions; MELLM~\cite{zhao25mellm} instead makes subtle facial motion explicit through enhanced optical-flow maps. Both concentrate processing on cues that may be obscured in the original sequence. ActFER~\cite{liu26actfer} makes evidence acquisition adaptive through facial-region tool calls, allowing observations to respond to the model's information needs. Signal reformulation extends this treatment to physiological evidence, from wavelet-compressed EEG prompts~\cite{chen25eegcopil} to modality-specific observations organized for causal reasoning~\cite{liu26phemo}. MERMAID~\cite{yang25mermaid} further expands the available evidence through emotion-conditioned visual variants, complementing selection and reformulation with generative augmentation.

\textbf{Encoder-level Mapping.}
Inputs establish the informational basis for affective perception, and encoders shape its realization in learned representations. Facial priors~\cite{li24faba,xing26emollama}, segmentation-guided features~\cite{li25faceinsigh}, and AU cues~\cite{xu25emochat} preserve details that general visual encoders may overlook. This specialization extends beyond local appearance through global and local micro-dynamics in MicroEmo~\cite{zhang24microem} and disentangled motion experts in DEFT-LLM~\cite{zhang25deftllm}. Scene-level expertise~\cite{luo25omnisila} adds contextual information to these localized cues. Emotion-LLaMA~\cite{cheng24emotion} brings facial, temporal, and acoustic encoding together. Omni-Emotion~\cite{yang25omniemot} similarly enriches general video features with dedicated facial representations. These complementary perceptual branches preserve a richer basis for subsequent affective interpretation.

\textbf{Adapter-level Mapping.}
Given the text-centric nature of LLMs, encoded multimodal features typically require adapters to align them with the language embedding space, providing another opportunity to preserve affective distinctions. Projection-based interfaces establish this alignment for visual~\cite{wang25ekevllm} and acoustic features~\cite{lin24e3}. Building on this connection, text-guided pseudo tokens~\cite{yang25mseadapt} introduce semantic guidance into the transfer. More selective interfaces use emotion-disentangled queries~\cite{xu24secap} or speech-aware queries~\cite{yang25emoq} to extract relevant acoustic information. EAA~\cite{du25eaa} connects semantic and acoustic features through dual cross-attention, with temporal specificity receiving greater emphasis in the utterance-aware queries of AcoustEmo~\cite{zhang26acouste}.

This selectivity also extends to preserving distinct sources of affective information. HumanOmni~\cite{zhao25humanomn} employs separate face, body, and interaction projectors. Emotion-Qwen~\cite{huang25emoqwen} adopts a complementary design that combines emotion-oriented and general visual compression. For neural signals, reprogramming~\cite{babu26reprogra} and semantic EEG-token alignment~\cite{zeng25wavemind} provide analogous bridges. Across these designs, the interface balances selective information transfer with compatibility between perceptual encoders and the LLM.

\textbf{Token-level Mapping.}
Organizing perceptual representations into individual tokens enables further finer-grained control. Attention-guided pruning~\cite{fang25catch} and emotion-sensitive selection and reweighting~\cite{liu26wetr} reduce redundancy without additional training. Beyond regulating existing tokens, Prototype-as-Prompt~\cite{zhao26prototyp} introduces sentiment-aware prototypes as interpretive anchors, and NeuroLM~\cite{jiang25neurolm} maps neural signals into text-aligned discrete tokens. These designs give the token sequence more explicit semantic structure. Complementing such structural choices, learning strategies shape the supervision and evolution of affective representations. 

\textbf{Multitask Learning.}
Broader supervision encourages representations to retain information useful across related tasks. MELLM~\cite{zhao25mellm} combines recognition and descriptive understanding, linking categorical decisions with richer accounts of facial motion. In speech, CG SLLM~\cite{morais25elcces} and SpeechER~\cite{zhang25speechr} connect semantic and emotion-related learning. SeaLLMs-Audio~\cite{liu25seallmsau} broadens this scope through a wider mixture of audio instructions, and UniMind~\cite{lu25unimind} extends multitask instruction tuning to EEG with task-aware guidance. These approaches expose affective representations to complementary demands beyond a single prediction format.

\textbf{Auxiliary-Objective Optimization.}
Beyond diversifying tasks, auxiliary objectives impose targeted constraints on representations and predictions. SECap~\cite{xu24secap} promotes emotion-disentangled acoustic representations through mutual-information and contrastive objectives. EmoQ~\cite{yang25emoq} combines supervised contrastive and focal losses to improve emotion discrimination. Logical constraints in FaceInsight~\cite{li25faceinsigh} and dimensional-affect consistency in ProFocus~\cite{zhang26profoc} further regulate the relationships captured by the model. Distillation provides another source of guidance: ES4R~\cite{gao26es4r} couples cross-path supervision with distribution matching, PL-Distill~\cite{yang26pldistil} aligns projector representations and output logits, and Light-MER~\cite{zheng26lightme} combines hidden-state distillation with reward-based optimization. Such objectives make specific representational properties explicit learning targets.

\textbf{Curriculum-Based Adaptation.}
These learning demands can also be distributed across successive stages, establishing perceptual foundations before introducing more demanding objectives. BLSP-Emo~\cite{wang24blspemo} progresses from semantic alignment toward emotion-aware speech understanding. A related progression connects EEG pretraining, cross-modal alignment, and instruction tuning in E$^2$-LLM~\cite{ma26e2llm}. Nano-EmoX~\cite{huang26nanoemo} extends staged learning across modality alignment, fusion, and multitask adaptation, and Emotion-LLaMAv2~\cite{peng26emotionl} emphasizes progression from perception to cognition. The sequence of supervision thus allows later affective capabilities to build on established representations.

Together, perceptual interventions preserve and expose affective cues, and learning strategies complementarily refine their discriminability and semantic organization, providing a more informative basis for subsequent multimodal coordination and reasoning.

\begin{table*}[t]
\centering
\caption{Multimodal Affective Coordination methods with their key attributes. Coordination Site identifies the interaction location, Interaction Type describes the mechanism, and Control indicates how the interaction is guided.}
\label{tab:5-4}
\vskip -0.1in
\begin{adjustbox}{max width=\textwidth}
\footnotesize
\renewcommand{\arraystretch}{1.1}
\setlength{\tabcolsep}{2pt}
\begin{tabular}{@{}lccccccc@{}}
\toprule
\textbf{Method} &
\textbf{Year} &
\textbf{Task} &
\textbf{Modality} &
\textbf{Coordination Site} &
\textbf{Interaction Type} &
\textbf{Control} &
\textbf{Key Design} \\
\midrule
% Emotion-LLaMAv2~\cite{peng26emotionl}{,} Conv-Attention~\cite{cheng24convatt}{,} AffectGPT~\cite{lian25affectgp}{,} MSE-Adapter~\cite{yang25mseadapt}{,} M$^{3}$F~\cite{wu25m3f}{,} ES4R~\cite{gao26es4r}{,} EGMF~\cite{qiao26egmf}{,} MEERA~\cite{niu26meera}{,} Light-MER~\cite{zheng26lightme}{,} Nano-EmoX~\cite{huang26nanoemo}{,} Omni-Emotion~\cite{yang25omniemot}{,} AA-SLLM~\cite{mai25aasllm}{,} BeMERC~\cite{fu25bemerc}{,} SoMeLVLM~\cite{zhang24somelvl}{,} E$^{3}$ Emotion-LLaMa~\cite{lin24e3}{,} MoSEAR~\cite{han25camer}{,} CHASE~\cite{sun26emomm}{,} SABER-LLM~\cite{zhao26saber}{,} MiSTER-E~\cite{dutta26mistere}

% Feature / Attention or Gating

Emotion-LLaMAv2~\cite{peng26emotionl} &
2026 &
FER, AER, CMER &
A,V,T &
Feature &
Attention / Gating &
Joint &
Local-Global Audiovisual Prefusion \\

Conv-Attention~\cite{cheng24convatt} &
2024 &
CMER &
A,V,T &
Feature &
Attention / Gating &
Joint &
Parallel Convolution-Attention Fusion \\

AffectGPT~\cite{lian25affectgp} &
2025 &
CMER &
A,V,T &
Feature &
Attention / Gating &
Query / Adaptive &
Query- or Attention-Based Prefusion \\

MSE-Adapter~\cite{yang25mseadapt} &
2025 &
CMER &
A,V,T &
Feature &
Attention / Gating &
Modality &
Text-Gated Multiscale Fusion \\

M$^{3}$F~\cite{wu25m3f} &
2025 &
CMER, MSU &
A,V,T &
Feature &
Attention / Gating &
Query &
Query-Based Audiovisual Memory \\

ES4R~\cite{gao26es4r} &
2026 &
MESC, ESS &
A,T &
Feature &
Attention / Gating &
Modality &
Speech-Guided Semantic Fusion \\

% Feature / Routing or Reweighting

EGMF~\cite{qiao26egmf} &
2026 &
CMER &
A,V,T &
Feature &
Routing / Reweighting &
Adaptive &
Hierarchical Expert Gating \\

MEERA~\cite{niu26meera} &
2026 &
CMER &
A,V,T &
Feature &
Routing / Reweighting &
Adaptive &
Instruction-Guided Expert Weighting \\

Light-MER~\cite{zheng26lightme} &
2026 &
CMER &
A,V,T &
Feature &
Routing / Reweighting &
Adaptive &
Adaptive Audiovisual Weighting \\

Nano-EmoX~\cite{huang26nanoemo} &
2026 &
CMER, MSU, MESC &
A,V,T &
Feature &
Routing / Reweighting &
Adaptive &
Gated Multiscale Audiovisual Experts \\

% Input Token / Aggregation

\rowcolor{gray!8}
Omni-Emotion~\cite{yang25omniemot} &
2025 &
FER, CMER &
A,V &
Input Token &
Aggregation &
Joint &
Audiovisual Token Concatenation \\

\rowcolor{gray!8}
AA-SLLM~\cite{mai25aasllm} &
2025 &
AER &
A,T &
Input Token &
Aggregation &
Joint &
Speech Tokens with Acoustic Prompts \\

\rowcolor{gray!8}
BeMERC~\cite{fu25bemerc} &
2025 &
CMER &
A,V,T &
Input Token &
Aggregation &
Joint &
Behavior-Aligned Multimodal Tokens \\

% Input Token / Projection

\rowcolor{gray!8}
SoMeLVLM~\cite{zhang24somelvl} &
2024 &
ITER, MSU &
I,T &
Input Token &
Projection &
Joint &
Image-Text Connector Adaptation \\

\rowcolor{gray!8}
E$^{3}$ Emotion-LLaMa~\cite{lin24e3} &
2024 &
MSU &
A,V,T &
Input Token &
Projection &
Joint &
Projected Audiovisual Tokens \\

% LLM Attention / Routing or Reweighting

MoSEAR~\cite{han25camer} &
2025 &
FER, CMER &
A,V,T &
LLM Attention &
Routing / Reweighting &
Adaptive &
Modality-Balanced Attention \\

CHASE~\cite{sun26emomm} &
2026 &
CMER &
A,V,T &
LLM Attention &
Routing / Reweighting &
Adaptive &
Conflict-Aware Attention Steering \\

% Output / Aggregation

\rowcolor{gray!8}
SABER-LLM~\cite{zhao26saber} &
2026 &
CMER, MSU &
A,V &
Output &
Aggregation &
Joint &
Structured Audiovisual Evidence Fusion \\

% Output / Routing or Reweighting

\rowcolor{gray!8}
MiSTER-E~\cite{dutta26mistere} &
2026 &
CMER &
A,T &
Output &
Routing / Reweighting &
Adaptive &
Gated Speech-Text Expert Logits \\

\bottomrule
\end{tabular}
\end{adjustbox}
\vskip -0.1in
\end{table*}

\subsubsection{\textbf{Multimodal Affective Coordination}} \label{sec:5_2_2}
In multimodal settings, the \textit{Inter-Modal Affective Gap} makes coordination across heterogeneous cues an essential complement to intra-modal representation. \textit{Multimodal Affective Coordination} thereby builds on these representations to strengthen complementary evidence, reduce redundancy, and reconcile conflicting signals. \Cref{tab:5-4} groups these methods by coordination site, with representative mechanisms illustrated in \cref{fig:5-3} (b).

\textbf{Feature-level Coordination.}
Coordination can begin before LLM processing, fusing modality-specific features into complementary unity. Conv-Attention~\cite{cheng24convatt} combines convolutional interactions with global attention, a local-global design also adopted for audiovisual prefusion in Emotion-LLaMAv2~\cite{peng26emotionl}. More selective fusion uses learned queries or modality-specific guidance. AffectGPT~\cite{lian25affectgp} explores both query-based extraction and attention-weighted audiovisual fusion, and M$^{3}$F~\cite{wu25m3f} extends query-based retrieval to long-term audiovisual memory. Modality-guided designs introduce an explicit conditioning source: MSE-Adapter~\cite{yang25mseadapt} uses textual features to gate multiscale audiovisual fusion, whereas ES4R~\cite{gao26es4r} uses speech representations to query contextual semantics. These mechanisms condition integration on informative queries or complementary modality cues.

The varying relevance of different sources also motivates adaptive weighting and expert selection. Light-MER~\cite{zheng26lightme} learns audiovisual contribution weights through a compact fusion module. Expert-based approaches make this allocation more specialized: EGMF~\cite{qiao26egmf} gates local, cross-modal, and global experts, and Nano-EmoX~\cite{huang26nanoemo} combines audio-guided interactions with gated multiscale visual expertise. MEERA~\cite{niu26meera} further conditions perceptual expert weighting on instruction semantics. Together, these designs make feature integration responsive to the input and task, reducing reliance on uniformly combining all available representations.

\textbf{Input-Token-level Coordination.}
Heterogeneous evidence can also be aligned and assembled within the LLM's input sequence. Projection-based interfaces connect audiovisual features with language tokens in E$^{3}$ Emotion-LLaMa~\cite{lin24e3}, and connector adaptation supports joint image-text processing in SoMeLVLM~\cite{zhang24somelvl}. The resulting sequence can preserve distinct sources of affective evidence. Omni-Emotion~\cite{yang25omniemot} concatenates general visual, facial, and acoustic representations, while BeMERC~\cite{fu25bemerc} assembles projected audiovisual tokens with text after behavior-oriented alignment. AA-SLLM~\cite{mai25aasllm} instead complements projected speech features with textual descriptions of acoustic properties. Such interfaces provide an alternative to prior compression by bringing heterogeneous cues into a shared processing context.

\textbf{LLM-Attention-level Coordination.}
Organizing multimodal tokens enables convenient processing but offers limited control over cross-modal interactions, motivating attention-level coordination within the LLM. MoSEAR~\cite{han25camer} combines modality-specialized experts with attention reallocation to mitigate modality bias. CHASE~\cite{sun26emomm} introduces conflict-sensitive head steering through a learned router, making evidence integration responsive to cross-modal disagreement.

\textbf{Output-level Coordination.}
Coordination can additionally operate on modality-specific predictions or generated evidence, making their contributions explicit at the decision stage. MiSTER-E~\cite{dutta26mistere} dynamically weights logits from speech, text, and cross-modal experts, allowing specialized judgments to contribute differently across utterances. SABER-LLM~\cite{zhao26saber} instead follows a perceive-then-reason structure, aggregating visual and acoustic evidence descriptions into a joint interpretation. The former combines predictive scores, whereas the latter integrates articulated observations, providing distinct routes from separate assessments to a multimodal conclusion.

Across these sites, coordination ranges from representation fusion and shared token interfaces to attention regulation and evidence aggregation. Collectively, perceptual mapping and multimodal coordination bridge the gap between heterogeneous sensory signals and affective meaning, providing LLMs with emotionally informative representations and a coherent evidential basis for reasoning.

\subsection{Affective Reasoning and Explanation}
\label{sec:5_3}

\begin{figure*}
    \centering
    \includegraphics[width=1\linewidth]{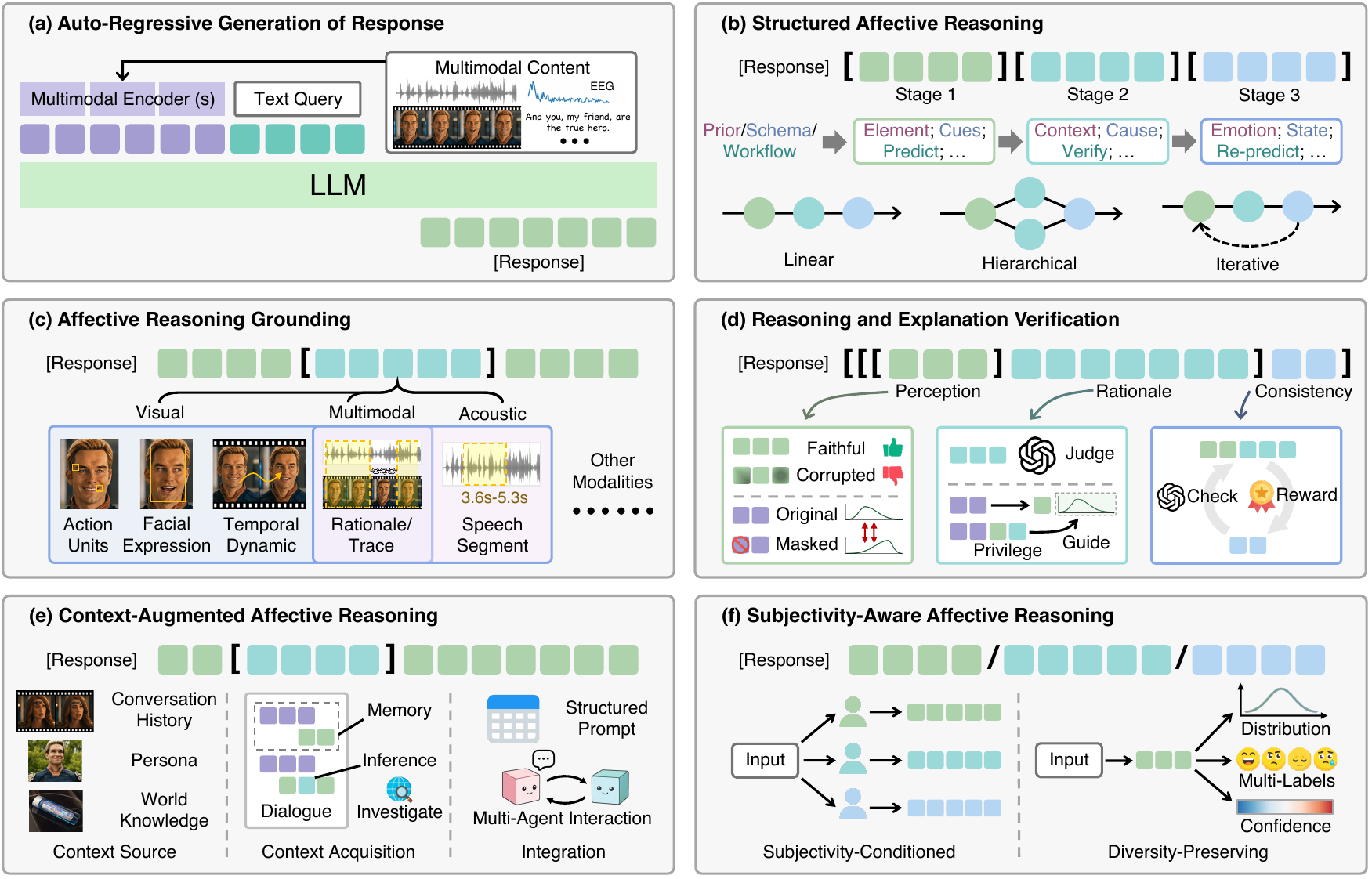}
    \vskip -0.1in
    \caption{Illustration of autoregressive response generation (a) and five branches of \textit{Affective Reasoning and Explanation}. \textit{Structured Affective Reasoning} (b) decomposes affective inference into organized reasoning steps. \textit{Affective Reasoning Grounding} (c) connects interpretations with supporting evidence, and \textit{Reasoning and Explanation Verification} (d) examines their reliability. \textit{Context-Augmented Affective Reasoning} (e) enriches inference with contextual information, while \textit{Subjectivity-Aware Affective Reasoning} (f) accounts for variation in emotional perspectives and interpretations.}
    \label{fig:5-4}
    \vskip -0.14in
\end{figure*}

Given informative affective representations, LLMs process them through successive Transformer layers and generate textual responses autoregressively. This language-based interface offers a natural platform for articulating the connection between evidence and judgment, motivating methods collectively termed \textit{Affective Reasoning and Explanation} to address \textit{Affective Process Opacity}. According to their primary focus, we distinguish five complementary branches: \textit{Structured Affective Reasoning}, \textit{Affective Reasoning Grounding}, \textit{Reasoning and Explanation Verification}, \textit{Context-Augmented Affective Reasoning}, and \textit{Subjectivity-Aware Affective Reasoning}, as illustrated in \cref{fig:5-4}.

\begin{table*}[t]
\centering
\caption{Structured Affective Reasoning methods with their structural sources, reasoning topologies, and representative flows. Prior, Schema, and Workflow denote theory-guided organization, explicit formats, and procedural orchestration, respectively.}
\label{tab:5-5}
\vskip -0.1in
\begin{adjustbox}{max width=\textwidth}
\footnotesize
\renewcommand{\arraystretch}{1.1}
\setlength{\tabcolsep}{3pt}
\begin{tabular}{@{}lcccccc@{}}
\toprule
\textbf{Method} &
\textbf{Year} &
\textbf{Task} &
\textbf{Modality} &
\textbf{Structure Source} &
\textbf{Topology} &
\textbf{Reasoning Flow} \\
\midrule
% EmoCaliber~\cite{wu26emocaliber}{,} OneEmo~\cite{huang26oneemo}{,} EmoToM~\cite{etesam24emotio}{,} CoDAR~\cite{xu26moca}{,} PH-EMO~\cite{liu26phemo}{,} AG-CTR$^{2}$~\cite{chu26agctr2}{,} TMPO~\cite{luo26hitemotio}{,} VidEmo~\cite{zhang25videmo}{,} ProFocus~\cite{zhang26profoc}{,} Affective-CoT~\cite{huang25affecti}{,} Agent-MER~\cite{lai25agentmer}{,} OSIRIS~\cite{abdelf26osmo}{,} ExpLLM~\cite{lan25expllm}{,} Facial-R1~\cite{wu26facialr1}{,} TAG~\cite{lin26tag}{,} MELLM~\cite{zhao25mellm}{,} EmoThinker~\cite{xu26emothinker}{,} EmotionThinker~\cite{wang26emotiont}{,} Comparative Reasoning~\cite{naini26compara}{,} CogAudio-LLM~\cite{zhao26cogaudio}{,} PALLM~\cite{kim26pallm}{,} IEAT~\cite{wang26ieat}{,} ELENA~\cite{saim25anatomy}{,} EMO-R3~\cite{fang26emor3}{,} EEmo-Logic~\cite{gao26eemologic}{,} ParalinGPT~\cite{lin24parallm}{,} AffectGPT-R1~\cite{lian25affr1}{,} EmotionReasoner~\cite{zhao26emotionr}{,} SABER-LLM~\cite{zhao26saber}{,} AffectOmni~\cite{wang26affomni}{,} MulCoT-RD~\cite{shangg26mulcot}{,} SSRG Framework~\cite{kang25ssrgssrd}{,} SarcasmMiner~\cite{li26sarcasmmin}{,} E$^{2}$-LLM~\cite{ma26e2llm}{,} GAT~\cite{she26aicabench}{,} AffectAgent~\cite{wang26affagt}{,} CofiPara~\cite{chen24cofipara}{,} SC-CoQ~\cite{ouyang25punch}{,} GraphEMO~\cite{tu26graphemo}{,} RAER~\cite{wen25raer}{,} PEP-MEK~\cite{xing26emotionh}{,} V-AlphaSocial~\cite{lin25valpha}{,} ActFER~\cite{liu26actfer}{,} EmoAgent-R1~\cite{fang26emoagent}{,} FIRES~\cite{wang25fires}{.}
% Prior / Linear

EmoCaliber~\cite{wu26emocaliber} &
2025 & VCER & I &
Prior & Linear &
Elements $\rightarrow$ Context $\rightarrow$ Emotion \\

OneEmo~\cite{huang26oneemo} &
2026 & CMER, MSU, MESC & A,V,T &
Prior & Linear &
Cues $\rightarrow$ State/Cause $\rightarrow$ Emotion/Response \\

EmoToM~\cite{etesam24emotio} &
2023 & MSU & I &
Prior & Linear &
Narrative $\rightarrow$ Contextual Reasoning $\rightarrow$ Emotion \\

CoDAR~\cite{xu26moca} &
2026 & MSU & I,V,T &
Prior & Linear &
Observation $\rightarrow$ Expectation Conflict $\rightarrow$ Hidden State \\

PH-EMO~\cite{liu26phemo} &
2026 & PER & A,V,P &
Prior & Linear &
Trigger $\rightarrow$ Physiology $\rightarrow$ Expression $\rightarrow$ Emotion \\

AG-CTR$^{2}$~\cite{chu26agctr2} &
2026 & MESC & V,T &
Prior & Linear &
Event $\rightarrow$ Appraisal $\rightarrow$ Coping $\rightarrow$ Emotion \\

TMPO~\cite{luo26hitemotio} &
2026 &
All MER Tasks except PER &
I,A,V,T &
Prior & Linear &
Cues $\rightarrow$ Mental-State Attribution $\rightarrow$ Judgment \\

% Prior / Hierarchical

VidEmo~\cite{zhang25videmo} &
2025 & FER, VCER & V,T &
Prior & Hierarchical &
Attributes $\rightarrow$ Expressions $\rightarrow$ Emotion \\

ProFocus~\cite{zhang26profoc} &
2026 & VCER & I &
Prior & Hierarchical &
Atmosphere $\rightarrow$ Narrative $\rightarrow$ Details $\rightarrow$ Emotion \\

% Pending full-text verification.
Affective-CoT~\cite{huang25affecti} &
2025 & CMER & A,V,T &
Prior & Hierarchical &
Perceptual Evidence $\rightarrow$ Cognitive Hierarchy $\rightarrow$ Emotion \\

Agent-MER~\cite{lai25agentmer} &
2025 & CMER & A,V,T &
Prior & Hierarchical &
Coarse Categories $\rightarrow$ Fine Categories $\rightarrow$ Voting \\

% Prior / Iterative

OSIRIS~\cite{abdelf26osmo} &
2026 & MSU & A,V,T &
Prior & Iterative &
Emotion History $\rightarrow$ Current Cues $\rightarrow$ State Update \\

% Schema / Linear

\rowcolor{gray!8}
ExpLLM~\cite{lan25expllm} &
2025 & FER & I &
Schema & Linear &
Action Units $\rightarrow$ Interpretation $\rightarrow$ Emotion \\

\rowcolor{gray!8}
Facial-R1~\cite{wu26facialr1} &
2026 & FER & I &
Schema & Linear &
AU-Informed Reasoning $\rightarrow$ Emotion \\

\rowcolor{gray!8}
TAG~\cite{lin26tag} &
2026 & FER & I &
Schema & Linear &
Action Units $\rightarrow$ Grounded Reasoning $\rightarrow$ Emotion \\

\rowcolor{gray!8}
MELLM~\cite{zhao25mellm} &
2025 & FER & V &
Schema & Linear &
Motion/AUs $\rightarrow$ Micro-Expression $\rightarrow$ Report \\

\rowcolor{gray!8}
EmoThinker~\cite{xu26emothinker} &
2026 & FER, CMER, MSU & A,V &
Schema & Linear &
Evidence $\rightarrow$ Conflict Resolution $\rightarrow$ Emotion \\

\rowcolor{gray!8}
EmotionThinker~\cite{wang26emotiont} &
2026 & AER & A &
Schema & Linear &
Prosodic/Semantic Cues $\rightarrow$ Reasoning $\rightarrow$ Emotion \\

\rowcolor{gray!8}
Comparative Reasoning~\cite{naini26compara} &
2026 & AER & A &
Schema & Linear &
Audio Pair $\rightarrow$ Attribute Comparison $\rightarrow$ Verdict \\

\rowcolor{gray!8}
CogAudio-LLM~\cite{zhao26cogaudio} &
2026 & AER, MESC & A &
Schema & Linear &
Emotion $\rightarrow$ Intent $\rightarrow$ Mental State $\rightarrow$ Strategy \\

\rowcolor{gray!8}
PALLM~\cite{kim26pallm} &
2026 & AER, MESC & A &
Schema & Linear &
Prosodic Reasoning $\rightarrow$ Sentiment/Response \\

\rowcolor{gray!8}
IEAT~\cite{wang26ieat} &
2026 & AER, MESC, ESS & A,T &
Schema & Linear &
Emotional State/Cause $\rightarrow$ Text/Speech Response \\

\rowcolor{gray!8}
ELENA~\cite{saim25anatomy} &
2025 & VCER & I &
Schema & Linear &
Body Parts $\rightarrow$ Embodied Narrative $\rightarrow$ Emotion \\

\rowcolor{gray!8}
EMO-R3~\cite{fang26emor3} &
2026 & VCER & I &
Schema & Linear &
Trigger $\rightarrow$ Human Reflection $\rightarrow$ Polarity/Arousal \\

\rowcolor{gray!8}
EEmo-Logic~\cite{gao26eemologic} &
2026 & VCER & I &
Schema & Linear &
Visual Reasoning $\rightarrow$ Task-Specific Assessment \\

\rowcolor{gray!8}
ParalinGPT~\cite{lin24parallm} &
2024 & CMER & A,T &
Schema & Linear &
Current Sentiment $\rightarrow$ Response Sentiment $\rightarrow$ Response \\

\rowcolor{gray!8}
AffectGPT-R1~\cite{lian25affr1} &
2025 & CMER & A,V,T &
Schema & Linear &
Multimodal Reasoning $\rightarrow$ Emotion Set \\

\rowcolor{gray!8}
EmotionReasoner~\cite{zhao26emotionr} &
2026 & CMER & A,V,T &
Schema & Linear &
Evidence $\rightarrow$ Explanation $\rightarrow$ Emotion \\

\rowcolor{gray!8}
SABER-LLM~\cite{zhao26saber} &
2026 & CMER, MSU & A,V &
Schema & Linear &
Audiovisual Evidence $\rightarrow$ Causal Reasoning $\rightarrow$ Emotion \\

\rowcolor{gray!8}
AffectOmni~\cite{wang26affomni} &
2026 & CMER, MSU & A,V,T &
Schema & Linear &
Context $\rightarrow$ Reasoning $\rightarrow$ Answer \\

\rowcolor{gray!8}
MulCoT-RD~\cite{shangg26mulcot} &
2026 & ITER & I,T &
Schema & Linear &
Text/Image Evidence $\rightarrow$ Conflict Resolution $\rightarrow$ Sentiment \\

\rowcolor{gray!8}
SSRG Framework~\cite{kang25ssrgssrd} &
2025 & MSU & A,V,T &
Schema & Linear &
Multimodal Cues $\rightarrow$ Sarcasm Subtype $\rightarrow$ Rationale \\

\rowcolor{gray!8}
SarcasmMiner~\cite{li26sarcasmmin} &
2026 & MSU & A,V,T &
Schema & Linear &
Multimodal Evidence $\rightarrow$ Reasoning $\rightarrow$ Sarcasm Judgment \\

\rowcolor{gray!8}
E$^{2}$-LLM~\cite{ma26e2llm} &
2026 & PER & P,T &
Schema & Linear &
EEG Description $\rightarrow$ Comparison $\rightarrow$ Justification \\

% Schema / Hierarchical

\rowcolor{gray!8}
GAT~\cite{she26aicabench} &
2026 & VCER & I &
Schema & Hierarchical &
Visual Grounding $\rightarrow$ Affective Tree $\rightarrow$ Answer \\

% Workflow / Linear

AffectAgent~\cite{wang26affagt} &
2026 & CMER & A,V,T &
Workflow & Linear &
Query Planning $\rightarrow$ Evidence Filtering $\rightarrow$ Emotion \\

CofiPara~\cite{chen24cofipara} &
2024 & MSU & I,T &
Workflow & Linear &
Contrasting Rationales $\rightarrow$ Sarcasm Target Identification \\

SC-CoQ~\cite{ouyang25punch} &
2025 & MSU & I,T &
Workflow & Linear &
Simple Questions $\rightarrow$ Complex Questions $\rightarrow$ Punchline \\

% Workflow / Hierarchical
% Pending full-text verification.

GraphEMO~\cite{tu26graphemo} &
2026 & CMER, MESC & A,V,T &
Workflow & Hierarchical &
Emotion-Cause Graph $\rightarrow$ Hierarchical Context $\rightarrow$ Support \\

% Workflow / Iterative

RAER~\cite{wen25raer} &
2025 & FER, CMER & A,V,T &
Workflow & Iterative &
Initial Reasoning $\rightarrow$ Retrieval $\rightarrow$ Refinement \\

PEP-MEK~\cite{xing26emotionh} &
2026 & AER, VCER, CMER & I,A,V,T &
Workflow & Iterative &
Predict $\rightarrow$ Verify $\rightarrow$ Re-predict \\

V-AlphaSocial~\cite{lin25valpha} &
2025 & MSU & I,T &
Workflow & Iterative &
Observation $\rightarrow$ Reasoning $\rightarrow$ Reflection $\rightarrow$ Revision \\

% Workflow / Adaptive

ActFER~\cite{liu26actfer} &
2026 & FER & I &
Workflow & Adaptive &
Face $\rightarrow$ Adaptive ROI Inspection $\rightarrow$ Emotion \\

EmoAgent-R1~\cite{fang26emoagent} &
2026 & CMER & A,V,T &
Workflow & Adaptive &
Input-Dependent Routing $\rightarrow$ Specialist $\rightarrow$ Emotion \\

FIRES~\cite{wang25fires} &
2025 & CMER, MESC & V,T &
Workflow & Adaptive &
Aspect Selection $\rightarrow$ Focused Reasoning $\rightarrow$ Response \\

\bottomrule
\end{tabular}
\end{adjustbox}
\vskip -0.1in
\end{table*}

\subsubsection{\textbf{Structured Affective Reasoning}} \label{sec:5_3_1}
Structured reasoning is a well-established technique that guides models through explicit constraints rather than free exploration. Its adoption in multimodal affective computing has inspired methods collectively termed \textit{Structured Affective Reasoning}, summarized in \cref{tab:5-5}. Theoretical priors, reasoning schemas, and procedural workflows provide different foundations for organizing inference into diverse structures, with representative forms illustrated in \cref{fig:5-4} (b).

\textbf{Prior-Based Reasoning.}
Psychological priors provide a conceptual framework for connecting observable cues with emotional experience. EmoCaliber~\cite{wu26emocaliber} builds emotional interpretations from salient elements and their contextual relationships, and EmoToM~\cite{etesam24emotio} combines perception with contextual reasoning. Mental-state attribution guides recognition across tasks in TMPO~\cite{luo26hitemotio} and connects emotional understanding with supportive responses in OneEmo~\cite{huang26oneemo}. More specialized approaches explain affect through expectation conflicts~\cite{xu26moca}, appraisal and coping~\cite{chu26agctr2}, or physiological responses~\cite{liu26phemo}, grounding intermediate reasoning in distinct accounts of emotional experience.

More varied structures are also supported. Hierarchical approaches connect visual attributes with expressions~\cite{zhang25videmo}, relate global atmosphere to local details~\cite{zhang26profoc}, or organize interpretation across cognitive levels~\cite{huang25affecti}. Agent-MER~\cite{lai25agentmer} instead draws on a hierarchy of emotion categories, refining coarse predictions into finer distinctions and combining the results through voting. OSIRIS~\cite{abdelf26osmo} extends reasoning over time, updating emotional interpretations as new observations are considered alongside prior emotional states.

\textbf{Schema-Based Reasoning.}
Beyond theory-driven guidance, reasoning schemas organize affective inference through heuristically designed steps that have also demonstrated empirical benefits. A common strategy is to establish an explicit evidential basis before reaching a judgment. Facial AUs~\cite{lan25expllm,wu26facialr1,lin26tag}, subtle motion~\cite{zhao25mellm}, bodily expressions~\cite{saim25anatomy}, and emotional triggers~\cite{fang26emor3} provide intermediate anchors for visual interpretation. Acoustic and physiological approaches follow a similar logic through prosodic and semantic analysis~\cite{wang26emotiont}, comparisons between recordings~\cite{naini26compara}, or EEG descriptions~\cite{ma26e2llm}. With multiple modalities, the emphasis shifts toward reconciling evidence through causal interpretation~\cite{zhao26saber} and conflict resolution~\cite{xu26emothinker,shangg26mulcot}, including the incongruity underlying sarcasm~\cite{kang25ssrgssrd,li26sarcasmmin}. 

From another perspective, the resulting interpretations can also support different outputs, from explanatory judgments~\cite{zhao26emotionr} and emotion sets~\cite{lian25affr1} to task-specific assessments~\cite{gao26eemologic}. AffectOmni~\cite{wang26affomni} separates context, reasoning, and answers within a shared format, whereas GAT~\cite{she26aicabench} organizes visually grounded reasoning into a tree. For interactive tasks, an emotional judgment becomes an intermediate step toward response generation. ParalinGPT~\cite{lin24parallm} connects current and response sentiments before producing a reply, and CogAudio-LLM~\cite{zhao26cogaudio} elaborates this transition through intent, mental state, and response strategy. Related approaches incorporate prosodic reasoning~\cite{kim26pallm} or emotional-state and cause analysis~\cite{wang26ieat}, extending interpretation from explaining an observed emotion to informing an appropriate response.

\textbf{Workflow-Based Reasoning.}
Workflow-based reasoning shifts the emphasis from organizing reasoning content to orchestrating inference operations, specifying their sequence, dependencies, and information flow. AffectAgent~\cite{wang26affagt} assigns query planning, evidence filtering, and emotion prediction to successive stages. Contrasting rationales~\cite{chen24cofipara} and progressively complex questions~\cite{ouyang25punch} introduce alternative procedures for social interpretation. Relational organization is also possible, as in GraphEMO~\cite{tu26graphemo}, which incorporates emotion-cause relations into hierarchical context for support generation.

Revisiting intermediate results makes these procedures responsive to additional evidence. RAER~\cite{wen25raer} refines initial reasoning through retrieval, PEP-MEK~\cite{xing26emotionh} returns to predictions after verification, and V-AlphaSocial~\cite{lin25valpha} couples reflection with revision. Flexibility can also be enhanced through the selection of operations: ActFER~\cite{liu26actfer} selectively inspects facial regions, EmoAgent-R1~\cite{fang26emoagent} routes inputs to specialists, and FIRES~\cite{wang25fires} focuses reasoning on selected aspects of an interaction. Across these designs, explicit intermediate steps give affective inference structure, and their organization accommodates both predefined paths and input-dependent execution.

\begin{table*}[t]
\centering
\caption{Affective Reasoning Grounding methods with their key attributes. Grounding Scope distinguishes evidence within individual modalities from relationships across modalities, with some methods addressing both.}
\label{tab:5-6}
\vskip -0.1in
\begin{adjustbox}{max width=\textwidth}
\footnotesize
\renewcommand{\arraystretch}{1.1}
\setlength{\tabcolsep}{4pt}
\begin{tabular}{@{}lcccccc@{}}
\toprule
\textbf{Method} &
\textbf{Year} &
\textbf{Task} &
\textbf{Modality} &
\textbf{Grounding Scope} &
\textbf{Evidence Content} &
\textbf{Key Design} \\
\midrule
% FEALLM~\cite{hu25feallm}{,} Facial-R1~\cite{wu26facialr1}{,} ActFER~\cite{liu26actfer}{,} TAG~\cite{lin26tag}{,} VidEmo~\cite{zhang25videmo}{,} AlignCap~\cite{liang24alignca}{,} PALLM~\cite{kim26pallm}{,} ELENA~\cite{saim25anatomy}{,} MERMAID~\cite{yang25mermaid}{,} ProFocus~\cite{zhang26profoc}{,} EmoDETective~\cite{huang25emodete}{,} ICM~\cite{luo25omnisila}{,} GRASP~\cite{kim26grasp}{,} VToM~\cite{chen25vtom}{,} ARSENAL~\cite{agarwa24mememq}{,} EmoThinker~\cite{xu26emothinker}{,} ArtECulture-RAG~\cite{chen26artecult}{,} EmoOmni~\cite{zhao26emoomni}{,} SABER-LLM~\cite{zhao26saber}{,} AffectOmni~\cite{wang26affomni}{,} XMEI-Qwen~\cite{huang26jx4mei}{,} CofiPara~\cite{chen24cofipara}{,} S$^{3}$-MSD~\cite{zhu26s3msd}{,} SSRG Framework~\cite{kang25ssrgssrd}{,} MoLE~\cite{lee26smilenext}
% Intra-Modal

FEALLM~\cite{hu25feallm} &
2025 & FER & I &
Intra-Modal &
Action Units, Facial Expressions &
Local Facial Cues Explain Emotion \\

Facial-R1~\cite{wu26facialr1} &
2026 & FER & I &
Intra-Modal &
Action Units &
AU-Label-Guided Facial Reasoning \\

ActFER~\cite{liu26actfer} &
2026 & FER & I &
Intra-Modal &
Action Units, Facial Regions &
Active Local-Evidence Acquisition \\

TAG~\cite{lin26tag} &
2026 & FER & I &
Intra-Modal &
Action Units, Facial Regions &
AU-Grounded Facial Reasoning \\

VidEmo~\cite{zhang25videmo} &
2025 & FER, VCER & V &
Intra-Modal &
Facial Attributes, Expressions &
Affective-Tree Rationale Grounding \\

AlignCap~\cite{liang24alignca} &
2024 & AER & A &
Intra-Modal &
Pitch, Intonation, Speech Rate &
Acoustic-Cue-Grounded Emotion Captions \\

PALLM~\cite{kim26pallm} &
2026 & AER, MESC & A &
Intra-Modal &
Pauses, Pitch, Prosody &
Audio-Grounded Affective Reasoning \\

ELENA~\cite{saim25anatomy} &
2025 & VCER & I &
Intra-Modal &
Body Parts, Posture, Gestures &
Body-Part-Grounded Emotion Narratives \\

MERMAID~\cite{yang25mermaid} &
2025 & VCER & I &
Intra-Modal &
Facial Expressions, Posture, Scene &
Visual Reflection and Verification \\

ProFocus~\cite{zhang26profoc} &
2026 & VCER & I &
Intra-Modal &
Art Style, Subjects, Visual Details &
Progressive Art-Evidence Focusing \\

EmoDETective~\cite{huang25emodete} &
2025 & VCER & A,V &
Intra-Modal &
Prosody, Objects, Composition, Shot Scale &
Attribute-Wise Cause Facts and Analyses \\

ICM~\cite{luo25omnisila} &
2025 & VCER & V &
Intra-Modal &
Expressions, Actions, Object Relations &
Temporal Localization and Attribution \\

GRASP~\cite{kim26grasp} &
2026 & MSU & V &
Intra-Modal &
Gaze Targets, Pointing Gestures &
Identity-Aware Gaze and Gesture Grounding \\

VToM~\cite{chen25vtom} &
2024 & MSU & V,T &
Intra-Modal &
Facial Expressions, Gestures, Actions &
Question-Guided Frame Localization \\

% Inter-Modal

\rowcolor{gray!8}
ARSENAL~\cite{agarwa24mememq} &
2024 & MSU & I,T &
Inter-Modal &
Visual Entities, Meme Text &
Image-Text Entity-Role Explanations \\

% Intra- & Inter-Modal
\rowcolor{gray!8}
EmoThinker~\cite{xu26emothinker} &
2026 & FER, CMER, MSU & A,V &
Intra- \& Inter-Modal &
Facial Expressions, Prosody, Speech Content &
Cross-Modal Cue and Conflict Reasoning \\

\rowcolor{gray!8}
ArtECulture-RAG~\cite{chen26artecult} &
2026 & VCER & I,T &
Intra- \& Inter-Modal &
Visual Concepts, Cultural Rationales &
Concept-Grounded Cultural Evidence \\

\rowcolor{gray!8}
EmoOmni~\cite{zhao26emoomni} &
2026 & CMER & A,V,T &
Intra- \& Inter-Modal &
Facial, Acoustic, and Textual Cues &
Guided-Token Evidence Extraction \\

\rowcolor{gray!8}
SABER-LLM~\cite{zhao26saber} &
2026 & CMER, MSU & A,V &
Intra- \& Inter-Modal &
Facial Expressions, Gestures, Prosody &
Conflict-Aware Evidence Decomposition \\

\rowcolor{gray!8}
AffectOmni~\cite{wang26affomni} &
2026 & CMER, MSU & A,V &
Intra- \& Inter-Modal &
Expressions, Body Actions, Temporal Spans &
Rationale-to-Region Grounding \\

\rowcolor{gray!8}
XMEI-Qwen~\cite{huang26jx4mei} &
2026 & CMER, MSU & A,V,T &
Intra- \& Inter-Modal &
Facial Expressions, Prosody, Utterances &
Cue-Grounded Emotion-Intent Explanations \\

\rowcolor{gray!8}
CofiPara~\cite{chen24cofipara} &
2024 & MSU & I,T &
Intra- \& Inter-Modal &
Visual Entities, Textual Targets &
Rationale-Guided Target Identification \\

\rowcolor{gray!8}
S$^{3}$-MSD~\cite{zhu26s3msd} &
2026 & MSU & I,T &
Intra- \& Inter-Modal &
Visual Entities, Textual Entities &
Entity-Corruption-Guided Explanations \\

\rowcolor{gray!8}
SSRG Framework~\cite{kang25ssrgssrd} &
2025 & MSU & A,V,T &
Intra- \& Inter-Modal &
Action Units, Prosody, Dialogue Context &
Facial-Prosodic Incongruity Explanations \\

\rowcolor{gray!8}
MoLE~\cite{lee26smilenext} &
2026 & MSU & A,V,T &
Intra- \& Inter-Modal &
Action Units, Prosody, Utterances &
Textualized Cues Explain Laughter \\

\bottomrule
\end{tabular}
\end{adjustbox}
\vskip -0.1in
\end{table*}

\subsubsection{\textbf{Affective Reasoning Grounding}} ~\label{sec:5_3_2}
Accurate affective comprehension depends on the precise perception of multimodal evidence. The branch of \textit{Affective Reasoning Grounding} methods explicates this process by linking interpretations to cues within individual modalities and to relationships across modalities. We follow these two perspectives in the discussion, with representative methods summarized in \cref{tab:5-6} and mechanisms illustrated in \cref{fig:5-4} (c).

\textbf{Intra-Modal Grounding.}
Within individual modalities, grounding connects abstract emotional interpretations with concrete perceptual cues. AU-based explanations associate facial movements with emotion~\cite{hu25feallm,wu26facialr1}, extending to facial attributes and expressions in video~\cite{zhang25videmo}. Bodily expressions~\cite{saim25anatomy}, artistic details~\cite{zhang26profoc}, and emotion-related video attributes~\cite{huang25emodete} broaden the available evidence beyond the face. Acoustic approaches similarly anchor captions and reasoning in vocal characteristics~\cite{liang24alignca,kim26pallm}. These descriptions make the perceptual basis of an interpretation explicit without necessarily locating each cue in the input.

Explicit localization provides a more precise connection. ActFER~\cite{liu26actfer} actively acquires local facial observations, whereas TAG~\cite{lin26tag} produces AU-related bounding boxes alongside its reasoning. Temporal localization associates sentiment with relevant video intervals~\cite{luo25omnisila} or retrieves frames supporting mental-state judgments~\cite{chen25vtom}. GRASP~\cite{kim26grasp} further preserves participant identities when grounding gaze and gestures, resolving the individuals and interaction targets involved. MERMAID~\cite{yang25mermaid} complements these localization strategies through visual reflection and verification, revisiting image evidence to refine ambiguous interpretations.

\textbf{Inter-Modal Grounding.}
Grounding also extends to the inter-modal setting. For instance, ARSENAL~\cite{agarwa24mememq} explains entity roles through image-text rationales. Most approaches combine the inter- and intra-modal perspectives. CofiPara~\cite{chen24cofipara} uses rationales to guide visual and textual target identification, and S$^{3}$-MSD~\cite{zhu26s3msd} strengthens reliance on relevant visual entities through targeted perturbations. Facial-prosodic incongruity~\cite{kang25ssrgssrd} and textualized cues for laughter~\cite{lee26smilenext} extend this combined perspective to spoken interactions.

Explicitly distinguishing the contributing modalities further clarifies the basis of an interpretation. EmoOmni~\cite{zhao26emoomni} marks evidence extraction with modality-specific tokens, complemented by cue decomposition and conflict reasoning~\cite{xu26emothinker,zhao26saber}. Such evidence also supports joint emotion-intent explanations~\cite{huang26jx4mei}, and visual concepts anchor retrieved cultural rationales~\cite{chen26artecult}. AffectOmni~\cite{wang26affomni} links the resulting explanations back to perception by translating generated rationales into instructions for visual-region grounding. Together, these methods make affective explanations more traceable to their supporting observations.

\begin{table*}[t]
\centering
\caption{Reasoning and Explanation Verification methods with their key attributes. Evaluator Scoring encompasses rule-based and model-based assessments of the specified verification target.}
\label{tab:5-7}
\vskip -0.1in
\begin{adjustbox}{max width=\textwidth}
\footnotesize
\renewcommand{\arraystretch}{1.1}
\setlength{\tabcolsep}{4pt}
\begin{tabular}{@{}lcccccc@{}}
\toprule
\textbf{Method} &
\textbf{Year} &
\textbf{Task} &
\textbf{Modality} &
\textbf{Verification Target} &
\textbf{Verification Approach} &
\textbf{Key Design} \\
\midrule
% Facial-R1~\cite{wu26facialr1}{,} ActFER~\cite{liu26actfer}{,} TAG~\cite{lin26tag}{,} EmoOmni~\cite{zhao26emoomni}{,} AVEm-DPO~\cite{chaube26avere}{,} OPPO~\cite{han26oppo}{,} S$^{3}$-MSD~\cite{zhu26s3msd}{,} SEKE~\cite{wang25ekevllm}{,} VidEmo~\cite{zhang25videmo}{,} TMPO~\cite{luo26hitemotio}{,} AlignCap~\cite{liang24alignca}{,} EmoCaliber~\cite{wu26emocaliber}{,} EmotionReasoner~\cite{zhao26emotionr}{,} AffectOmni~\cite{wang26affomni}{,} CAIR~\cite{zhang26cair}{,} OneEmo~\cite{huang26oneemo}{,} SarcasmMiner~\cite{li26sarcasmmin}{,} EMO-R3~\cite{fang26emor3}{,} ERV~\cite{rha26erv}{,} EmObserver~\cite{wu26emobserver}{,} AffectGPT-R1~\cite{lian25affr1}{,} PEP-MEK~\cite{xing26emotionh}{,} MulCoT-RD~\cite{shangg26mulcot}
% Perception / Evaluator Scoring

Facial-R1~\cite{wu26facialr1} &
2026 &
FER &
I &
Perception &
Evaluator Scoring &
AU and Emotion-Label Reward Checks \\

ActFER~\cite{liu26actfer} &
2026 &
FER &
I &
Perception &
Evaluator Scoring &
Multilevel AU and Emotion Rewards \\

TAG~\cite{lin26tag} &
2026 &
FER &
I &
Perception &
Evaluator Scoring &
AU-Aware Spatial Grounding Rewards \\

EmoOmni~\cite{zhao26emoomni} &
2026 &
CMER &
A,V,T &
Perception &
Evaluator Scoring &
LLM-Based Evidence Verification \\

% Perception / Contrastive Checking

AVEm-DPO~\cite{chaube26avere} &
2026 &
FER, AER, CMER &
A,V &
Perception &
Contrastive Checking &
Faithful-Hallucinated Cue Preferences \\

OPPO~\cite{han26oppo} &
2026 &
CMER, MSU &
A,V,T &
Perception &
Contrastive Checking &
Cue Coverage and Modality-Masking Checks \\

S$^{3}$-MSD~\cite{zhu26s3msd} &
2026 &
MSU &
I,T &
Perception &
Contrastive Checking &
Entity-Perturbation Preference Pairs \\

% Perception / Self-Verification

SEKE~\cite{wang25ekevllm} &
2025 &
FER &
I &
Perception &
Self-Verification &
Uncertainty-Aware Annotation Checks \\

% Rationale / Evaluator Scoring

\rowcolor{gray!8}
VidEmo~\cite{zhang25videmo} &
2025 &
FER, VCER &
V &
Rationale &
Evaluator Scoring &
Tree-Edit and Generative Caption Rewards \\

\rowcolor{gray!8}
TMPO~\cite{luo26hitemotio} &
2026 &
All MER Tasks except PER &
I,A,V,T &
Rationale &
Evaluator Scoring &
Mental-State and Coherence Rewards \\

\rowcolor{gray!8}
AlignCap~\cite{liang24alignca} &
2024 &
AER &
A &
Rationale &
Evaluator Scoring &
Model-Scored Caption Preferences \\

\rowcolor{gray!8}
EmoCaliber~\cite{wu26emocaliber} &
2025 &
VCER &
I &
Rationale &
Evaluator Scoring &
Hallucination and Logic Screening \\

\rowcolor{gray!8}
EmotionReasoner~\cite{zhao26emotionr} &
2026 &
CMER &
A,V,T &
Rationale &
Evaluator Scoring &
Explanation Fidelity and Format Rewards \\

\rowcolor{gray!8}
AffectOmni~\cite{wang26affomni} &
2026 &
CMER, MSU &
A,V &
Rationale &
Evaluator Scoring &
People-Focus and Temporal-Order Rewards \\

\rowcolor{gray!8}
CAIR~\cite{zhang26cair} &
2026 &
CMER, MSU &
A,V,T &
Rationale &
Evaluator Scoring &
Causal-Mediation Utility Rewards \\

\rowcolor{gray!8}
OneEmo~\cite{huang26oneemo} &
2026 &
CMER, MSU, MESC &
A,V,T &
Rationale &
Evaluator Scoring &
Reverse Verification and Rubric Rewards \\

\rowcolor{gray!8}
SarcasmMiner~\cite{li26sarcasmmin} &
2026 &
MSU &
A,V,T &
Rationale &
Evaluator Scoring &
Generative-Judge Reasoning Rewards \\

% Rationale / Self-Verification

\rowcolor{gray!8}
EMO-R3~\cite{fang26emor3} &
2026 &
VCER &
I &
Rationale &
Self-Verification &
Visual and Emotional Self-Checks \\

% Consistency / Evaluator Scoring

ERV~\cite{rha26erv} &
2026 &
FER &
A,V &
Consistency &
Evaluator Scoring &
Sentence-Level Emotion Alignment \\

EmObserver~\cite{wu26emobserver} &
2026 &
VCER &
I,T &
Consistency &
Evaluator Scoring &
Analysis-Answer Consistency Rewards \\

AffectGPT-R1~\cite{lian25affr1} &
2025 &
CMER &
A,V,T &
Consistency &
Evaluator Scoring &
Reasoning-Emotion Alignment Rewards \\

% Consistency / Self-Verification

PEP-MEK~\cite{xing26emotionh} &
2026 &
AER, VCER, CMER &
I,A,V,T &
Consistency &
Self-Verification &
Knowledge-Guided Predict-Explain-Predict \\

MulCoT-RD~\cite{shangg26mulcot} &
2026 &
ITER &
I,T &
Consistency &
Self-Verification &
Label-Aware Filtering and Adaptive Retry \\

\bottomrule
\end{tabular}
\end{adjustbox}
\vskip -0.1in
\end{table*}

\subsubsection{\textbf{Reasoning and Explanation Verification}} \label{sec:5_3_3}
Building on \textit{Affective Reasoning Grounding}, \textit{Reasoning and Explanation Verification} further strengthens the reliability through explicit checks. These checks target perceptual accuracy, rationale validity, or explanation-prediction consistency, depending on the component under scrutiny. \Cref{fig:5-4} (d) illustrates the verification mechanisms, with representative methods detailed in \cref{tab:5-7}.

\textbf{Perception Verification.}
Verification begins with the reliability of the perceptual cues. Three approaches provide complementary checks: evaluator scoring assesses evidence against explicit criteria, contrastive checking compares different evidence conditions, and self-verification uses the model's own assessments to identify unreliable outputs. Facial-R1~\cite{wu26facialr1} and ActFER~\cite{liu26actfer} combine AU and emotion-level rewards, with the latter additionally evaluating the utility of local inspection. TAG~\cite{lin26tag} introduces spatial checks by overlapping predicted regions with AU-related boxes. For multimodal evidence, EmoOmni~\cite{zhao26emoomni} uses an LLM-based verifier to filter extracted cues before constructing reasoning traces.

Comparisons provide a complementary interface. AVEm-DPO~\cite{chaube26avere} constructs preferences between faithful and hallucinated descriptions. OPPO~\cite{han26oppo} combines cue-coverage rewards with comparisons under modality masking, discouraging claims unsupported by their source modality. Targeted visual-entity perturbations serve a related purpose in S$^{3}$-MSD~\cite{zhu26s3msd}, reducing reliance on textual shortcuts. Verification can precede model training as well: SEKE~\cite{wang25ekevllm} combines affective knowledge with uncertainty-aware sampling to improve the reliability of generated facial annotations.

\textbf{Rationale Verification.}
Accurate cues alone do not ensure a sound explanation. Rationale-oriented checks therefore assess the structure, factual support, and logic of the resulting interpretation. VidEmo~\cite{zhang25videmo} evaluates captions through affective-tree comparisons and scoring, and AlignCap~\cite{liang24alignca} converts model-scored caption quality into preference supervision. Verification further extends to the factual and logical validity of reasoning, for example through hallucination screening~\cite{wu26emocaliber}, coherence checks on mental-state reasoning~\cite{luo26hitemotio}, and explanation-fidelity assessment~\cite{zhao26emotionr}.

More targeted evaluators examine people-centric and temporal evidence~\cite{wang26affomni}, factual and reasoning coherence~\cite{huang26oneemo}, or fabricated audiovisual support~\cite{li26sarcasmmin}. CAIR~\cite{zhang26cair} takes a functional perspective, rewarding rationale for predicting the correct emotion through a dual-path comparison. EMO-R3~\cite{fang26emor3} instead feeds generated reasoning back into the model to check its visual support and emotional coherence, incorporating self-assessment into reward construction.

\textbf{Consistency Verification.}
A further requirement is agreement between the explanation and the predicted emotion, as inconsistent rationales can mislead rather than clarify. ERV~\cite{rha26erv} evaluates emotionally salient sentences against the target emotion, providing finer-grained feedback than whole-response scoring. AffectGPT-R1~\cite{lian25affr1} compares emotion concepts extracted from reasoning with the final prediction, and EmObserver~\cite{wu26emobserver} uses an LLM judge to assess analysis coherence and alignment with the final verdict.

Such checks can guide revision as well as scoring. PEP-MEK~\cite{xing26emotionh} revisits an initial prediction after generating an explanation informed by modality and emotion knowledge. MulCoT-RD~\cite{shangg26mulcot} applies label-aware filtering and adaptive retries when constructing reasoning demonstrations. Consistency verification thus influences both the examples from which models learn and the revisions made during inference.

\begin{table*}[t]
\centering
\caption{Context-Augmented Affective Reasoning methods with their key attributes. Context Source identifies the contextual information used, and Context Acquisition indicates how it is obtained.}
\label{tab:context-augmented-reasoning}
\vskip -0.1in
\begin{adjustbox}{max width=\textwidth}
\footnotesize
\renewcommand{\arraystretch}{1.1}
\setlength{\tabcolsep}{3pt}
\begin{tabular}{@{}lccccccc@{}}
\toprule
\textbf{Method} &
\textbf{Year} &
\textbf{Task} &
\textbf{Modality} &
\textbf{Context Source} &
\textbf{Context Acquisition} &
\textbf{Integration} &
\textbf{Key Design} \\
\midrule
% ParalinGPT~\cite{lin24parallm}{,} M$^{3}$F~\cite{wu25m3f}{,} OSIRIS~\cite{abdelf26osmo}{,} E3RG~\cite{lin25e3rg}{,} MTMEUR-Agent~\cite{hu25mtmeur}{,} Pers-ICL~\cite{takaha24persic}{,} MLLM-PIEA~\cite{gao25mllmpiea}{,} CMTD~\cite{nguyen26cmtd}{,} EmoCommonSense~\cite{xenos25emocom}{,} WisdoM~\cite{wang24wisdom}{,} RAER~\cite{wen25raer}{,} ArtECulture-RAG~\cite{chen26artecult}{,} AffectAgent~\cite{wang26affagt}{,} MIND~\cite{liu25mind}{,} PrismAgent~\cite{ding26prismage}{,} AffectVerse~\cite{zhao26affverse}{,} BIP-ALM~\cite{jin24mmtomqa}{,} Resonant Minds~\cite{shangg26resona}
% Conversation History / Memory

ParalinGPT~\cite{lin24parallm} &
2024 &
CMER &
A,T &
Conversation History &
Memory &
Structured Prompt &
Speech-Text-Sentiment History \\

M$^{3}$F~\cite{wu25m3f} &
2025 &
CMER, MSU &
A,V,T &
Conversation History &
Memory &
Feature Fusion &
Audiovisual Memory Attention \\

OSIRIS~\cite{abdelf26osmo} &
2026 &
MSU &
A,V,T &
Conversation History &
Memory &
Feature Fusion &
History-Based Emotion Tracking \\

E3RG~\cite{lin25e3rg} &
2025 &
MESC, ESS, EVGE &
A,V,T &
Conversation History &
Memory &
Structured Prompt &
History and Identity Conditioning \\

% Conversation History / Inference

MTMEUR-Agent~\cite{hu25mtmeur} &
2025 &
CMER, MSU &
V,T &
Conversation History &
Inference &
Multi-Agent Interaction &
Context and Character Analysis \\

% Persona / Memory

\rowcolor{gray!8}
Pers-ICL~\cite{takaha24persic} &
2024 &
VCER &
I,T &
Persona &
Memory &
Structured Prompt &
User-Specific Emotion Examples \\

% Persona / Inference

\rowcolor{gray!8}
MLLM-PIEA~\cite{gao25mllmpiea} &
2025 &
VCER, MSU &
I,T &
Persona &
Inference &
Structured Prompt &
History-Derived Persona Profiles \\

\rowcolor{gray!8}
CMTD~\cite{nguyen26cmtd} &
2026 &
CMER &
A,V,T &
Persona &
Inference &
Multi-Agent Interaction &
Trait and Distortion Agents \\

% World Knowledge / Inference

EmoCommonSense~\cite{xenos25emocom} &
2024 &
VCER &
I,V &
World Knowledge &
Inference &
Feature Fusion &
Commonsense-Visual Fusion \\

WisdoM~\cite{wang24wisdom} &
2024 &
ITER &
I,T &
World Knowledge &
Inference &
Probabilistic Integration &
Selective Contextual Fusion \\

% World Knowledge / Investigation

RAER~\cite{wen25raer} &
2025 &
FER, CMER &
A,V,T &
World Knowledge &
Investigation &
Structured Prompt &
Retrieved Reasoning Exemplars \\

ArtECulture-RAG~\cite{chen26artecult} &
2026 &
VCER &
I,T &
World Knowledge &
Investigation &
Structured Prompt &
Cultural Concept Retrieval \\

AffectAgent~\cite{wang26affagt} &
2026 &
CMER &
A,V,T &
World Knowledge &
Investigation &
Multi-Agent Interaction &
Evidence Retrieval and Filtering \\

MIND~\cite{liu25mind} &
2025 &
MSU &
I,T &
World Knowledge &
Investigation &
Multi-Agent Interaction &
Retrieval and Bidirectional Reasoning \\

PrismAgent~\cite{ding26prismage} &
2026 &
MSU &
I,T &
World Knowledge &
Investigation &
Multi-Agent Interaction &
Retrieval and Adversarial Judgment \\

% Latent State / Inference

\rowcolor{gray!8}
AffectVerse~\cite{zhao26affverse} &
2026 &
CMER &
A,V,T &
Latent State &
Inference &
Token Injection &
Predictive Affective Belief Tokens \\

\rowcolor{gray!8}
BIP-ALM~\cite{jin24mmtomqa} &
2024 &
MSU &
V,T &
Latent State &
Inference &
Probabilistic Integration &
Bayesian Mental-State Inference \\

\rowcolor{gray!8}
Resonant Minds~\cite{shangg26resona} &
2026 &
MSU, EVGE &
A,V,T &
Latent State &
Inference &
Multi-Agent Interaction &
Mental-State-Guided Interaction \\

\bottomrule
\end{tabular}
\end{adjustbox}
\vskip -0.1in
\end{table*}

\subsubsection{\textbf{Context-Augmented Affective Reasoning}} \label{sec:5_3_4}
Even when emotional cues within a conversation are precisely interpreted, their significance can vary considerably with the participants’ prior experiences, personal characteristics, and broader social or cultural context. \textit{Context-Augmented Affective Reasoning} thereby enriches their interpretation with information recalled from memory, inferred from observations, or retrieved from external resources. We organize the following discussion by these context sources. \Cref{tab:context-augmented-reasoning} summarizes the corresponding methods, complementing the overview in \cref{fig:5-4} (e).

\textbf{Conversation History.}
Emotional meaning develops over an interaction, making preceding exchanges an important reference for interpreting the present. Historical information can be retained explicitly in the input or compressed into representations. ParalinGPT~\cite{lin24parallm} serializes speech, text, and sentiment history, preserving their conversational sequence. M$^{3}$F~\cite{wu25m3f} instead incorporates historical audiovisual features through memory-bank attention, and OSIRIS~\cite{abdelf26osmo} combines prior descriptions with multimodal memory to support continuous self-emotion tracking. These designs connect current observations with accumulated experience at different representational levels. Beyond retaining history, MTMEUR-Agent~\cite{hu25mtmeur} assigns contextual interpretation to specialists in background, character dynamics, and events. The same continuity supports generation in E3RG~\cite{lin25e3rg}, which combines dialogue history with identity profiles and historical audiovisual references to maintain emotionally and personally consistent responses.

\textbf{Persona.}
Accumulated interactions reveal not only changing emotional states but relatively persistent individual differences. Persona-based methods use these differences to personalize interpretations of otherwise similar stimuli. User-specific examples provide a direct reference in Pers-ICL~\cite{takaha24persic}, and MLLM-PIEA~\cite{gao25mllmpiea} develops a more explicit account by deriving personality profiles from historical information. This progression makes personal context available beyond individual demonstrations. CMTD~\cite{nguyen26cmtd} combines stable personality traits with dynamic cognitive distortions through specialized agents, connecting enduring dispositions with situation-dependent patterns of interpretation.

\textbf{World Knowledge.}
Personal and conversational context may still leave cultural references, implicit associations, or situational meanings unresolved. One response is to elicit relevant knowledge already encoded in pretrained models. EmoCommonSense~\cite{xenos25emocom} generates commonsense descriptions to complement visual features, translating contextual relations into linguistic evidence. WisdoM~\cite{wang24wisdom} further accounts for the potential noise in such generated context by selectively combining context-conditioned predictions for difficult samples. Their shared contribution is to make background knowledge operational without requiring an external retrieval step.

Retrieval extends this knowledge base with explicit reference material. ArtECulture-RAG~\cite{chen26artecult} retrieves cultural concepts for artwork interpretation, and RAER~\cite{wen25raer} draws on affective exemplars to refine reasoning. Agent-based designs place greater emphasis on evaluating and consolidating the retrieved material: AffectAgent~\cite{wang26affagt} links retrieval with specialist evidence filtering, MIND~\cite{liu25mind} derives insights through bidirectional analysis of related memes, and PrismAgent~\cite{ding26prismage} subjects retrieved evidence to contrasting interpretations and staged judgment. External context therefore serves not merely as supplementary prompt content but as material for comparison, selection, and deliberation.

\textbf{Latent State.}
Some explanatory context concerns unobserved states inferred from behavior and is difficult to obtain through retrieval. BIP-ALM~\cite{jin24mmtomqa} evaluates mental-state hypotheses through action likelihoods, using Bayesian inverse planning to connect observed behavior with underlying beliefs and goals. AffectVerse~\cite{zhao26affverse} adopts a predictive representation-based approach: imagined future audiovisual features are aggregated into belief tokens and injected into the LLM, introducing expectations about affective dynamics into current interpretation. These tokens encode predictive context rather than explicit psychological categories. Resonant Minds~\cite{shangg26resona} brings mental-state inference into reciprocal interaction, using inferred states to guide emotionally expressive avatar responses. Context augmentation thus situates affective reasoning within an evolving account of individuals, their experiences, and their interactions.

\begin{table*}[t]
\centering
\caption{Subjectivity-Aware Affective Reasoning methods with their key attributes. Decision Treatment characterizes personalized or conditioned judgments, emotion distributions, multiple labels, and calibrated confidence.}
\label{tab:subjectivity-aware-reasoning}
\vskip -0.1in
\begin{adjustbox}{max width=\textwidth}
\footnotesize
\renewcommand{\arraystretch}{1.12}
\setlength{\tabcolsep}{5pt}
\begin{tabular}{@{}lcccccc@{}}
\toprule
\textbf{Method} &
\textbf{Year} &
\textbf{Task} &
\textbf{Modality} &
\textbf{Reasoning Type} &
\textbf{Decision Treatment} &
\textbf{Key Design} \\
\midrule
% PersSER~\cite{ihori25persser}{,} Pers-ICL~\cite{takaha24persic}{,} MLLM-PIEA~\cite{gao25mllmpiea}{,} CMTD~\cite{nguyen26cmtd}{,} OSIRIS~\cite{abdelf26osmo}{,} EmObserver~\cite{wu26emobserver}{,} ArtECulture-RAG~\cite{chen26artecult}{,} DiME-Aug~\cite{zhang26dimeaug}{,} Emotion-LLaMAv2~\cite{peng26emotionl}{,} EEmo-Logic~\cite{gao26eemologic}{,} EmoGist~\cite{seoh25emogist}{,} AffectGPT~\cite{lian25affectgp}{,} AffectGPT-R1~\cite{lian25affr1}{,} Agent-MER~\cite{lai25agentmer}{,} AcoustEmo~\cite{zhang26acouste}{,} EmotionReasoner~\cite{zhao26emotionr}{,} OneEmo~\cite{huang26oneemo}{,} EmoCaliber~\cite{wu26emocaliber}{.}
% Subjectivity-Conditioned / Personalized Prediction

PersSER~\cite{ihori25persser} &
2025 &
AER &
A,T &
Subjectivity-Conditioned &
Personalized Prediction &
Speaker-Specific Examples \\

Pers-ICL~\cite{takaha24persic} &
2024 &
VCER &
I,T &
Subjectivity-Conditioned &
Personalized Prediction &
User-Specific Demonstrations \\

MLLM-PIEA~\cite{gao25mllmpiea} &
2025 &
VCER, MSU &
I,T &
Subjectivity-Conditioned &
Personalized Prediction &
Personality-Guided Prediction \\

CMTD~\cite{nguyen26cmtd} &
2026 &
CMER &
A,V,T &
Subjectivity-Conditioned &
Personalized Prediction &
Trait and Distortion Modeling \\

OSIRIS~\cite{abdelf26osmo} &
2026 &
MSU &
A,V,T &
Subjectivity-Conditioned &
Personalized Prediction &
History-Based Emotion Tracking \\

% Subjectivity-Conditioned / Conditioned Judgment

EmObserver~\cite{wu26emobserver} &
2026 &
VCER &
I,T &
Subjectivity-Conditioned &
Conditioned Judgment &
Observer-Conditioned Judgment \\

ArtECulture-RAG~\cite{chen26artecult} &
2026 &
VCER &
I,T &
Subjectivity-Conditioned &
Conditioned Judgment &
Culture-Conditioned Explanations \\

% Diversity-Preserving / Distribution Prediction

\rowcolor{gray!8}
DiME-Aug~\cite{zhang26dimeaug} &
2026 &
AER &
A,T &
Diversity-Preserving &
Distribution Prediction &
Human-Synthetic Distributions \\

% Diversity-Preserving / Multi-Label Prediction

\rowcolor{gray!8}
Emotion-LLaMAv2~\cite{peng26emotionl} &
2026 &
FER, AER, CMER &
A,V,T &
Diversity-Preserving &
Multi-Label Prediction &
Multi-Label Emotion Prediction \\

\rowcolor{gray!8}
EEmo-Logic~\cite{gao26eemologic} &
2026 &
VCER &
I &
Diversity-Preserving &
Multi-Label Prediction &
Top-Three Emotion Ranking \\

\rowcolor{gray!8}
EmoGist~\cite{seoh25emogist} &
2025 &
VCER&
I,T &
Diversity-Preserving &
Multi-Label Prediction &
Description-Guided Label Decisions \\

\rowcolor{gray!8}
AffectGPT~\cite{lian25affectgp} &
2025 &
CMER &
A,V,T &
Diversity-Preserving &
Multi-Label Prediction &
Description-Based Emotion Sets \\

\rowcolor{gray!8}
AffectGPT-R1~\cite{lian25affr1} &
2025 &
CMER &
A,V,T &
Diversity-Preserving &
Multi-Label Prediction &
Emotion-Set Reward Optimization \\

\rowcolor{gray!8}
Agent-MER~\cite{lai25agentmer} &
2025 &
CMER &
A,V,T &
Diversity-Preserving &
Multi-Label Prediction &
Hierarchical Emotion Search \\

\rowcolor{gray!8}
AcoustEmo~\cite{zhang26acouste} &
2026 &
CMER &
A,V,T &
Diversity-Preserving &
Multi-Label Prediction &
Utterance-Aware Emotion Reasoning \\

\rowcolor{gray!8}
EmotionReasoner~\cite{zhao26emotionr} &
2026 &
CMER &
A,V,T &
Diversity-Preserving &
Multi-Label Prediction &
Emotion-Wheel Set Rewards \\

\rowcolor{gray!8}
OneEmo~\cite{huang26oneemo} &
2026 &
CMER, MSU, MESC &
A,V,T &
Diversity-Preserving &
Multi-Label Prediction &
Unified Emotion-Set Reasoning \\

% Diversity-Preserving / Calibrated Confidence

\rowcolor{gray!8}
EmoCaliber~\cite{wu26emocaliber} &
2025 &
VCER &
I &
Diversity-Preserving &
Calibrated Confidence &
Verbalized Confidence Calibration \\

\bottomrule
\end{tabular}
\end{adjustbox}
\vskip -0.1in
\end{table*}

\subsubsection{\textbf{Subjectivity-Aware Affective Reasoning}} \label{sec:5_3_5}
Additional context can clarify affective evidence without resolving all differences in its interpretation. \textit{Subjectivity-Aware Affective Reasoning} accommodates this variability by conditioning judgments on individual perspectives or preserving diverse outcomes and predictive uncertainty. These two branches, \textit{Subjectivity-Conditioned} and \textit{Diversity-Preserving} reasoning, organize the methods in \cref{tab:subjectivity-aware-reasoning} and are illustrated in \cref{fig:5-4} (f).

\textbf{Subjectivity-Conditioned Reasoning.}
Individual differences affect both emotional expression and the perception of affective stimuli. Example-based personalization accommodates these differences through individual reference examples: PersSER~\cite{ihori25persser} conditions speech emotion recognition on enrollment utterances from the target speaker, and Pers-ICL~\cite{takaha24persic} uses user-specific demonstrations to personalize visual emotion prediction. Beyond direct demonstrations, personal experience can be distilled into profiles or retained as an evolving emotional history. MLLM-PIEA~\cite{gao25mllmpiea} derives personality profiles from historical experiences to guide image emotion analysis. CMTD~\cite{nguyen26cmtd} combines stable traits with dynamic cognitive distortions, linking enduring dispositions to situation-dependent interpretations. OSIRIS~\cite{abdelf26osmo} emphasizes temporal continuity by interpreting current self-emotion in relation to prior emotional states and multimodal history.

Observer characteristics and cultural background provide further anchors for subjective judgments. EmObserver~\cite{wu26emobserver} incorporates specific conditions into emotional statements and assesses their validity from the specified perspective. ArtECulture-RAG~\cite{chen26artecult} grounds artwork emotion judgments and explanations in retrieved cultural concepts, extending individualized interpretation to culturally shared associations.

\textbf{Diversity-Preserving Reasoning.}
Unlike subjectivity-conditioned methods, which tailor judgments to particular perspectives, diversity-preserving methods represent alternatives or uncertainty in the predictions themselves. DiME-Aug~\cite{zhang26dimeaug} combines human and synthetic annotations into emotion distributions, preserving disagreement through distributional supervision and supporting class-balanced learning through augmentation. Multiple-label prediction accommodates affective diversity without requiring a full probability distribution. Emotion-LLaMAv2~\cite{peng26emotionl} supports multi-label recognition across heterogeneous taxonomies, EEmo-Logic~\cite{gao26eemologic} ranks the top three evoked emotions, and EmoGist~\cite{seoh25emogist} evaluates candidate labels independently using retrieved descriptions. These formulations retain complementary aspects of emotional interpretation, including relative prevalence, ranked plausibility, and label co-occurrence.

Open-vocabulary recognition further expands the range of expressible emotions. Descriptive supervision in AffectGPT~\cite{lian25affectgp} supports nuanced emotion sets, complemented by utterance-aware acoustic representations in AcoustEmo~\cite{zhang26acouste}. Agent-MER~\cite{lai25agentmer} broadens fine-grained coverage through hierarchical search and self-consistent voting. Emotion-set rewards directly support optimization in AffectGPT-R1~\cite{lian25affr1} and EmotionReasoner~\cite{zhao26emotionr}, with OneEmo~\cite{huang26oneemo} integrating this capability into a broader recognition-and-interaction framework. Alongside expanded label coverage, EmoCaliber~\cite{wu26emocaliber} introduces calibrated verbalized confidence to qualify the reliability of a selected interpretation, including single-emotion predictions. The distinction is consequential: distributions preserve disagreement among judgments, multiple labels accommodate affective multiplicity, and confidence expresses uncertainty in the model's own decision.

\begin{figure*}[t]
    \centering
    \includegraphics[width=1\linewidth]{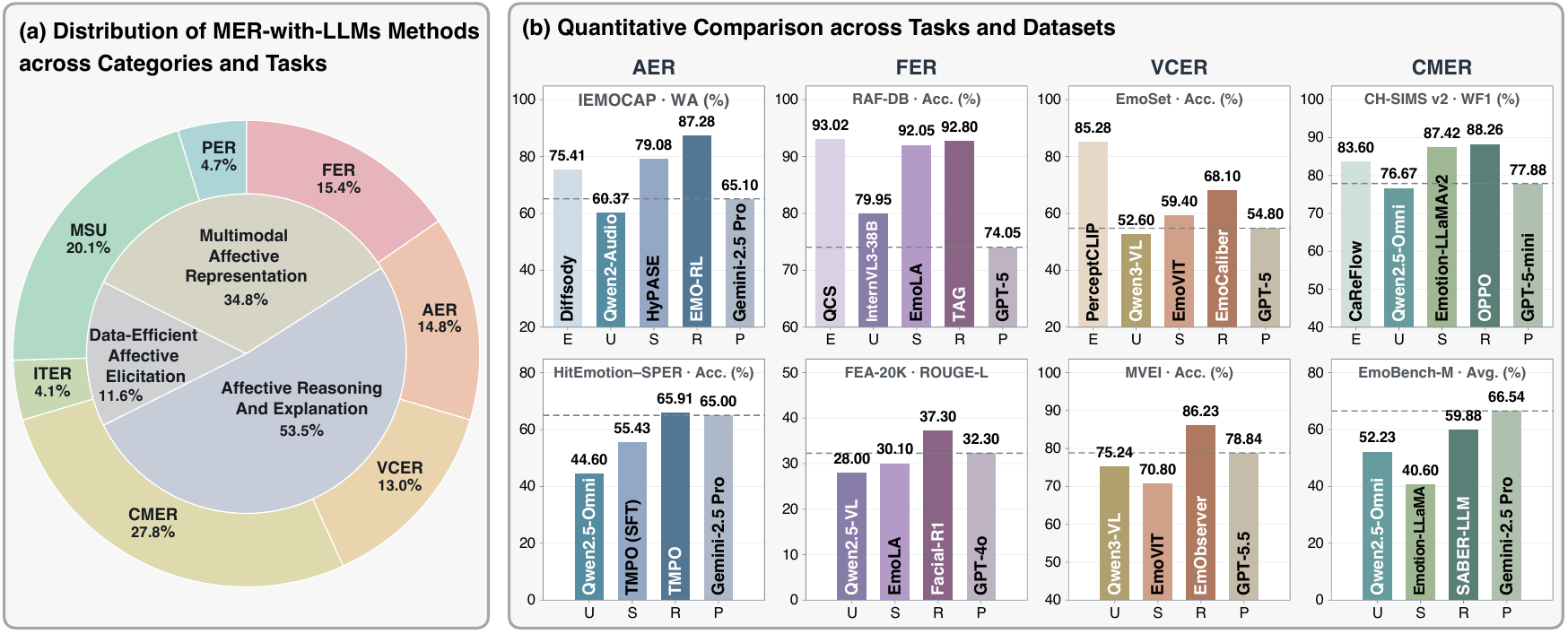}
    \vskip -0.1in
    \caption{Overview of the distribution and performance of MER-with-LLMs methods. The right panel compares representative expert small models (E), untuned open-source MLLMs (U), SFT-optimized MER-with-LLMs methods (S), RL-optimized MER-with-LLMs methods (R), and proprietary MLLMs (P) across four widely studied tasks. The top row presents results on conventional datasets, while the bottom row reports results on LLM-oriented benchmarks. Quantitative results are compiled from the reviewed literature with established comparability across evaluation protocols~\cite{qu25diffsody,qwen2audio,jin26hypase,li25emorl,gemini25,wang25qcs,internvl3,li24faba,lin26tag,gpt5,zalcher25perceptclip,qwen3vl,xie24emovit,wu26emocaliber,mai26careflow,qwen25omni,peng26emotionl,han26oppo,gpt4o,luo26hitemotio,qwen25vl,wu26facialr1,wu26emobserver,gpt55,cheng24emotion,zhao26saber,ge26synser,huang26oneemo}. Unless specified, we report the 7--8B variants for models available at multiple scales.}
    \label{fig:5-5}
    \vskip -0.1in
\end{figure*}

\subsection{Discussion}

\subsubsection{\textbf{Quantitative Analysis}}
The preceding discussion organizes MER-with-LLMs into three categories and nine methodological branches spanning affective elicitation, representation, and reasoning. \Cref{fig:5-5} (a) summarizes their distribution across categories and tasks. \textit{Affective Reasoning and Explanation} accounts for over half of the methods, highlighting substantial interest in explicit interpretation beyond emotion prediction. This emphasis accompanies broad task coverage, with particularly strong representation of CMER and MSU connecting affective recognition with multimodal and social understanding.

\Cref{fig:5-5} (b) complements this methodological overview with performance comparisons on four extensively studied tasks: AER, FER, VCER, and CMER. On conventional datasets, the illustrated proprietary models consistently underperform expert small models, exposing a gap between general multimodal competence and specialized affective recognition. On LLM-oriented benchmarks, however, proprietary models remain competitive even against affectively optimized alternatives. This contrast suggests that performance on conventional datasets partly reflects adaptation to predefined categories and annotation conventions, rather than affective competence that transfers uniformly across evaluation settings. Consistent with this distinction, some SFT models underperform untuned open models on MVEI and EmoBench-M, indicating that specialized supervision does not invariably improve broader affective understanding. Nevertheless, post-training offers a viable route to narrowing these gaps: both SFT and RL can benefit affective recognition, with the illustrated RL-based methods approaching or exceeding expert models on several conventional datasets and proprietary models on LLM-oriented benchmarks. Although differences in backbones and training conditions limit causal attribution, these results position targeted post-training as an effective complement to general model capabilities, supporting the progression from specialized emotion recognition toward broader affective interpretation.

% Muted traffic-light colors.
\definecolor{merGood}{HTML}{DCEBDD}
\definecolor{merMiddle}{HTML}{FFF0CD}
\definecolor{merPoor}{HTML}{F2DADA}
\definecolor{merGroup}{HTML}{EBEBEB}

% Dependency and cost: lower is preferable.
\newcommand{\merCostLow}{\cellcolor{merGood}\textbf{Low}}
\newcommand{\merCostMid}{\cellcolor{merMiddle}\textbf{Middle}}
\newcommand{\merCostHigh}{\cellcolor{merPoor}\textbf{High}}

% Capabilities: higher is preferable.
\newcommand{\merCapLow}{\cellcolor{merPoor}\textbf{Low}}
\newcommand{\merCapMid}{\cellcolor{merMiddle}\textbf{Middle}}
\newcommand{\merCapHigh}{\cellcolor{merGood}\textbf{High}}

\begin{table*}[t]
\centering
\caption{Qualitative comparison of MER-with-LLMs method categories. Generalization denotes anticipated transfer beyond the adaptation distribution, with (w.) or without (w/o.) task-specific SFT. Green, yellow, and red indicate favorable, intermediate, and unfavorable assessments, respectively.}
\label{tab:sec5dis}
\vskip -0.1in
\begin{adjustbox}{max width=\textwidth}
\scriptsize
\renewcommand{\arraystretch}{1.30}
\setlength{\tabcolsep}{4pt}
\begin{tabular}{@{}l*{8}{>{\centering\arraybackslash}m{0.52in}}}
\toprule
\multirow{2}{*}{\textbf{Method}} & \multicolumn{2}{c}{\textbf{Supervision Dependency}} & \multicolumn{2}{c}{\textbf{Computational Cost}} & \multicolumn{2}{c}{\textbf{Generalization}} & \multirow{2}{*}{\textbf{Robustness}} & \multirow{2}{*}{\textbf{Explain.}} \\
\cmidrule(lr){2-3}\cmidrule(lr){4-5}\cmidrule(lr){6-7}
& \textbf{Labels} & \textbf{CoT} & \textbf{Training} & \textbf{Inference} & \textbf{w/o. SFT} & \textbf{w. SFT} & & \\
\midrule

\rowcolor{merGroup}\multicolumn{9}{c}{\textbf{(\S\ref{sec:5_1}) Data-Efficient Affective Elicitation}} \\
(\S\ref{sec:5_1_1}) Inference-Time Affective Intervention & \merCostLow & \merCostLow & \merCostLow & \merCostMid & \merCapMid & \merCapMid & \merCapMid & \merCapMid \\
(\S\ref{sec:5_1_2}) Example-Guided Affective Adaptation & \merCostLow & \merCostLow & \merCostLow & \merCostMid & \merCapHigh & \merCapLow & \merCapLow & \merCapMid \\

\rowcolor{merGroup}\multicolumn{9}{c}{\textbf{(\S\ref{sec:5_2}) Multimodal Affective Representation}} \\
(\S\ref{sec:5_2_1}) Perceptual Affective Mapping & \merCostHigh & \merCostLow & \merCostMid & \merCostLow & \merCapLow & \merCapMid & \merCapMid & \merCapLow \\
(\S\ref{sec:5_2_2}) Multimodal Affective Coordination & \merCostHigh & \merCostLow & \merCostHigh & \merCostMid & \merCapLow & \merCapMid & \merCapMid & \merCapLow \\

\rowcolor{merGroup}\multicolumn{9}{c}{\textbf{(\S\ref{sec:5_3}) Affective Reasoning and Explanation}} \\
(\S\ref{sec:5_3_1}) Structured Affective Reasoning & \merCostMid & \merCostHigh & \merCostMid & \merCostHigh & \merCapMid & \merCapHigh & \merCapMid & \merCapHigh \\
(\S\ref{sec:5_3_2}) Affective Reasoning Grounding & \merCostMid & \merCostHigh & \merCostMid & \merCostHigh & \merCapMid & \merCapHigh & \merCapHigh & \merCapHigh \\
(\S\ref{sec:5_3_3}) Reasoning and Explanation Verification & \merCostMid & \merCostMid & \merCostHigh & \merCostHigh & \merCapMid & \merCapHigh & \merCapHigh & \merCapHigh \\
(\S\ref{sec:5_3_4}) Context-Augmented Affective Reasoning & \merCostMid & \merCostMid & \merCostHigh & \merCostHigh & \merCapHigh & \merCapHigh & \merCapHigh & \merCapHigh \\
(\S\ref{sec:5_3_5}) Subjectivity-Aware Affective Reasoning & \merCostMid & \merCostMid & \merCostHigh & \merCostHigh & \merCapMid & \merCapHigh & \merCapHigh & \merCapMid \\

\bottomrule
\end{tabular}
\end{adjustbox}
\vskip -0.1in
\end{table*}

\subsubsection{\textbf{Qualitative Analysis}}
Complementing these performance comparisons, \cref{tab:sec5dis} compares the methodological branches across five dimensions: supervision dependency, computational cost, generalization, robustness, and explainability. The ratings reflect a heuristic synthesis of the design choices and trade-offs commonly observed in representative methods.

\textbf{Supervision Dependency.}
The branches differ first in the amount and form of supervision they require, from task labels to CoT annotations. \textit{Data-Efficient Affective Elicitation} keeps both requirements modest by drawing on pretrained capabilities through targeted interventions or limited examples. Adapting the perceptual interface introduces greater demands: \textit{Multimodal Affective Representation} relies more heavily on labeled data to develop emotion-sensitive features and align specialized modules with the language backbone. For \textit{Affective Reasoning and Explanation}, pretrained semantic knowledge provides a stronger starting point, shifting the supervision emphasis toward how evidence supports a judgment. Training-based implementations of \textit{Structured Affective Reasoning} and \textit{Affective Reasoning Grounding} often make extensive use of CoT annotations to learn reasoning procedures and evidence associations. Other reasoning branches can use sparser feedback on outcomes or selected intermediate steps, often through RL, reducing their reliance on complete CoT annotations.

\textbf{Computational Cost.}
Computational cost refers to the additional resources required to train and run affective methods on top of a pretrained backbone. \textit{Data-Efficient Affective Elicitation} generally avoids extensive parameter updates, placing most additional cost in inference-time retrieval, input augmentation, or feature manipulation. Within \textit{Multimodal Affective Representation}, lightweight implementations of \textit{Perceptual Affective Mapping} update a projector or adapter while keeping the encoder largely frozen, limiting both training costs and inference overhead. By comparison, \textit{Multimodal Affective Coordination} introduces more elaborate fusion, routing, or cross-modal attention modules, increasing training demands and per-input computation. For \textit{Affective Reasoning and Explanation}, a major computational expense comes from generating extended reasoning through token-by-token autoregressive decoding, with longer traces incurring greater inference latency. RL carries this expense into training through the generation and evaluation of reasoning rollouts, adding substantial computation beyond parameter updates.

\textbf{Generalization.}
Generalization refers to a model's ability to transfer learned capabilities to unseen data, domains, or tasks, assessed here with and without task-specific SFT. Without SFT, \textit{Example-Guided Affective Adaptation} and \textit{Context-Augmented Affective Reasoning} often support flexible transfer by updating the examples or contextual evidence supplied to a largely unchanged pretrained backbone. The transferability of \textit{Inference-Time Affective Intervention} depends more closely on how broadly its intervention assumptions apply across settings. Newly introduced components in \textit{Multimodal Affective Representation}, meanwhile, typically require alignment with the backbone before they can effectively exploit its pretrained capabilities. SFT changes this balance by establishing the required alignment and supporting transfer across related settings. For \textit{Example-Guided Affective Adaptation}, narrowly targeted SFT can instead encourage format specialization and weaken the ability to learn from demonstrations on other tasks~\cite{wang24promot}. Reasoning-oriented SFT offers a potentially more transferable form of specialization by teaching intermediate procedures that can be reused and composed across problems~\cite{yin25composable}.

\textbf{Robustness.}
Robustness refers to the ability to maintain reliable affective predictions under input perturbations and domain shifts. As with generalization, the robustness of \textit{Inference-Time Affective Intervention} depends on the applicability of its underlying assumptions. \textit{Example-Guided Affective Adaptation} introduces an additional source of sensitivity through demonstration selection, presentation, and label distribution~\cite{wu25iclconfigu}. For \textit{Multimodal Affective Representation}, task-specific adaptation can improve tolerance to variations represented in the training data, leaving stability under unfamiliar conditions dependent on the coverage of that supervision. Reasoning-oriented methods offer more explicit ways to respond to changing evidence. \textit{Affective Reasoning Grounding} ties judgments to the observations available, and \textit{Reasoning and Explanation Verification} provides opportunities to detect inconsistencies. \textit{Context-Augmented Affective Reasoning} and \textit{Subjectivity-Aware Affective Reasoning} further accommodate differences in circumstances and interpretation. These mechanisms underpin their generally favorable assessment. Within this group, \textit{Structured Affective Reasoning} can be more restrictive, as a prescribed schema may encourage models to impose familiar reasoning patterns on situations that call for a different interpretation.

\textbf{Explainability.}
Explainability is another important property, reflecting how readily people can understand the basis of an affective judgment. \textit{Data-Efficient Affective Elicitation} often builds on explicit insights into relevant cues or informative examples, making the purpose of an intervention understandable even when its effect on a particular prediction remains implicit. \textit{Multimodal Affective Representation} offers less visibility into this process, as affective knowledge is encoded in learned parameters and feature interactions that are difficult to interpret directly. \textit{Affective Reasoning and Explanation} makes the basis of a judgment more accessible through intermediate reasoning. Despite the lack of a guarantee that explanations causally reflect the model’s decisions, data-driven learning or explicit optimization often promotes consistency between the two, generally improving explainability. Within this category, \textit{Subjectivity-Aware Affective Reasoning} presents a distinct challenge: explaining emotion distributions or confidence estimates requires accounting for the relative plausibility of different interpretations, beyond justifying a single prediction. The added nuance of these outputs is therefore not always matched by equally informative explanations.

Taken together, these comparisons reveal persistent trade-offs: specialized representations require substantial adaptation, while explicit reasoning improves inspectability at the cost of longer inference. Evidence that humans can recognize many emotional expressions rapidly and accurately~\cite{tracy08automaticity} motivates a broader search for architectures that better reconcile these capabilities. Two directions are particularly promising. LLM-based decision models, exemplified by UEmo-D1~\cite{wu26uemod1}, replace explicit CoT generation with direct probability estimates over candidate options, offering a route to broader task coverage with lower inference costs. A complementary direction emphasizes adaptiveness, allowing models first to assess whether an input requires additional reasoning, contextual retrieval, or verification. Tailoring these operations to the uncertainty and evidential demands of each input could strengthen robustness and explainability.

\section{MEG-with-LLMs}
\label{sec:6}

\begin{figure*}[t]
    \centering
    \includegraphics[width=1\linewidth]{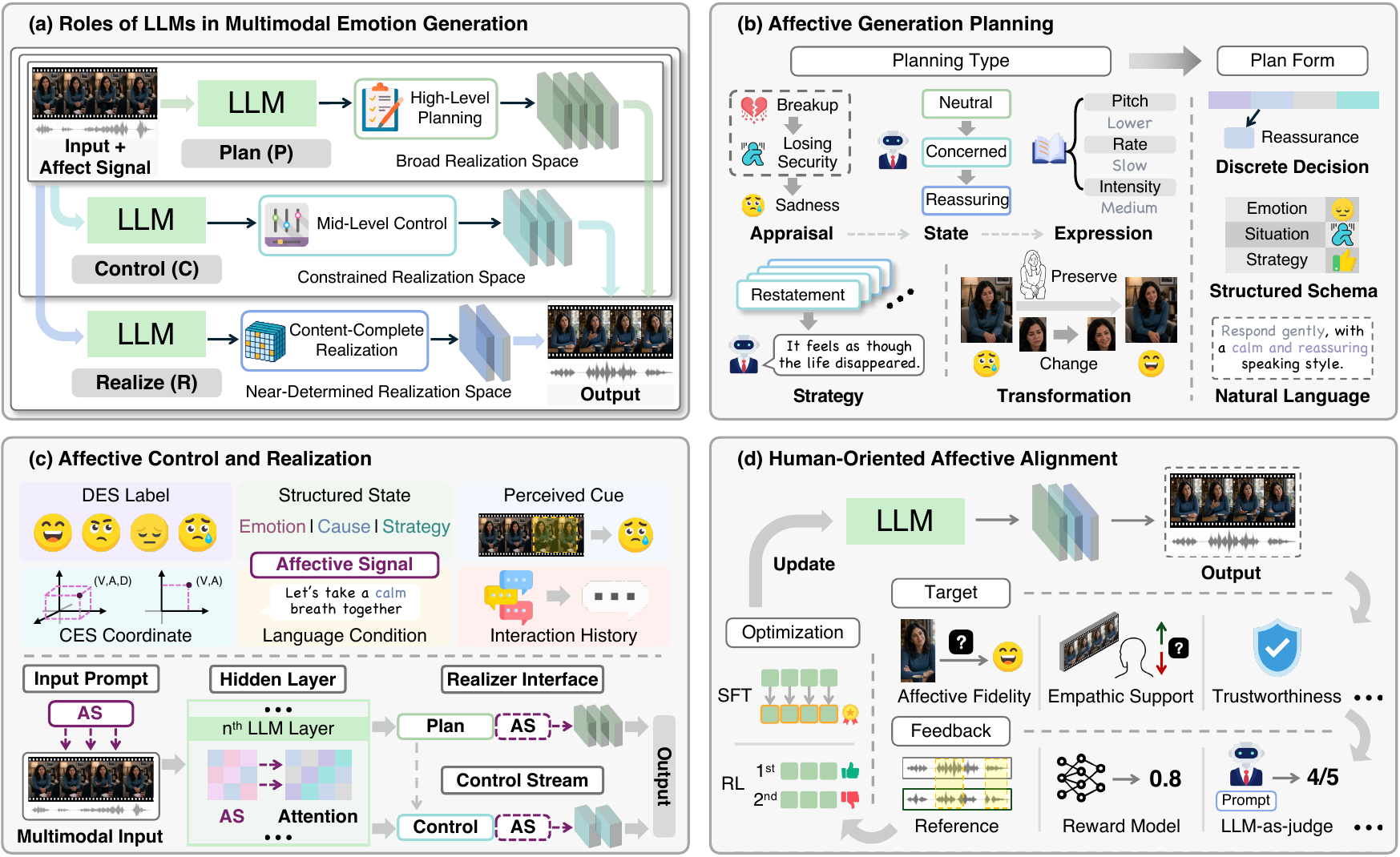}
    \vskip -0.1in
    \caption{Illustration of LLM roles (a) and three branches of MEG-with-LLMs. An LLM may produce a high-level plan (\textit{P}), a mid-level control sequence (\textit{C}), or a content-complete realization (\textit{R}). \textit{Affective Generation Planning} (b) specifies intended affective behavior through decisions, structured schemas, or natural-language plans. \textit{Affective Control and Realization} (c) incorporates affective signals (abbreviated as AS) at different model interfaces to guide output generation. \textit{Human-Oriented Affective Alignment} (d) uses feedback and optimization to improve affective fidelity, empathic support, and trustworthiness.}
    \vskip -0.1in
    \label{fig:6-1}
\end{figure*}

While understanding serves as a critical component of multimodal affective intelligence, generation determines how that intelligence is expressed and applied. Correct affective interpretation alone does not ensure an appropriate response: its realization may still involve unsuitable support strategies, prosody, or visual behavior. A growing body of work therefore harnesses the broad knowledge and representational capabilities of LLMs to generate behavior conditioned on intended affect, interaction context, and content requirements, forming the paradigm we term \textit{MEG-with-LLMs}.

Unlike recognition, whose outputs typically remain textual, generation extends to acoustic and visual forms with distinct requirements. MESC selects and expresses appropriate support; ESS realizes vocal expression while preserving linguistic content; and EVGE modifies visual appearance or motion while maintaining relevant content and identity. These outputs generally require additional trainable modules to translate LLM-generated representations into modality-specific signals. Accordingly, \cref{fig:6-1} (a) distinguishes three roles for LLMs: \textit{Plan}, specifying affective intentions or response strategies while leaving multiple realizations possible; \textit{Control}, supplying executor-facing constraints that narrow the realization space; and \textit{Realize}, producing complete responses or multimodal token sequences for downstream decoding. These roles can coexist within one system, connecting high-level decisions with increasingly specific representations.

This role-based view exposes the central \textit{Intention-to-Realization
Opacity} of MEG: an inappropriate intention, inadequate control, or imperfect rendering can each compromise the resulting behavior. Following our challenge-oriented perspective, we organize existing methods into \textit{Affective Generation Planning}, \textit{Affective Control and Realization}, and \textit{Human-Oriented Affective Alignment}. These complementary families respectively handle the organization of intended behavior, its translation into observable expression, and its alignment with human expectations. The following subsections examine their contributions closely.

\begin{table*}[t]
\centering
\caption{Affective Generation Planning methods with their key attributes. Input$^{*}$ indicates the primary content modalities, excluding affective instructions. P, C, and R indicate LLM participation in planning, control, and realization, respectively.}
\label{tab:6-1}
\vskip -0.1in
\begin{adjustbox}{max width=\textwidth}
\footnotesize
\renewcommand{\arraystretch}{1.1}
\setlength{\tabcolsep}{4pt}
\begin{tabular}{@{}lcccccccccc@{}}
\toprule
\textbf{Method} &
\textbf{Year} &
\textbf{Task} &
\textbf{Input$^{*}$} &
\textbf{Output} &
\textbf{P} &
\textbf{C} &
\textbf{R} &
\textbf{Planning Type} &
\textbf{Plan Form} &
\textbf{Key Design} \\
\midrule
% AG-CTR$^{2}$~\cite{chu26agctr2}{,} IEAT~\cite{wang26ieat}{,} Self-EmoQ~\cite{zhao26selfemoq}{,} E3RG~\cite{lin25e3rg}{,} Resonant Minds~\cite{shangg26resona}{,} AUHead~\cite{lyu26auhead}{,} PRISM~\cite{zhang26prism}{,} Dopamine Audiobook~\cite{rong25dopamine}{,} BatonVoice~\cite{wang26batonvoi}{,} TED-TTS~\cite{liang26tedtts}{,} OV-InstructTTS-TEP~\cite{ren26ovtts}{,} Kling-Avatar~\cite{ding25klingavatar}{,} CogPortrait~\cite{feng26cogportrait}{,} PerceptiveAgent~\cite{yan24percept}{,} OSUM-EChat~\cite{geng25osumecha}{,} LPE~\cite{xie25lpe}{,} Chain-Talker~\cite{hu25chaintalk}{,} Think-Before-Draw~\cite{shi25thinkdraw}{,} PEER~\cite{wang25peer}{,} MOCC-R1~\cite{zheng26moccr1}{,} SMES~\cite{chu25mesc}{,} GraphEMO~\cite{tu26graphemo}{,} FIRES~\cite{wang25fires}{,} OneEmo~\cite{huang26oneemo}{,} ReEmpathy~\cite{jia26reempathy}{,} Moodifier~\cite{ye25moodifier}{,} EPEM~\cite{luo26epemlavc}{.}
% Appraisal - White

AG-CTR$^{2}$~\cite{chu26agctr2} &
2026 & MESC & V,T & T &
\checkmark & - & \checkmark &
Appraisal & Structured &
Appraisal-Guided Reasoning and Retrieval \\

IEAT~\cite{wang26ieat} &
2026 & MESC, ESS, AER & A,T & A,T &
\checkmark & - & \checkmark &
Appraisal & Structured &
Emotional Attribution and Cause Reasoning \\

% State - Gray

\rowcolor{gray!8}
Self-EmoQ~\cite{zhao26selfemoq} &
2026 & ESS & T & A,T &
\checkmark & \checkmark & \checkmark &
State & Discrete &
Long-Horizon Self-Emotion Planning \\

\rowcolor{gray!8}
E3RG~\cite{lin25e3rg} &
2025 & MESC, ESS, EVGE & A,V,T & A,V,T &
\checkmark & \checkmark & \checkmark &
State & Structured &
Emotion- and Identity-Aware Responses \\

\rowcolor{gray!8}
Resonant Minds~\cite{shangg26resona} &
2026 & EVGE, MSU & A,V,T & A,V,T &
\checkmark & \checkmark & \checkmark &
State & Structured &
Theory-of-Mind-Guided Avatar States \\

% Expression - White

AUHead~\cite{lyu26auhead} &
2026 & EVGE & I,A & V &
\checkmark & \checkmark & - &
Expression & Discrete &
Audio-Aligned AU Planning \\

PRISM~\cite{zhang26prism} &
2026 & MESC, ESS & A & A,T &
\checkmark & \checkmark & \checkmark &
Expression & Structured &
Prosody-Aware Multi-Agent Planning \\

Dopamine Audiobook~\cite{rong25dopamine} &
2025 & ESS & I,T & A &
\checkmark & \checkmark & - &
Expression & Structured &
JSON Prosody and Audio-Event Plans \\

BatonVoice~\cite{wang26batonvoi} &
2026 & ESS & T & A &
\checkmark & \checkmark & \checkmark &
Expression & Structured &
JSON Vocal-Feature Planning \\

TED-TTS~\cite{liang26tedtts} &
2026 & ESS & T & A &
\checkmark & \checkmark & - &
Expression & Structured &
Segment-Level Emotion-Duration Plans \\

OV-InstructTTS-TEP~\cite{ren26ovtts} &
2026 & ESS & T & A,T &
\checkmark & \checkmark & \checkmark &
Expression & Structured &
Emotion and Acoustic Attribute Reasoning \\

Kling-Avatar~\cite{ding25klingavatar} &
2025 & EVGE & I,A & V &
\checkmark & \checkmark & - &
Expression & Structured &
Storyline-Guided Blueprint Generation \\

CogPortrait~\cite{feng26cogportrait} &
2026 & EVGE & I,A & V &
\checkmark & \checkmark & - &
Expression & Structured &
Hierarchical Agents for Keypoint Control \\

PerceptiveAgent~\cite{yan24percept} &
2024 & MESC, ESS & A & A,T &
\checkmark & \checkmark & \checkmark &
Expression & Textual &
Acoustic-to-Speaking-Style Planning \\

OSUM-EChat~\cite{geng25osumecha} &
2025 & MESC, ESS & A & A,T &
\checkmark & - & \checkmark &
Expression & Textual &
Dual Linguistic-Paralinguistic Thinking \\

LPE~\cite{xie25lpe} &
2025 & MESC, AER & A & T &
\checkmark & - & \checkmark &
Expression & Textual &
Listen-Perceive-Express Reasoning \\

Chain-Talker~\cite{hu25chaintalk} &
2025 & ESS & A,T & A &
\checkmark & \checkmark & \checkmark &
Expression & Textual &
Empathetic Captions to Speech Codes \\

Think-Before-Draw~\cite{shi25thinkdraw} &
2025 & EVGE & I,A & V &
\checkmark & \checkmark & - &
Expression & Textual &
Chain-of-Thought Facial-Muscle Planning \\

% Strategy - Gray

\rowcolor{gray!8}
PEER~\cite{wang25peer} &
2025 & MESC & I,T & T &
\checkmark & - & \checkmark &
Strategy & Structured &
Process-Outcome-Aligned Empathy \\

\rowcolor{gray!8}
MOCC-R1~\cite{zheng26moccr1} &
2026 & MESC & V,T & T &
\checkmark & - & \checkmark &
Strategy & Structured &
Grounded Plan Coherence and Execution \\

\rowcolor{gray!8}
SMES~\cite{chu25mesc} &
2025 & MESC, CMER & A,V,T & T &
\checkmark & - & \checkmark &
Strategy & Structured &
Emotion-Strategy-Response Reasoning \\

\rowcolor{gray!8}
GraphEMO~\cite{tu26graphemo} &
2026 & MESC, CMER & A,V,T & T &
\checkmark & - & \checkmark &
Strategy & Structured &
Hierarchical Emotion-Cause-Support Chain \\

\rowcolor{gray!8}
FIRES~\cite{wang25fires} &
2025 & MESC, CMER & V,T & T &
\checkmark & - & \checkmark &
Strategy & Structured &
Adaptive Aspect and Strategy Selection \\

\rowcolor{gray!8}
OneEmo~\cite{huang26oneemo} &
2026 & MESC, CMER, MSU & A,V,T & T &
\checkmark & - & \checkmark &
Strategy & Structured &
State-Cause-Goal-Strategy Reasoning \\

\rowcolor{gray!8}
ReEmpathy~\cite{jia26reempathy} &
2026 & MESC, ESS & A & A,T &
\checkmark & - & \checkmark &
Strategy & Textual &
Reflective Guidance for Spoken Responses \\

% Transformation - White

Moodifier~\cite{ye25moodifier} &
2025 & EVGE & I & I &
\checkmark & \checkmark & - &
Transformation & Textual &
Localized Emotion-Edit Specifications \\

EPEM~\cite{luo26epemlavc} &
2026 & EVGE & I & I &
\checkmark & \checkmark & - &
Transformation & Textual &
Emotion Conversion with Content Retention \\

\bottomrule
\end{tabular}
\end{adjustbox}
\vskip -0.1in
\end{table*}

\subsection{Affective Generation Planning} \label{sec:6_1}
Affective instructions are often abstract, leaving their concrete realization underspecified and contributing to \textit{Intention-to-Realization Opacity}. To bridge this gap, \textit{Affective Generation Planning} produces intermediate representations, typically expressed in language, that translate intended affect into actionable guidance for generation. \Cref{tab:6-1} groups representative methods into appraisal, state, expression, strategy, and transformation planning, with their principal forms illustrated in \cref{fig:6-1} (b). These approaches address complementary planning needs, from interpreting an emotional situation to specifying the behavior or transformation to be realized.

\textbf{Appraisal Planning.}
Appraisal planning begins with the personal significance of an emotional situation, relating observed events to the individual's concerns, expectations, and coping possibilities. This interpretation provides a conceptual basis for selecting an appropriate response. AG-CTR$^{2}$~\cite{chu26agctr2} turns appraisal-guided reasoning into retrieval queries, connecting the interpretation of the situation with relevant emotional-support knowledge. IEAT~\cite{wang26ieat} incorporates emotional attribution and cause reasoning into spoken interaction, grounding empathetic generation in an account of the experience underlying the perceived emotion.

\textbf{State Planning.}
The resulting response need not reproduce the user's emotion. State planning instead selects an affective state for the system to express, potentially considering its development over successive turns. Self-EmoQ~\cite{zhao26selfemoq} formulates self-emotion selection as a long-horizon decision, using discrete states to guide response generation and speech synthesis. The emphasis shifts from matching an isolated utterance to maintaining appropriate emotional continuity.

For multimodal responses, a planned state can provide a shared reference across output channels. E3RG~\cite{lin25e3rg} combines emotion and identity information to coordinate textual responses with audiovisual expression. Resonant Minds~\cite{shangg26resona} places this coordination within a reciprocal interaction, using Theory-of-Mind inference and conversational history to inform avatar responses. Here, affective state selection is embedded in an ongoing model of the interaction, extending beyond the choice of a vocal emotion.

\textbf{Expression Planning.}
A planned emotional state leaves considerable freedom in its observable expression. Expression planning translates that state into more specific choices about speaking style or visual behavior before rendering. PerceptiveAgent~\cite{yan24percept} derives speaking style from acoustic evidence, and LPE~\cite{xie25lpe} connects listening and affect perception with empathetic textual expression. OSUM-EChat~\cite{geng25osumecha} jointly considers response content and vocal delivery through linguistic and paralinguistic reasoning. PRISM~\cite{zhang26prism} coordinates these decisions through cooperating agents for perception, response planning, and vocal realization.

The resulting guidance ranges from descriptive intentions to detailed synthesis specifications. Chain-Talker~\cite{hu25chaintalk} uses empathetic captions to guide semantic speech-code generation, and OV-InstructTTS-TEP~\cite{ren26ovtts} develops the intended delivery through emotional and acoustic attribute reasoning. Structured plans make individual expressive properties more explicit, covering prosody and audio events in Dopamine Audiobook~\cite{rong25dopamine} and vocal features in BatonVoice~\cite{wang26batonvoi}. TED-TTS~\cite{liang26tedtts} adds temporal precision through segment-level schedules, supporting expressive variation within an utterance.

Visual generation requires an analogous translation into facial and bodily behavior. Audio-aligned AU sequences~\cite{lyu26auhead} and physiologically informed facial-muscle descriptions~\cite{shi25thinkdraw} give talking-head generators concrete guidance beyond an emotion label. Longer or more complex performances benefit from temporal organization. CogPortrait~\cite{feng26cogportrait} combines event planning with semantic-physiological constraints through hierarchical agents. Kling-Avatar~\cite{ding25klingavatar} operates at a broader directorial level, using an MLLM-generated storyline to guide blueprint-video construction and subsequent animation.

\textbf{Strategy Planning.}
In emotional support, generation aims not only to express affect but also to influence the user's emotional experience constructively. This calls for an appropriate support strategy, such as validation or reframing, grounded in the user's circumstances. SMES~\cite{chu25mesc} explicitly connects emotion interpretation, strategy selection, and response generation. GraphEMO~\cite{tu26graphemo} grounds this choice in hierarchical emotion-cause relations, and OneEmo~\cite{huang26oneemo} incorporates the response goal into state-cause-goal-strategy reasoning.

Such planning can also be adaptive. FIRES~\cite{wang25fires} selectively considers relevant aspects and support strategies according to the input. ReEmpathy~\cite{jia26reempathy} interleaves reflection with spoken responses, making planning an ongoing part of generation. Training can further align strategy selection with its execution. PEER~\cite{wang25peer} combines supervision of history analysis, state inference, and strategy selection with feedback on response quality. MOCC-R1~\cite{zheng26moccr1} explicitly rewards grounded plan coherence and execution, encouraging a contextually justified response intent and a reply that fulfills it.

\textbf{Transformation Planning.}
In affective editing, generation aims to reshape the emotional impression of existing content, requiring a plan that reconciles the intended change with source preservation. Moodifier~\cite{ye25moodifier} translates target affect into localized image-edit specifications. EPEM~\cite{luo26epemlavc} similarly combines emotion conversion with content retention, directing the transformation toward affect-relevant changes. Across these approaches, planning makes affective intent actionable by specifying the state to convey, the support strategy to adopt, or the content to transform, while different levels of specificity balance explicit guidance with flexibility in realization.

\begin{table*}[t]
\centering
\caption{Affective Control and Realization methods with their key attributes. DES and CES denote discrete and continuous emotion spaces, respectively. Affective Signal specifies the conditioning information, and Conditioning Site identifies its point of application in generation.}
\label{tab:6-2}
\vskip -0.1in
\begin{adjustbox}{max width=\textwidth}
\footnotesize
\renewcommand{\arraystretch}{1.1}
\setlength{\tabcolsep}{4pt}
\begin{tabular}{@{}lcccccccccc@{}}
\toprule
\textbf{Method} &
\textbf{Year} &
\textbf{Task} &
\textbf{Input$^{*}$} &
\textbf{Output} &
\textbf{P} &
\textbf{C} &
\textbf{R} &
\textbf{Affective Signal} &
\textbf{Conditioning Site} &
\textbf{Key Design} \\
\midrule
% Emo-DPO~\cite{gao25emodpo}{,} PUE~\cite{gao25pue}{,} PROEMO~\cite{zhang25proemo}{,} UDDETTS~\cite{liu25uddetts}{,} Sympatheia~\cite{dindar26sympat}{,} EmoTra-TTS~\cite{liu26emotra}{,} EmoStyle~\cite{hong26emostyle}{,} AG-CTR$^{2}$~\cite{chu26agctr2}{,} SMES~\cite{chu25mesc}{,} GraphEMO~\cite{tu26graphemo}{,} InstructDubber~\cite{zhang26instdub}{,} BatonVoice~\cite{wang26batonvoi}{,} AUHead~\cite{lyu26auhead}{,} Chain-Talker~\cite{hu25chaintalk}{,} VoxInstruct~\cite{zhou24voxinstruct}{,} Affectron~\cite{cho26affectron}{,} ZS-TTS-PT~\cite{gao26posttts}{,} EmoVoice~\cite{yang25emovoice}{,} Diary T2I~\cite{cho26emodiary}{,} EmoInstruct-TTS~\cite{wu26emotts}{,} Moodifier~\cite{ye25moodifier}{,} EPEM~\cite{luo26epemlavc}{,} Emotion-Director~\cite{jia25emotiondirector}{,} OV-InstructTTS-TEP~\cite{ren26ovtts}{,} Neuron Steering~\cite{zhao26esn}{,} SAE Steering~\cite{du26saetts}{,} EmoShift~\cite{zhou26emoshift}{,} Task-Vector Arithmetic~\cite{brito26taskvector}{,} Emo-FiLM~\cite{wang25emofilm}{,} DeSRPA~\cite{tang26desrpa}{,} EmoWorld~\cite{wang26emoworld}{,} BLSP-Emo~\cite{wang24blspemo}{,} Dual-Info SLM~\cite{wang25dualinfo}{,} IndexTTS2~\cite{zhou26indextts}{,} Gesture2Speech~\cite{kumar26gesture}{,} EmoAVLM~\cite{tan25emoavlm}{,} EATS-Speech~\cite{xing25eatsspee}{,} ES4R~\cite{gao26es4r}{,} OpenOmni~\cite{luo25openomni}{,} USDM~\cite{kim24usdm}{,} AuEmoChat~\cite{jia26auemochat}{.}
% DES Label -- White

Emo-DPO~\cite{gao25emodpo} &
2025 & ESS & A,T & A &
-- & -- & \checkmark &
DES Label & Input Prompt &
Preference-Tuned Emotional Speech \\

PUE~\cite{gao25pue} &
2025 & ESS & T & A &
-- & -- & \checkmark &
DES Label & Input Prompt &
Mixed and Unseen Emotion Control \\

PROEMO~\cite{zhang25proemo} &
2025 & ESS & T & A &
-- & \checkmark & \checkmark &
DES Label & Control Stream &
Word-Level Emotion and Intensity \\

% CES Coordinate -- Gray

\rowcolor{gray!8}
UDDETTS~\cite{liu25uddetts} &
2025 & ESS & A,T & A &
-- & -- & \checkmark &
CES Coordinate & Input Prompt &
Label and Dimensional Control \\

\rowcolor{gray!8}
Sympatheia~\cite{dindar26sympat} &
2026 & MESC, ESS & A & A &
-- & -- & \checkmark &
CES Coordinate & Hidden Layer &
Valence--Arousal Conditioning \\

\rowcolor{gray!8}
EmoTra-TTS~\cite{liu26emotra} &
2026 & ESS & A,T & A &
-- & -- & \checkmark &
CES Coordinate & Control Stream &
Temporal VAD and Frame-Level Cues \\

% Structured State -- White

EmoStyle~\cite{hong26emostyle} &
2026 & EVGE & T & I &
\checkmark & \checkmark & -- &
Structured State & Hidden Layer &
Affective Fields and Style Routing \\

AG-CTR$^{2}$~\cite{chu26agctr2} &
2026 & MESC & V,T & T &
\checkmark & -- & \checkmark &
Structured State & Realizer Interface &
Appraisal-Guided Support Responses \\

SMES~\cite{chu25mesc} &
2025 & MESC, CMER & A,V,T & T &
\checkmark & -- & \checkmark &
Structured State & Realizer Interface &
Strategy-Guided Support Responses \\

GraphEMO~\cite{tu26graphemo} &
2026 & MESC, CMER & A,V,T & T &
\checkmark & -- & \checkmark &
Structured State & Realizer Interface &
Graph-Conditioned Support Responses \\

InstructDubber~\cite{zhang26instdub} &
2026 & ESS & V,T & A &
-- & \checkmark & -- &
Structured State & Realizer Interface &
Instruction-Guided Dubbing \\

BatonVoice~\cite{wang26batonvoi} &
2026 & ESS & T & A &
\checkmark & \checkmark & \checkmark &
Structured State & Realizer Interface &
JSON Vocal-Feature Control \\

AUHead~\cite{lyu26auhead} &
2026 & EVGE & I,A & V &
\checkmark & \checkmark & -- &
Structured State & Realizer Interface &
AU-Guided Talking Heads \\

Chain-Talker~\cite{hu25chaintalk} &
2025 & ESS & A,T & A &
\checkmark & \checkmark & \checkmark &
Structured State & Control Stream &
Understanding--Rendering Streams \\

% Language Condition -- Gray

\rowcolor{gray!8}
VoxInstruct~\cite{zhou24voxinstruct} &
2024 & ESS & A,T & A &
-- & \checkmark & \checkmark &
Language Condition & Input Prompt &
Instruction-to-Speech Tokens \\

\rowcolor{gray!8}
Affectron~\cite{cho26affectron} &
2026 & ESS & A,T & A &
-- & -- & \checkmark &
Language Condition & Input Prompt &
Contextual Nonverbal Vocalizations \\

\rowcolor{gray!8}
ZS-TTS-PT~\cite{gao26posttts} &
2026 & ESS & A,T & A &
-- & -- & \checkmark &
Language Condition & Input Prompt &
Segment Emotion--Duration Control \\

\rowcolor{gray!8}
EmoVoice~\cite{yang25emovoice} &
2025 & ESS & T & A &
-- & -- & \checkmark &
Language Condition & Input Prompt &
Free-Form Emotional Prompting \\

\rowcolor{gray!8}
Diary T2I~\cite{cho26emodiary} &
2026 & EVGE & T & I &
-- & \checkmark & -- &
Language Condition & Input Prompt &
Sentiment-to-Prompt Translation \\

\rowcolor{gray!8}
EmoInstruct-TTS~\cite{wu26emotts} &
2026 & ESS & A,T & A &
-- & -- & \checkmark &
Language Condition & Realizer Interface &
Instruction-to-Emotion Embeddings \\

\rowcolor{gray!8}
Moodifier~\cite{ye25moodifier} &
2025 & EVGE & I & I &
\checkmark & \checkmark & -- &
Language Condition & Realizer Interface &
Prompt- and Mask-Guided Editing \\

\rowcolor{gray!8}
EPEM~\cite{luo26epemlavc} &
2026 & EVGE & I & I &
\checkmark & \checkmark & -- &
Language Condition & Realizer Interface &
Content-Preserving Emotion Editing \\

\rowcolor{gray!8}
Emotion-Director~\cite{jia25emotiondirector} &
2025 & EVGE & T & I &
\checkmark & \checkmark & -- &
Language Condition & Realizer Interface &
Prompt Rewriting and Visual Cues \\

\rowcolor{gray!8}
OV-InstructTTS-TEP~\cite{ren26ovtts} &
2026 & ESS & T & A,T &
\checkmark & \checkmark & \checkmark &
Language Condition & Control Stream &
Open-Vocabulary Speech Instructions \\

% Latent Steering -- White

Neuron Steering~\cite{zhao26esn} &
2026 & ESS & A & A &
-- & -- & \checkmark &
Latent Steering & Hidden Layer &
Emotion-Sensitive Neuron Steering \\

SAE Steering~\cite{du26saetts} &
2026 & ESS & A,T & A &
-- & -- & \checkmark &
Latent Steering & Hidden Layer &
Sparse-Feature Emotion Steering \\

EmoShift~\cite{zhou26emoshift} &
2026 & ESS & A,T & A &
-- & -- & \checkmark &
Latent Steering & Hidden Layer &
Learned Output-Embedding Steering \\

Task-Vector Arithmetic~\cite{brito26taskvector} &
2026 & ESS & A,T & A &
-- & -- & \checkmark &
Latent Steering & Hidden Layer &
Speaker-Vector Emotion Arithmetic \\

Emo-FiLM~\cite{wang25emofilm} &
2025 & ESS & T & A &
-- & -- & \checkmark &
Latent Steering & Hidden Layer &
Word-Level Emotion Modulation \\

DeSRPA~\cite{tang26desrpa} &
2026 & ESS & T & A,T &
-- & \checkmark & \checkmark &
Latent Steering & Hidden Layer &
Cognitive--Vocal Vector Steering \\

EmoWorld~\cite{wang26emoworld} &
2026 & EVGE & I,T & V &
-- & \checkmark & -- &
Latent Steering & Hidden Layer &
Latent and Language-Cue Steering \\

% Perceived Cue -- Gray

\rowcolor{gray!8}
BLSP-Emo~\cite{wang24blspemo} &
2024 & MESC, AER & A & T &
-- & -- & \checkmark &
Perceived Cue & Hidden Layer &
Emotion-Aware Empathetic Responses \\

\rowcolor{gray!8}
Dual-Info SLM~\cite{wang25dualinfo} &
2025 & MESC, AER & A & T &
-- & -- & \checkmark &
Perceived Cue & Hidden Layer &
Dual Speech-Information Adapters \\

\rowcolor{gray!8}
IndexTTS2~\cite{zhou26indextts} &
2026 & ESS & A,T & A &
-- & -- & \checkmark &
Perceived Cue & Hidden Layer &
Reference-Guided Emotion Control \\

\rowcolor{gray!8}
Gesture2Speech~\cite{kumar26gesture} &
2026 & ESS & A,V,T & A &
-- & -- & \checkmark &
Perceived Cue & Hidden Layer &
Gesture-Guided Prosody Fusion \\

\rowcolor{gray!8}
EmoAVLM~\cite{tan25emoavlm} &
2025 & ESS, CMER & A,V,T & A &
-- & -- & \checkmark &
Perceived Cue & Hidden Layer &
Audiovisual-Guided Speech Tokens \\

\rowcolor{gray!8}
EATS-Speech~\cite{xing25eatsspee} &
2025 & ESS & A,T & A &
-- & \checkmark & -- &
Perceived Cue & Realizer Interface &
Text-Adaptive Reference Emotion \\

% Interaction History -- White

ES4R~\cite{gao26es4r} &
2026 & MESC, ESS & A,T & A,T &
-- & -- & \checkmark &
Interaction History & Hidden Layer &
Turn--Dialogue Affect Fusion \\

OpenOmni~\cite{luo25openomni} &
2025 & ESS & I,A,T & A,T &
-- & -- & \checkmark &
Interaction History & Hidden Layer &
Low-Latency Omnimodal Speech \\

USDM~\cite{kim24usdm} &
2024 & ESS & A,T & A,T &
-- & -- & \checkmark &
Interaction History & Hidden Layer &
Paralinguistic Speech Dialogue \\

AuEmoChat~\cite{jia26auemochat} &
2026 & ESS & A,T & A &
-- & \checkmark & \checkmark &
Interaction History & Control Stream &
Emotion-Preserving History Tokens \\

\bottomrule
\end{tabular}
\end{adjustbox}
\vskip -0.1in
\end{table*}

\subsection{Affective Control and Realization}
\label{sec:6_2}

Even explicit instructions do not readily translate into accurate multimodal expression due to the \textit{Multimodal Affective Gap}. This challenge motivates \textit{Affective Control and Realization} methods that introduce conditioning signals at different architectural sites to constrain generation toward the intended purpose. \Cref{tab:6-2} summarizes representative methods by affective signal and conditioning site, as illustrated in \cref{fig:6-1} (c). The following discussion is organized by signal type.

\textbf{Discrete Emotion Labels.}
Discrete labels provide a compact interface for specifying emotional categories and supervising their realization. Emo-DPO~\cite{gao25emodpo} strengthens the correspondence between requested emotion and generated speech through same-content preference comparisons, isolating emotional delivery from linguistic content. PUE~\cite{gao25pue} expands the range of requests to mixed and unseen emotions. PROEMO~\cite{zhang25proemo} introduces finer temporal control by assigning emotion and intensity at the word level, allowing different parts of an utterance to receive distinct expressive emphasis.

\textbf{Continuous Emotion Coordinates.}
Continuous coordinates extend categorical control to graded affective variation. UDDETTS~\cite{liu25uddetts} connects the two representations within a unified synthesis framework, supporting both discrete labels and VAD specifications. This shared interface accommodates familiar emotion categories alongside dimensional adjustments to the generated expression. The coordinates can represent an interaction-level condition or a temporally evolving signal. Sympatheia~\cite{dindar26sympat} uses valence--arousal conditioning to guide an emotionally adaptive voice assistant, combining affect inferred from speech with explicit specifications when available. EmoTra-TTS~\cite{liu26emotra} develops the temporal form through affect tokens at the language-model stage and frame-level conditioning in the acoustic decoder. Coordinating these stages allows emotional transitions to shape both speech-token generation and acoustic rendering within an utterance.

\textbf{Structured States.}
More fine-grained generation requirements involve several related decisions, such as an emotional interpretation, a response strategy, and a desired delivery. Structured states retain these distinctions as explicit conditions. In emotional support, SMES~\cite{chu25mesc} carries emotion and strategy information into the reply, AG-CTR$^{2}$~\cite{chu26agctr2} provides appraisal-grounded context, and GraphEMO~\cite{tu26graphemo} incorporates organized emotion--cause relations. Their structures connect the preceding interpretation with the semantic content of the response.

For audiovisual realization, the same principle yields conditions closer to execution. JSON vocal-feature specifications guide synthesis in BatonVoice~\cite{wang26batonvoi}, and duration and emotion controls connect visual context with expressive dubbing in InstructDubber~\cite{zhang26instdub}. AUHead~\cite{lyu26auhead} supplies AU sequences for talking-head generation, translating affective intent into coordinated facial movement.

These structures can enter the generator through different interfaces. EmoStyle~\cite{hong26emostyle} encodes LLM-predicted affective fields for hidden-layer modulation and combines them with style-specialist routing. Chain-Talker~\cite{hu25chaintalk} connects an empathetic description with semantic speech codes within an autoregressive stream. Structured semantics thus guide either intermediate feature processing or the progression from interpretation to rendered output.

\textbf{Language Conditions.}
Natural-language conditions accommodate expressive requests beyond a predefined set of fields. EmoVoice~\cite{yang25emovoice} supports free-form emotional descriptions, and VoxInstruct~\cite{zhou24voxinstruct} combines content and delivery requirements within a unified instruction-to-speech interface. Optional reference speech supplies additional voice information. Affectron~\cite{cho26affectron} extends the expressive range to contextually appropriate nonverbal vocalizations, incorporating affective sounds alongside spoken language.

Connecting these descriptions with acoustic behavior motivates several complementary designs. OV-InstructTTS-TEP~\cite{ren26ovtts} develops emotional and acoustic attributes through intermediate reasoning before speech generation. EmoInstruct-TTS~\cite{wu26emotts} instead constructs an explicit representational bridge, using an instruction-conditioned flow model to derive semantic--acoustic emotion embeddings for synthesis. ZS-TTS-PT~\cite{gao26posttts} strengthens adherence within the pretrained generator through post-training, targeting segment-specific emotion and duration requirements without introducing additional inference-time control modules.

In visual generation, language translates emotional meaning into scene attributes. Diary T2I~\cite{cho26emodiary} converts sentiment inferred from diary entries into prompts containing trigger words. Emotion-Director~\cite{jia25emotiondirector} enriches prompt rewriting with retrieved visual cues, supported by preference optimization that distinguishes emotional effects under similar scene semantics. The language interface thereby connects an abstract emotional request with concrete visual content and presentation. Language conditions can further guide affective changes to existing images. Moodifier~\cite{ye25moodifier} combines emotion-specific prompts with spatial masks to localize the transformation. EPEM~\cite{luo26epemlavc} couples an edited semantic description with the original image, coordinating conversion with retention.

\textbf{Latent Steering.}
Subtle affective qualities can be difficult to articulate through labels or language. Latent steering offers control within the model's own representational space, making such variation accessible through feature modulation. Emo-FiLM~\cite{wang25emofilm} modulates text embeddings with word-level emotion features, and EmoShift~\cite{zhou26emoshift} learns compact emotion-specific offsets with adjustable strength in the output embedding space. Control directions can also be extracted from pretrained representations. SAE-based steering~\cite{du26saetts} identifies sparse features for emotion induction and suppression, and neuron-level intervention~\cite{zhao26esn} selects emotion-sensitive units from successful, content-preserving generations. At the speaker-embedding interface, task-vector arithmetic~\cite{brito26taskvector} transfers neutral-to-emotion directions estimated from reference examples to unseen speakers.

Coordinated steering extends this control across different aspects of expression. DeSRPA~\cite{tang26desrpa} separates cognitive intervention in a frozen LLM from acoustic control, connecting character-consistent language with vocal emotion. EmoWorld~\cite{wang26emoworld} combines visual-atmosphere steering, semantic cues, and temporal modulation in a frozen video generator, with an MLLM-derived cue library linking feature-space interventions to scene-level affective changes.

\textbf{Perceived Cues.}
Beyond deliberately specified controls, affective guidance can emerge from the speech and behavior observed during interaction. BLSP-Emo~\cite{wang24blspemo} carries emotion-relevant speech information into language-based processing for empathetic textual responses. Dual-Info SLM~\cite{wang25dualinfo} separates linguistic and paralinguistic information through dedicated adapters, preserving their complementary contributions to response generation.

The same perceptual evidence can shape vocal delivery. EmoAVLM~\cite{tan25emoavlm} incorporates audiovisual cues into speech-token generation, and Gesture2Speech~\cite{kumar26gesture} connects bodily motion with prosody through multimodal feature fusion. These designs use observed behavior to guide expression across modalities. Reference-based synthesis provides a complementary form of transfer to new linguistic content. IndexTTS2~\cite{zhou26indextts} supports reference-conditioned emotional speech alongside duration control. EATS-Speech~\cite{xing25eatsspee} further adapts reference expression patterns to the target text through an LLM-based emotion converter, producing content-sensitive emotion tokens for synthesis.

\textbf{Interaction History.}
Accumulating these observations across turns provides a temporal context for affective realization. USDM~\cite{kim24usdm} incorporates paralinguistic information into speech dialogue, and OpenOmni~\cite{luo25openomni} extends contextual response generation to low-latency omnimodal interaction.

History can be organized at different scales or selectively compressed before generation. ES4R~\cite{gao26es4r} combines turn-level and dialogue-level affective context to guide empathetic responses and speech. AuEmoChat~\cite{jia26auemochat} retains emotion-relevant history through token merging and predicts a target token in a learned authentic-emotion space. The resulting token conditions speech rendering, carrying accumulated conversational evidence into the current expression.

Collectively, these methods connect affective intent and observed context with generation through complementary conditioning interfaces. Coupled with \textit{Affective Generation Planning}, they progressively translate abstract intentions into executable specifications and ultimately observable expression.

\begin{table*}[t]
\centering
\caption{Human-Oriented Affective Alignment methods with their key attributes. RL broadly covers reward- and preference-based optimization, including DPO~\cite{rafailov23dpo}, listwise ~\cite{liu25lipo} and anchored preference optimization~\cite{doosterlinck25apo}. TTA denotes Test-Time Adaptation, used here to include parameter-free reflection, prompt refinement, and candidate selection.}
\label{tab:6-3}
\vskip -0.1in
\begin{adjustbox}{max width=\textwidth}
\footnotesize
\renewcommand{\arraystretch}{1.1}
\setlength{\tabcolsep}{2pt}
\begin{tabular}{@{}lccccccccccc@{}}
\toprule
\textbf{Method} &
\textbf{Year} &
\textbf{Task} &
\textbf{Input$^{*}$} &
\textbf{Output} &
\textbf{P} &
\textbf{C} &
\textbf{R} &
\textbf{Alignment Target} &
\textbf{Alignment Feedback} &
\textbf{Optimization} &
\textbf{Key Design} \\
\midrule
% EPEM~\cite{luo26epemlavc}{,} Emo-DPO~\cite{gao25emodpo}{,} HybridEmo~\cite{zhou26hybemo}{,} BatonVoice~\cite{wang26batonvoi}{,} Emo-LiPO~\cite{lin26emolipo}{,} HPRO~\cite{nie26hpro}{,} DiffRO~\cite{gao25diffro}{,} RRPO~\cite{wang26rrpo}{,} ALPO~\cite{lin26alpo}{,} SA-SLM~\cite{wang26saslm}{,} RLAIF-SPA~\cite{yang25rlaifspa}{,} EmoFeedback$^{2}$~\cite{shu25emofeedback}{,} LPJ~\cite{guan26dlpj}{,} OSUM-EChat~\cite{geng25osumecha}{,} BLSP-Emo~\cite{wang24blspemo}{,} RE-LLM~\cite{chen26rellm}{,} PEER~\cite{wang25peer}{,} FIRES~\cite{wang25fires}{,} MOCC-R1~\cite{zheng26moccr1}{,} OneEmo~\cite{huang26oneemo}{,} ReEmpathy~\cite{jia26reempathy}{,} Affectron~\cite{cho26affectron}{,} Self-EmoQ~\cite{zhao26selfemoq}{,} MultiMood~\cite{le26multimood}{.}
% Affective Fidelity -- white

EPEM~\cite{luo26epemlavc} &
2026 & EVGE & I & I &
\checkmark & \checkmark & -- &
Affective Fidelity &
Reference &
SFT &
Emotion Editing with Content Retention \\

Emo-DPO~\cite{gao25emodpo} &
2025 & ESS & A,T & A &
-- & -- & \checkmark &
Affective Fidelity &
Reference &
SFT + RL &
Same-Content Emotion Preferences \\

HybridEmo~\cite{zhou26hybemo} &
2026 & ESS & A,T & A &
-- & -- & \checkmark &
Affective Fidelity &
Rule-Based Scoring &
SFT + RL &
Trajectory- and Blend-Aware Rewards \\

BatonVoice~\cite{wang26batonvoi} &
2026 & ESS & T & A &
\checkmark & \checkmark & \checkmark &
Affective Fidelity &
Rule-Based Scoring &
SFT + RL &
Acoustic-Feature Preference Tuning \\

Emo-LiPO~\cite{lin26emolipo} &
2026 & ESS & T & A &
-- & -- & \checkmark &
Affective Fidelity &
Rule-Based Scoring &
SFT + RL &
Listwise Emotion-Intensity Ranking \\

HPRO~\cite{nie26hpro} &
2026 & ESS & A,T & A &
-- & -- & \checkmark &
Affective Fidelity &
Reward Model &
RL &
Hierarchical Content--Style Rewards \\

DiffRO~\cite{gao25diffro} &
2025 & ESS & T & A &
-- & -- & \checkmark &
Affective Fidelity &
Reward Model &
RL &
Differentiable Codec-Level Rewards \\

RRPO~\cite{wang26rrpo} &
2026 & ESS & T & A &
-- & -- & \checkmark &
Affective Fidelity &
Reward Model &
RL &
Robust Rewards against Hacking \\

ALPO~\cite{lin26alpo} &
2026 & MESC, ESS & A & A,T &
-- & -- & \checkmark &
Affective Fidelity &
LLM-as-Judge &
SFT + RL &
Decoupled Text--Speech Advantages \\

SA-SLM~\cite{wang26saslm} &
2026 & ESS & A & A,T &
-- & -- & \checkmark &
Affective Fidelity &
LLM-as-Judge &
SFT + RL &
Self-Rewarded Intent--Speech Matching \\

RLAIF-SPA~\cite{yang25rlaifspa} &
2025 & ESS & T & A &
-- & -- & \checkmark &
Affective Fidelity &
LLM-as-Judge &
RL &
Prosodic and Semantic Reward Checks \\

EmoFeedback$^{2}$~\cite{shu25emofeedback} &
2025 & EVGE & T & I &
\checkmark & \checkmark & -- &
Affective Fidelity &
LLM-as-Judge &
RL + TTA &
Visual Rewards and Prompt Revision \\

LPJ~\cite{guan26dlpj} &
2026 & ESS & T & A &
-- & -- & -- &
Affective Fidelity &
LLM-as-Judge &
TTA &
Prosody-Based Candidate Selection \\

% Empathic Support -- gray

\rowcolor{gray!8}
OSUM-EChat~\cite{geng25osumecha} &
2025 & MESC, ESS & A & A,T &
\checkmark & -- & \checkmark &
Empathic Support &
Reference &
SFT &
Dual-Thinking Empathy Supervision \\

\rowcolor{gray!8}
BLSP-Emo~\cite{wang24blspemo} &
2024 & MESC, AER & A & T &
-- & -- & \checkmark &
Empathic Support &
Reference &
SFT &
Emotion-Aware Continuation Tuning \\

\rowcolor{gray!8}
RE-LLM~\cite{chen26rellm} &
2026 & MESC, AER & A & T &
-- & -- & \checkmark &
Empathic Support &
Reference &
SFT &
Emotion-Nuance Response Alignment \\

\rowcolor{gray!8}
PEER~\cite{wang25peer} &
2025 & MESC & I,T & T &
\checkmark & -- & \checkmark &
Empathic Support &
Reward Model &
SFT + RL &
Unified Process--Outcome Rewards \\

\rowcolor{gray!8}
FIRES~\cite{wang25fires} &
2025 & MESC, CMER & V,T & T &
\checkmark & -- & \checkmark &
Empathic Support &
Reward Model &
SFT + RL &
Support-Skill and Reference Rewards \\

\rowcolor{gray!8}
MOCC-R1~\cite{zheng26moccr1} &
2026 & MESC & V,T & T &
\checkmark & -- & \checkmark &
Empathic Support &
LLM-as-Judge &
SFT + RL &
Grounded Plan--Response Consistency \\

\rowcolor{gray!8}
OneEmo~\cite{huang26oneemo} &
2026 & MESC, CMER, MSU & A,V,T & T &
\checkmark & -- & \checkmark &
Empathic Support &
LLM-as-Judge &
SFT + RL &
Psychology-Grounded Response Rewards \\

\rowcolor{gray!8}
ReEmpathy~\cite{jia26reempathy} &
2026 & MESC, ESS & A & A,T &
\checkmark & -- & \checkmark &
Empathic Support &
LLM-as-Judge &
SFT + TTA &
Interleaved Empathy Critique \\

% Contextual Appropriateness -- white

Affectron~\cite{cho26affectron} &
2026 & ESS & A,T & A &
-- & -- & \checkmark &
Contextual Appropriateness &
Reference &
SFT &
Context-Aligned Vocal Events \\

Self-EmoQ~\cite{zhao26selfemoq} &
2026 & ESS & T & A,T &
\checkmark & \checkmark & \checkmark &
Contextual Appropriateness &
LLM-as-Judge &
RL &
Plutchik-Guided Emotion Planning \\

% Trustworthiness -- gray

\rowcolor{gray!8}
MultiMood~\cite{le26multimood} &
2026 & MESC, CMER & A,V,T & T &
\checkmark & -- & \checkmark &
Trustworthiness &
LLM-as-Judge &
SFT + RL &
Trust and Semantic-Similarity Rewards \\

\bottomrule
\end{tabular}
\end{adjustbox}
\vskip -0.1in
\end{table*}

\subsection{Human-Oriented Affective Alignment} \label{sec:6_3}

Due to \textit{Subjective and Contextual Variability}, misalignment may still exist between faithful execution and human expectations. This motivates \textit{Human-Oriented Affective Alignment}, which introduces explicit alignment objectives informed by reference examples, reward signals, or evaluative feedback. These objectives guide supervised learning, reward and preference optimization, and inference-time refinement. \Cref{tab:6-3} groups representative methods accordingly, as also illustrated in \cref{fig:6-1} (d).

\textbf{Affective Fidelity.}
An intuitive objective is affective fidelity, typically assessed through agreement between intended emotion and generated expression, alongside content and identity preservation. Reference-based supervision establishes this correspondence through converted and original versions in EPEM~\cite{luo26epemlavc}. Emo-DPO~\cite{gao25emodpo} uses same-content speech preferences to isolate emotional delivery from linguistic variation, directing optimization toward the requested expression.

Beyond reference-based supervision, rule-based scoring makes emotional progression, delivery, and intensity explicit alignment criteria. HybridEmo~\cite{zhou26hybemo} evaluates emotional trajectories and blends, capturing variation within an utterance. BatonVoice~\cite{wang26batonvoi} uses acoustic-feature preferences to align synthesized speech with vocal specifications, and Emo-LiPO~\cite{lin26emolipo} applies listwise preference optimization to preserve relative emotion-intensity ordering.

Complementing these predefined criteria, learned reward models provide more customized assessments of fidelity and preservation. DiffRO~\cite{gao25diffro} backpropagates differentiable codec-level rewards through speech generation. HPRO~\cite{nie26hpro} structures this feedback across multiple scales, separating content from style and progressively aligning frame-, word-, and sentence-level objectives. RRPO~\cite{wang26rrpo} addresses reward hacking by discouraging score increases that do not correspond to genuine improvements in speech quality.

LLM-based evaluators extend this feedback to combined semantic and expressive judgments, supporting both training and inference-time refinement. RLAIF-SPA~\cite{yang25rlaifspa} evaluates prosody alongside semantic fidelity. For joint generation, ALPO~\cite{lin26alpo} separates textual and acoustic advantages and assigns each to its corresponding token stream, providing targeted feedback for content and delivery. SA-SLM~\cite{wang26saslm} derives expressive intent from semantic generation states and aligns the resulting speech with that intent through realization-aware self-reward. At inference time, EmoFeedback$^{2}$~\cite{shu25emofeedback} supplements visual reward optimization with textual suggestions for prompt revision, allowing each generated image to inform the next attempt. LPJ~\cite{guan26dlpj} uses prosodic judgments to select among speech candidates without updating the generator.

\textbf{Empathic Support.}
Beyond conveying an intended emotion, supportive responses need to address the user's circumstances and emotional needs. Reference-supervised methods establish this connection by incorporating affective information into response learning. BLSP-Emo~\cite{wang24blspemo} preserves emotional speech cues during language-based continuation, and RE-LLM~\cite{chen26rellm} reinforces their influence through dimensional emotion representations and auxiliary supervision. OSUM-EChat~\cite{geng25osumecha} extends this approach to spoken responses, supervising linguistic and paralinguistic reasoning alongside empathetic expression.

\begin{figure*}[t]
    \centering
    \includegraphics[width=1\linewidth]{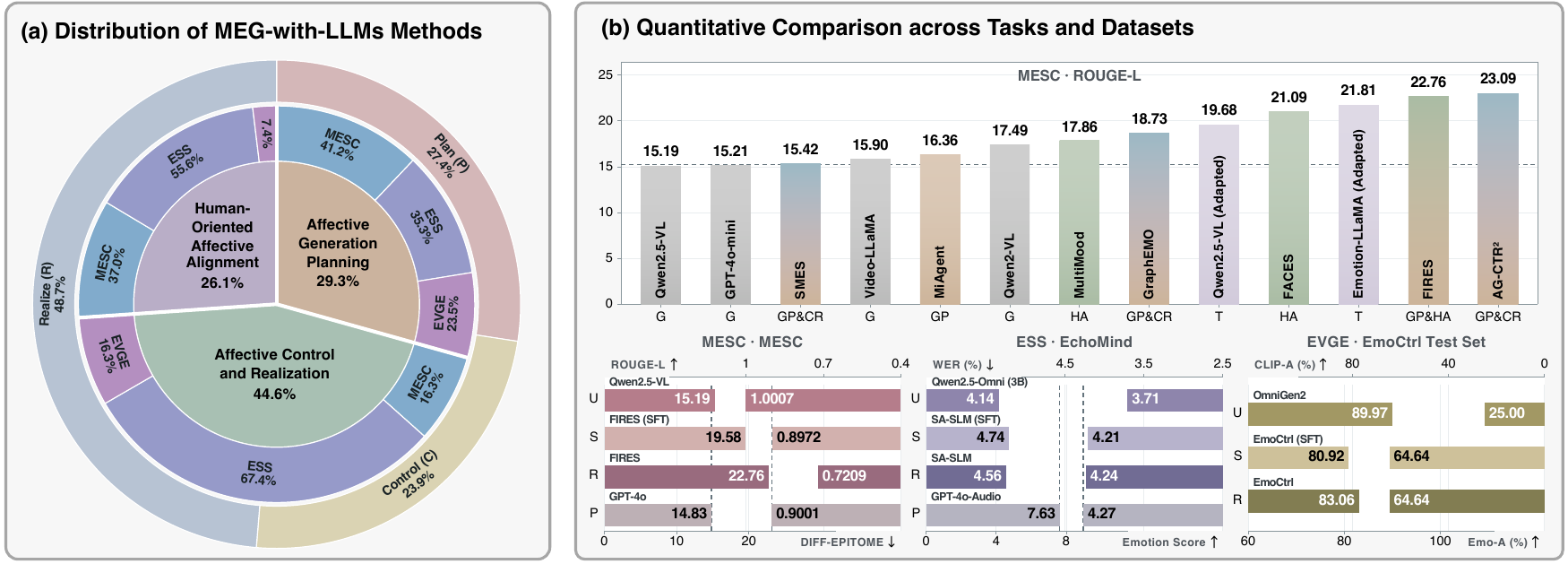}
    \vskip -0.1in
    \caption{Overview of the distribution and performance of MEG-with-LLMs methods. The upper-right panel compares general models (\textit{G}), methods for \textit{Affective Generation Planning} (\textit{GP}), \textit{Affective Control and Realization} (\textit{CR}), and \textit{Human-Oriented Affective Alignment} (\textit{HA}), and task-adapted models (\textit{T}) on MESC. The lower-right panels compare representative untuned open-source MLLMs (\textit{U}), SFT-optimized MEG-with-LLMs methods (\textit{S}), RL-optimized MEG-with-LLMs methods (\textit{R}), and proprietary MLLMs (\textit{P}) across MESC, ESS, and EVGE, pairing content-related and affective metrics for each task. Quantitative results are compiled from the reviewed literature with established comparability across evaluation protocols~\cite{qwen25vl,gpt4omini,chu25mesc,videollama,miagent,qwen2vl,le26multimood,tu26graphemo,faces,cheng24emotion,wang25fires,chu26agctr2,qwen25omni,wang26saslm,gpt4o,omnigen2,emoctrl}. Unless specified, for general-purpose open-source models available at multiple scales, we report the 7--8B variants.}
    \vskip -0.1in
    \label{fig:6-2}
\end{figure*}

Reward-based approaches make support quality an explicit objective, assessing both response decisions and their execution. FIRES~\cite{wang25fires} combines support-skill and reference-based rewards to guide adaptive reasoning and strategy selection. PEER~\cite{wang25peer} links history analysis, state inference, and strategy selection to response quality through unified process--outcome feedback. MOCC-R1~\cite{zheng26moccr1} places particular emphasis on grounded plan coherence and execution, rewarding response intentions supported by the inferred client state and replies that fulfill those intentions. OneEmo~\cite{huang26oneemo} incorporates psychology-grounded response rewards within its broader perception, understanding, and interaction framework.

Complementing scalar rewards, descriptive feedback provides finer-grained revision guidance. ReEmpathy~\cite{jia26reempathy} interleaves natural-language empathy critiques with spoken responses, allowing reflective guidance to inform subsequent content. This offers actionable feedback alongside the generation process as supplements to training-time supervision.

\textbf{Contextual Appropriateness.}
Contextual appropriateness broadens the focus from supportive content to the suitability of an emotional stance and its expression within an interaction. At the level of local vocal behavior, Affectron~\cite{cho26affectron} learns contextually aligned nonverbal vocalizations through reference supervision. Training these events together with the surrounding utterance makes their communicative placement part of the generation objective. At the emotional stance level, Self-EmoQ~\cite{zhao26selfemoq} uses Plutchik-guided feedback to support selection of the system's multiple emotional states before streaming speech. Alignment thus extends to the choice of emotion itself, complementing the fidelity objectives that assess expression against an already specified target.

\textbf{Trustworthiness.}
Appropriate and empathetic expression gains practical value when accompanied by dependable, contextually grounded responses. MultiMood~\cite{le26multimood} makes trustworthiness an explicit target, combining supervised initialization with trust-oriented model judgments and semantic-similarity feedback. Its optimization couples assessments of trust with response relevance, extending the alignment objective beyond perceived warmth or expressiveness. 

Collectively, these methods bring expressive fidelity, interpersonal sensitivity, and contextual suitability into the alignment objective, using supervision and evaluative feedback to better align generated behavior with human perceptions and expectations.

\begin{table*}[t]
\centering
\caption{Qualitative comparison of MEG-with-LLMs method categories. Targets and Process denote supervision on desired outcomes and intermediate planning or realization steps, respectively. Ratings summarize typical design tendencies, with transferability referring to the reuse of methodological mechanisms across settings.}
\label{tab:sec6dis}
\vskip -0.1in
\begin{adjustbox}{max width=\textwidth}
\scriptsize
\renewcommand{\arraystretch}{1.30}
\setlength{\tabcolsep}{4pt}
\begin{tabular}{@{}l*{8}{>{\centering\arraybackslash}m{0.64in}}}
\toprule
\multirow{2}{*}{\textbf{Method}} & \multicolumn{2}{c}{\textbf{Supervision Dependency}} & \multicolumn{2}{c}{\textbf{Computational Cost}} & \multirow{2}{*}{\textbf{Specificity}} & \multirow{2}{*}{\textbf{Flexibility}} & \multirow{2}{*}{\textbf{Inspectability}} & \multirow{2}{*}{\textbf{Transferability}} \\
\cmidrule(lr){2-3}\cmidrule(lr){4-5}
& \textbf{Targets} & \textbf{Process} & \textbf{Training} & \textbf{Inference} & & & & \\
\midrule
(\S\ref{sec:6_1}) Affective Generation Planning & \merCostLow & \merCostMid & \merCostMid & \merCostHigh & \merCapMid & \merCapHigh & \merCapHigh & \merCapHigh \\
(\S\ref{sec:6_2}) Affective Control and Realization & \merCostMid & \merCostHigh & \merCostHigh & \merCostMid & \merCapHigh & \merCapLow & \merCapMid & \merCapMid \\
(\S\ref{sec:6_3}) Human-Oriented Affective Alignment & \merCostMid & \merCostLow & \merCostHigh & \merCostMid & \merCapLow & \merCapHigh & \merCapMid & \merCapHigh \\
\bottomrule
\end{tabular}
\end{adjustbox}
\vskip -0.1in
\end{table*}

\subsection{Discussion}

\subsubsection{\textbf{Quantitative Analysis}}
Planning, control, and alignment together extend MEG-with-LLMs from specifying affective intentions to realizing and evaluating their expression. \Cref{fig:6-2} (a) summarizes this landscape, showing the greatest concentration of work in control and realization, followed by planning and alignment. Although all three branches cover generation tasks, planning is more evenly distributed across them. Across them, LLMs take on overlapping roles, with realization remaining their most common contribution.

The upper-right panel of \cref{fig:6-2} (b) connects this methodological breadth with performance on MESC. Dedicated MEG-with-LLMs methods generally outperform general-purpose models, indicating the value of explicitly modeling the affective demands of supportive interaction. Task-specific adaptation provides another effective route, with adapted Emotion-LLaMA outperforming the adapted general-purpose Qwen2.5-VL. This comparison suggests that affective understanding can provide a useful foundation for generation, linking advances in MER-with-LLMs with the production of supportive responses. Among dedicated methods, strong results arise from different methodological combinations, including human-oriented alignment in FACES, planning with alignment in FIRES, and planning with control and realization in AG-CTR$^{2}$. Their performance points to complementary routes toward effective affective generation.

The lower-right panels examine a further distinction between recognition and generation: an output needs to satisfy content requirements as well as convey appropriate affect. Each task therefore pairs a content-related metric on the left with an affective metric on the right. Unlike the broadly improving recognition scores discussed earlier, gains in generation do not necessarily occur on both dimensions simultaneously. For example, on EchoMind, SA-SLM shows a stronger emotion score but a higher Word Error Rate (WER) than Qwen2.5-Omni. EmoCtrl exhibits a similar trade-off, substantially improving emotion alignment over OmniGen2 at the cost of a lower CLIP \cite{radfor21clip} alignment score. Post-training can refine this balance: relative to its SFT variant, SA-SLM improves both WER and emotional quality, and EmoCtrl recovers some semantic alignment without sacrificing its emotion score. These comparisons involve different training conditions, but illustrate a recurring distinction: progress in affective generation concerns the balance between faithful content and effective emotional expression.

\subsubsection{\textbf{Qualitative Analysis}}
\Cref{tab:sec6dis} further summarizes the supervision requirements, computational costs, and functional characteristics of the three MEG-with-LLMs categories. As in \cref{sec:5}, the assessments reflect a heuristic synthesis of representative methods and their typical design trade-offs.

\textbf{Supervision Dependency.}
Supervision dependency describes the extent to which methods require annotations of desired outcomes and intermediate generation processes. \textit{Affective Generation Planning} places greater emphasis on the latter, with supervision ranging from appraisal and strategy demonstrations to detailed expression plans. Its requirements vary with the planning type and the extent to which pretrained capabilities can supply these intermediate decisions, motivating an overall moderate assessment of process supervision. \textit{Affective Control and Realization} often requires more detailed guidance to connect affective conditions with observable expression. Specialized control sequences and modality-specific interfaces may fall outside the backbone's pretraining, creating a need for additional attribute annotations, temporal correspondences, or control--output supervision. \textit{Human-Oriented Affective Alignment} instead emphasizes the qualities of the resulting behavior. Diverse, task-relevant human judgments or feedback approximating those judgments can specify desirable outcomes without prescribing the complete generation process, leading to lower dependence on process supervision.

\textbf{Computational Cost.}
Consistent with the earlier analysis, computational cost refers to the additional resources required for training and inference beyond pretrained components. Generating and optimizing multimodal outputs can make this overhead substantial, particularly when learning new realization interfaces or evaluating candidate outputs during alignment. \textit{Affective Generation Planning} can often build on existing language capabilities with relatively moderate training costs, while \textit{Affective Control and Realization} requires optimization of specialized mappings or generation modules. \textit{Human-Oriented Affective Alignment} can incur further costs through preference construction, reward evaluation, and repeated generation during training. At inference time, explicit planning adds serial computation before realization, especially when plans involve extended autoregressive reasoning or multiple planning stages. Control mechanisms are often integrated into the generation pipeline, and training-time alignment can be retained without repeating the optimization process. Their typical inference overhead is therefore more moderate, with additional costs arising when feedback or candidate selection is invoked during generation.

\textbf{Specificity.}
Specificity describes how precisely a method constrains the expression to be generated. The role distinction in \cref{fig:6-1} (a) illustrates how increasingly concrete specifications narrow the space of possible realizations. \textit{Affective Generation Planning} typically establishes an intended state, strategy, or expressive direction while leaving multiple realizations available. \textit{Affective Control and Realization} imposes more detailed constraints through attributes, temporal schedules, or modality-specific parameters, giving it higher specificity. \textit{Human-Oriented Affective Alignment} primarily defines what makes an output desirable, allowing different realizations to satisfy the same objective and consequently imposing fewer direct constraints on their form.

\textbf{Flexibility, Inspectability, and Transferability.}
Flexibility, inspectability, and transferability are also important properties that shape how readily a method can be used, examined, and adapted. Flexibility concerns the range of expressive requirements it can accommodate, inspectability describes how clearly its intermediate decisions and guidance can be examined, and transferability reflects how readily its mechanisms can be reused across settings. \textit{Affective Generation Planning} generally performs well across these dimensions because plans are often expressed in language or readable structured fields, making them adaptable, accessible for inspection, and relatively independent of a particular renderer. \textit{Affective Control and Realization} gains precision through more specialized interfaces, whose predefined attributes and implementation-specific representations can restrict expressive flexibility and reuse. Its inspectability varies between explicit controls and less transparent latent interventions. \textit{Human-Oriented Affective Alignment} leaves room for diverse outputs and offers objectives that can be reused across tasks or modalities, supporting flexibility and methodological transferability. Its feedback is only partially inspectable, however, as explicit rules and textual critiques coexist with learned reward functions whose judgments are less transparent.

Despite their distinct strengths and limitations, these categories are best understood as complementary functions that can work together within a single system. BatonVoice~\cite{wang26batonvoi}, for example, combines explicit vocal planning, synthesis-facing controls, and preference-based alignment, while PEER~\cite{wang25peer} connects strategy planning with supervision of both intermediate processes and response quality. Such integration allows plans to guide realization and feedback to refine their execution. Further work could build on these connections by jointly designing planning, control, and alignment to make better use of their complementary strengths.

\section{Perspectives and Outlook}
\label{sec:7}

The preceding sections reviewed the data resources and methodological developments. We now bring these strands together to examine broader patterns of development, discuss their emerging applications, and identify promising directions for future research.

\begin{figure*}[t]
    \centering
    \includegraphics[width=1\linewidth]{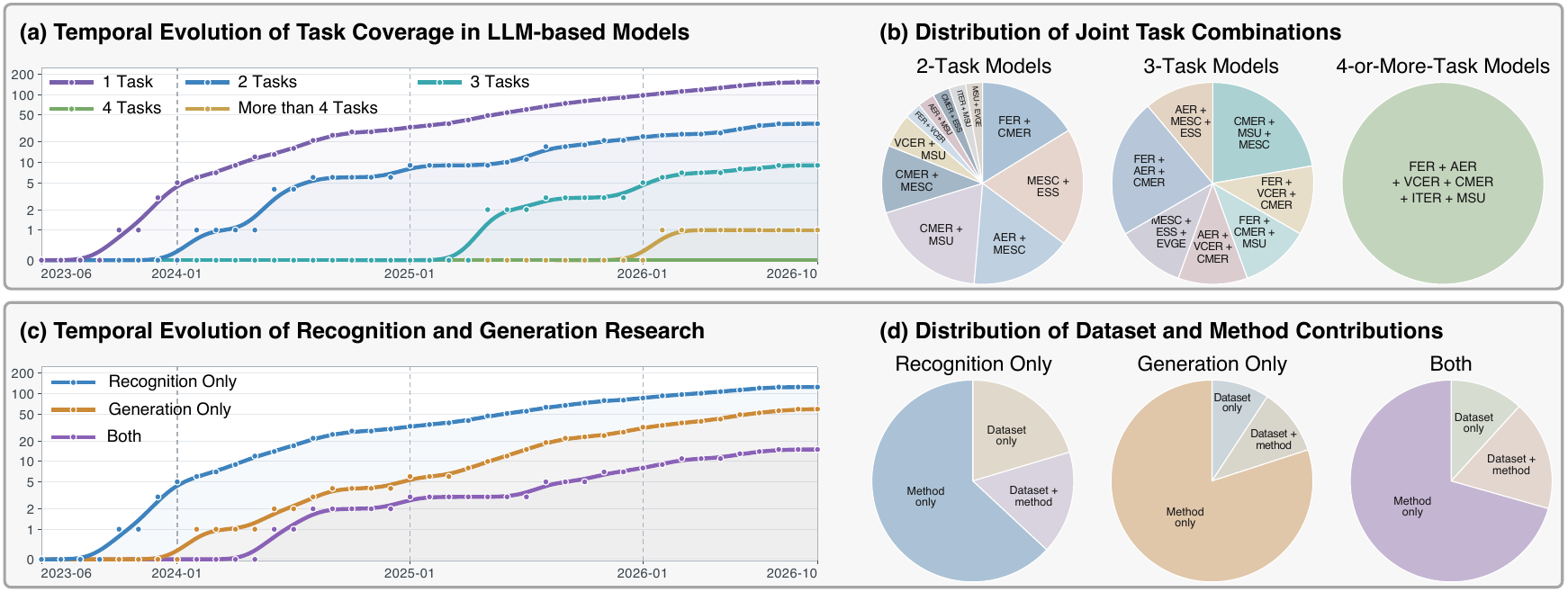}
    \vskip -0.1in
    \caption{Temporal evolution and task coverage of LLM-based affective research. (a)-(b) show cumulative paper counts by the number of tasks covered and the distribution of joint task combinations. (c)-(d) present cumulative paper counts for recognition, generation, and their intersection, alongside the proportions of dataset-only, method-only, and combined contributions within each category. The vertical axes in (a) and (c) use a symmetric logarithmic scale.}
    \vskip -0.1in
    \label{fig:7-1}
\end{figure*}

\subsection{Comprehensive Analysis}

The preceding sections examined methodological distributions and quantitative performance within specific settings. \Cref{fig:7-1} brings these observations together, tracing changes in task coverage and the development of recognition, generation, and their supporting data resources. Two trends stand out: methods increasingly address multiple tasks, and more work brings recognition and generation into a shared framework.

\Cref{fig:7-1} (a) shows steady growth in multi-task methods, with increasingly broad task coverage emerging over time. The combinations in \cref{fig:7-1} (b) reveal connections among related recognition tasks and between emotional support and speech generation. Broader combinations encompass several forms of affective understanding and extend into expressive output. Alongside the expanding task coverage discussed in the dataset section, these developments show a growing capacity to study and model affective capabilities within shared settings.

This expansion also brings understanding and expression into closer connection. \Cref{fig:7-1} (c) traces the growth of recognition research, the subsequent expansion of generation methods, and the increasing number of approaches spanning both. Although method-only papers form the largest share in each category, dedicated dataset contributions and work combining datasets with methods together account for a substantial proportion, as shown in \cref{fig:7-1} (d). This distribution reflects sustained attention to task formulation and data construction alongside methodological advances. Taken together, these developments point toward more unified modeling of affective understanding and generation, laying the groundwork for their coordinated use in real-world applications.

\subsection{Applications}

\textbf{Social Media Understanding and Moderation.}
Over the past two decades, social media has become woven into everyday life, providing spaces to share experiences, express opinions, and maintain relationships. Platforms such as Facebook, Instagram, WeChat, and Weibo collectively serve billions of users, offering both rich multimodal data and practical settings for affective applications~\cite{meta24results}. Understanding the feelings and intentions behind posts can help businesses respond to customer concerns, creators interpret audience reactions, and moderators distinguish playful exchanges from potentially abusive content. Such understanding can also inform direct interaction: Weibo's \textit{Comment Robert}~\cite{shi25commentrobert} is designed to recognize emotional cues and implicit meaning in users' posts, enabling replies that respond to their concerns rather than merely matching the topic. Beyond interpreting users' feelings, generative tools help convey them in the content they share. With Meta AI's video editor~\cite{meta25videoediting}, users can reshape a clip's visual mood to express the feeling behind a personal moment, giving a shared memory a playful, nostalgic, or reflective tone.

\textbf{Creative Media and Content Production.}
The expressive choices that shape individual social media posts take on a broader role in the production of films, advertisements, audiobooks, and animation. Emotion is central to these media, shaping atmosphere, giving characters depth, and drawing audiences into a story. Generative tools make it easier to experiment with expressive choices without recording or animating each version from scratch. ElevenLabs' Eleven v3~\cite{elevenlabs25audiotags} takes performance cues directly within a script to guide emotional tone, pacing, and nonverbal sounds. The same line can thus convey excitement, uncertainty, or quiet restraint, giving creators control over its delivery as well as its wording. NVIDIA's Audio2Face~\cite{nvidiaAudio2Face} extends this process to visual performance by generating facial animation from speech, helping align a character's expression with its voice. These capabilities fit naturally into an iterative production workflow, in which creators refine performances to suit a scene, a character, or a broader artistic vision.

\textbf{Emotional Support and Mental Well-Being.}
Affective understanding and communication are closely connected to mental health and well-being. More than one billion people worldwide live with mental health conditions, yet the global median workforce comprises only about 13 mental health workers per 100,000 people, highlighting a substantial gap between need and available care~\cite{who25mentalhealth}. Against this backdrop, as AI becomes part of everyday life, people are turning to conversational systems to share personal concerns and seek support~\cite{mcbain26chatbotuse}. Their ready availability offers an accessible point of contact when professional help is difficult to obtain, bringing AI's capacity to recognize distress and respond sensitively into sharper focus. Multimodal interaction adds an important dimension, since hesitation, vocal tension, and conversational context may convey feelings that users struggle to put into words. Jammy~\cite{wang25jammy} puts this idea into practice by using Hume's Empathic Voice Interface to inform music recommendations and reflective prompts from vocal cues. More clinically focused work has begun to evaluate whether supportive exchanges translate into measurable benefits: the text-based Therabot~\cite{heinz25therabot} reported greater symptom reductions than a waitlist control in a randomized trial with clinical oversight. These developments point toward multimodal support that combines an understanding of spoken concerns, vocal expression, and visible behavior with responses attuned to the individual's emotional needs.

\textbf{Embodied Interaction.}
Rapid advances in robotics are bringing AI beyond digital interfaces into the physical world. Service robots embody a familiar vision from fiction: machines that understand human intentions, respond naturally, and perform useful tasks. MER-with-LLMs and MEG-with-LLMs provide complementary foundations, from interpreting affective cues in audiovisual observations to generating expressive speech and behavior. Yet understanding people in recorded videos differs from interacting with them in a shared space. A robot must judge whether assistance is needed and whether its actions would be welcome, accounting for its own presence and social context. RobotEQ~\cite{fang26roboteq} examines this distinction through socially appropriate action judgments and spatial grounding without explicit user commands. RobotEQ-Video~\cite{che26roboteqvideo} extends the assessment to dynamic situations, drawing on temporal and acoustic cues to contextualize human needs. AffectLoop~\cite{pang26affectloop} brings these concerns into physical interaction on a Misty II robot, coordinating speech and expressive behavior through an evolving model of user and robot affect. These efforts mark an emerging research direction toward the long-envisioned prospect of socially and emotionally responsive robots in everyday life.

\subsection{Future Trends}
Progress in MER and MEG is bringing individual capabilities of affective intelligence to life, one after another. Yet gaps remain between these capabilities and their effective use in real-world applications. We outline several promising future directions for bridging these gaps below.

\textbf{End-to-End Task Formulation.}
Compelling task formulations and well-designed benchmarks can give a field both direction and momentum. In current research, many recognition tasks end with an emotion label or explanation, and generation tasks often evaluate expressive output without the broader purpose it serves. In practice, emotional understanding informs decisions, and emotional expression contributes to goals such as resolving confusion, sustaining engagement, or offering support. This gap calls for end-to-end tasks that connect perception, response selection, and generation to an observable outcome. MESC offers a useful example by bringing these capabilities into a shared context of supportive conversation, making their practical value assessable through a coherent task. Further opportunities lie in task frameworks organized around applications such as adaptive tutoring and collaborative assistance, or mental-state modeling processes such as tracking changing beliefs and anticipating emotional reactions to possible actions.

\textbf{Deeper Attribution of Emotional Experience.}
Another gap separates explanations of a model's decision from explanations of a person's emotional experience. Explainable MER can identify the facial, vocal, or contextual evidence supporting a judgment, making predictions easier to examine. Responding appropriately, however, may require a deeper understanding of the experiences that give rise to the emotion, beyond the cues supporting its recognition. Deeper attribution would connect emotional experience with personal goals, unmet expectations, relationships, and prior events, allowing similar expressions to inform different responses according to their underlying circumstances. Progress in this direction requires both a finer-grained taxonomy of emotional causes, organized into meaningful hierarchies, and a mapping from causes at different scales and levels of abstraction to observable indicators or explicitly defined concepts. These foundations could provide concrete supervision for aligning models with the underlying processes of emotional experience.

\textbf{Multi-Turn Affective Interaction and Role Simulation.}
Multi-turn affective interaction deserves particular attention within end-to-end task formulation, as it represents a broad spectrum of affective capabilities in use, from understanding emotional cues and planning responses to generating multimodal expression. Its difficulty lies in the evolving nature of the interaction, where each response reshapes the context for subsequent understanding and action. One promising route moves beyond imitating recorded conversations toward autonomous interaction with simulated users whose goals, emotional states, and willingness to engage evolve in response to the system. Fixed demonstrations cover a limited set of trajectories and may omit situations encountered by the learned policy. Interactive simulation can expose such situations, support recovery from misunderstandings, and enable RL with feedback on longer-term outcomes. This approach faces two substantial challenges: simulated users may not faithfully reflect human emotional responses, and extending their behavior beyond text to coherent vocal, facial, and bodily expression remains difficult. Beyond simulation, human-in-the-loop learning offers another potentially viable route, grounding adaptation in feedback from actual interactions.

\textbf{Mental World Modeling.} As AI moves from digital environments into the physical world, it is increasingly expected to interact with its surroundings as well as understand them. World models provide a promising foundation for this transition by predicting how actions alter the environment, with mental world modeling extending this predictive perspective to the people within it~\cite{fei26mentalworld}. The aim is to anticipate how people respond to different stimuli across modalities and time horizons, allowing a system to weigh the likely consequences of alternative responses before acting. In high-stakes interactions, this capability could help screen out responses that might aggravate psychological distress or reinforce self-harm intentions. In everyday exchanges, it could guide more individually appropriate choices, accounting for how personal habits and preferences shape the reception of an otherwise identical suggestion, reassurance, or gesture. Building such models remains challenging because many aspects of human cognition lack settled explanations, offering a less clearly established theoretical basis for predicting psychological responses than physical dynamics. Developing suitable evaluation frameworks, modeling architectures, and optimization algorithms under these conditions represents an important direction for future research.

\textbf{Personalized Affective Assistants.}
Looking further ahead, personalized affective assistants offer a path from occasional emotional understanding to sustained individual support. Personalization has long been a research concern in affective computing~\cite{zhao16personalized}, but recent developments in personal AI agents bring renewed promise to its practical realization. Meta's Muse, for example, combines personal memory with ongoing assistance across everyday tasks, illustrating an emerging setting for sustained adaptation~\cite{meta26muse}. One possible paradigm builds on a capable general backbone and progressively adapts its behavior as personal context accumulates. This adaptation may operate through memory, retrieval, and interaction policies in the surrounding agent harness, or through selective parameter updates that internalize experience within the model. These complementary routes create new opportunities and challenges for translating accumulated personal context into increasingly accurate and responsive affective understanding.

\textbf{Proactive Affective Interaction.}
Proactive affective interaction is another important but underexplored capability, allowing assistance to begin before a user explicitly requests it. Detecting frustration or hesitation could prompt a timely offer of help, a change in explanation, or a pause. Conventional prompt--response deployments provide no inherent mechanism for deciding when to initiate such actions, requiring additional observation, timing, and intervention policies. These decisions are especially delicate in affective settings: a correctly perceived emotion does not automatically justify a response, and an unsolicited intervention may interrupt or misread the situation. General-purpose approaches based on streaming video and event-triggered interaction offer possible foundations, but continuous processing introduces computational costs and privacy concerns. Dedicated tasks and datasets could make selective initiative measurable by distinguishing appropriate intervention from clarification or silence, with success judged by both timing and the consequences for the user.

\subsection{Ethical Considerations}
As affective computing is inseparable from human experience and interaction, ethical considerations are central to its research and application. Facial, vocal, and physiological data can reveal sensitive aspects of a person's emotional life, making informed consent and clear limits on collection, retention, and use essential. These safeguards should also extend to inferred emotional states, giving users meaningful control over how such information informs system behavior. For emotionally responsive dialogue systems, real-world deployment requires safeguards that ensure responses match users' circumstances, especially when they seek psychological support. Deployment also requires attention to context-specific legal restrictions, including the EU AI Act~\cite{eu24aiact}'s prohibition of biometric emotion inference in educational institutions and workplaces, subject to exceptions for medical or safety purposes. Applications such as affect-aware tutoring therefore require careful assessment of their intended use, data sources, and applicable safeguards. For generation beyond text, synthesizing identifiable voices and faces requires appropriate consent, respect for likeness rights, and transparent disclosure, alongside safeguards against impersonation and emotional manipulation. As AI continues to evolve, how people and AI coexist will remain an ongoing societal discussion, and the responsible use of affective capabilities should remain guided by applicable regulations and evolving ethical standards.

\section{Conclusion}
\label{sec:8}
Taken together, this paper provides a comprehensive review of the emerging MER-with-LLMs and MEG-with-LLMs paradigms. The review unfolds through a challenge-oriented perspective, organizing methods across tasks by the problems they address and examining them together to reveal shared principles and methodological connections. A field-wide analysis then brings these findings together, leading to a discussion of applications and future research directions. By synthesizing and categorizing existing work, identifying common principles, and clarifying methodological differences, this survey offers a structured foundation for further research, with the hope of advancing multimodal affective computing toward real-world deployment and applications.

%\clearpage
\FloatBarrier
\bibliographystyle{IEEEtran}
\bibliography{main}

\vfill

\end{document}